\documentclass[12pt,letterpaper]{article}
\usepackage[usenames, dvipsnames, table]{xcolor}
\usepackage[pagebackref=true]{hyperref}
\usepackage[margin=1.0in]{geometry}
\usepackage{cite}

\usepackage{tocloft}
\usepackage[]{todonotes}
\usepackage{bbold}
\usepackage{graphicx}
\usepackage{verbatim}
\usepackage[mathscr]{euscript}
\usepackage{slashed}
\usepackage{graphicx}
\usepackage{mathdots}
\usepackage{caption}

\usepackage[labelformat=parens, subrefformat=simple]{subcaption}

\usepackage{enumerate}
\usepackage{bbm}
\usepackage{psfrag}
\usepackage{subfiles}
\usepackage{psfrag}
\usepackage{relsize}
\usepackage{pgfplots}
\usepackage{tikzscale}

\usepackage{bbm} 					
\usepackage{slashed} 				
\usepackage{graphicx}				
\usepackage{subcaption}			
\usepackage{psfrag}				
\usepackage{tensor}				
\usepackage{fouridx}				
\usepackage{bm}					
\usepackage{mdframed}				
\usepackage{multirow}				
\usepackage{soul}					
\usepackage{bbold}				
\usepackage{multicol}				

\usepackage{amsmath}
\usepackage{amssymb}
\usepackage{amsthm}

\numberwithin{equation}{section}

\usepackage{mathtools}

\usepackage{feynmf}
\usepackage{marvosym}

\usepackage{import}

\newtheoremstyle{named}{}{}{\itshape}{}{\bfseries}{.}{.5em}{#3}
\theoremstyle{named}

\newtheoremstyle{unnamed}{}{}{\itshape}{}{\bfseries}{.}{.5em}{\thmname{#1}\thmnumber{ #2}}
\theoremstyle{unnamed}
\newtheorem{conj}{Conjecture}

\newtheoremstyle{namedandnumbered}{}{}{\itshape}{}{\bfseries}{.}{.5em}{\thmname{#1}\thmnumber{ #2} (\thmnote{#3})}
\theoremstyle{namedandnumbered}
\newtheorem{conjabc}{Conjecture}

\theoremstyle{definition}
\newtheorem*{definition}{Definition}
\newtheorem*{example}{Example}

\usepackage{diagbox}

\newcommand*\xoverline[2][0.75]{%
    \sbox{\myboxA}{$\m@th#2$}%
    \setbox\myboxB\null
    \ht\myboxB=\ht\myboxA%
    \dp\myboxB=\dp\myboxA%
    \wd\myboxB=#1\wd\myboxA
    \sbox\myboxB{$\m@th\overline{\copy\myboxB}$}
    \setlength\mylenA{\the\wd\myboxA}
    \addtolength\mylenA{-\the\wd\myboxB}%
    \ifdim\wd\myboxB<\wd\myboxA%
       \rlap{\hskip 0.5\mylenA\usebox\myboxB}{\usebox\myboxA}%
    \else
        \hskip -0.5\mylenA\rlap{\usebox\myboxA}{\hskip 0.5\mylenA\usebox\myboxB}%
    \fi}
\makeatother

\newcommand{\zel}{z_{\rm el}}
\newcommand{\zmag}{z_{\rm mag}}
\newcommand{\zvel}{{\vec z}_{\rm el}}
\newcommand{\zvmag}{{\vec z}_{\rm mag}}
\newcommand{\Tel}{T_{\rm el}}
\newcommand{\Tmag}{T_{\rm mag}}
\newcommand{\gel}{g_{\rm el}}
\newcommand{\gmag}{g_{\rm mag}}

\newcommand{\zvperp}{{\vec z}_\bot}
\newcommand{\dv}{{\vec d}}

\newcommand{\ZZ}{\mathbb{Z}}
\newcommand{\tA}{\tilde{A}}

\newcommand{\tK}{\tilde{K}}
\newcommand{\tq}{\tilde{q}}

\newcommand{\cC}{\mathcal{C}}

\newcommand{\cE}{\mathcal{E}}
\newcommand{\cF}{\mathcal{F}}

\newcommand{\cK}{\mathcal{K}}

\newcommand{\cM}{\mathcal{M}}

\newcommand{\cS}{\mathcal{S}}

\newcommand{\cV}{\mathcal{V}}

\newcommand{\vol}{\mathrm{vol}}

\newcommand{\LQG}{\Lambda_{\rm QG}}

\definecolor{cobalt}{RGB}{44, 98, 120}
\definecolor{celadon}{rgb}{0.67, 0.88, 0.69}
\definecolor{dm}{cmyk}{.20, 0, .30, 0}
\definecolor{burgundy}{rgb}{0.5, 0.0, 0.13}
\definecolor{plotBlue}{RGB}{94, 130, 181}

\newcommand{\rmd}{\textrm{d}}
\def\be{\begin{equation}}
\def\ee{\end{equation}}
\def\bea{\begin{eqnarray}}
\def\eea{\end{eqnarray}}

\hypersetup{
  colorlinks,
  citecolor=Violet,
  linkcolor=cobalt,
  urlcolor=Blue}

\newif\iffastcompile

\fastcompilefalse

\iffastcompile
\newcommand{\js}[1]{}
\newcommand{\jsi}[1]{}
\newcommand{\cl}[1]{}
\newcommand{\lm}[1]{}
\else
\newcommand{\js}[1]{\todo[color=cobalt!30,size=\scriptsize, bordercolor=cobalt!30]{JS: #1}}
\newcommand{\jsi}[1]{\todo[color=cobalt!30,size=\scriptsize, bordercolor=cobalt!30, inline]{JS: #1}}
\newcommand{\cl}[1]{\todo[color=burgundy!30, size=\scriptsize, bordercolor=burgundy!30]{CL: #1}}
\newcommand{\lm}[1]{\todo[color=dm!90, size=\scriptsize, bordercolor=dm!90]{LM: #1}}
\fi

\ProvideTextCommandDefault{\Dbar}{%
\leavevmode\lower.5ex\rlap{\hskip-.07em\accent"16}D%
}

\newcommand{\orthogcomplement}{\bot}

\renewcommand{\ker}{\operatorname{Ker}}

\newcommand{\brap}[1]{{\left( {#1} \right)}}
\newcommand{\bras}[1]{{\left[ {#1} \right]}}
\newcommand{\brac}[1]{{\left\{ {#1} \right\}}}

\newcommand{\brav}[1]{{\left\vert {#1} \right\vert}}
\newcommand{\bravv}[1]{{\left\Vert {#1} \right\Vert}}
\graphicspath{{./fig}} 

\newcommand{\cEperp}{\cE_{\text{perp}}}
\newcommand{\cEker}{\cE_{\text{ker}}}
\newcommand{\cElimdir}{\cE_{0}}

\newcommand{\cMtau}{\cM_{\text{parallel}}}
\newcommand{\cMkerorth}{\cM_{\text{ker}^\bot}}
\newcommand{\cMlight}{\cM_{\text{L}}}

\begin{document}
	\newcommand{\main}{.}
\begin{titlepage}

\setcounter{page}{1} \baselineskip=15.5pt \thispagestyle{empty}

\bigskip\

\bigskip\

\vspace{2cm}
\begin{center}
{\LARGE \bfseries Co-Scaling and Alignment of Electric and Magnetic Towers}

 \end{center}
\vspace{0.5cm}

\begin{center}
{\fontsize{14}{30}\selectfont Matthew Reece$^a$, Tom Rudelius$^b$, and Christopher Tudball$^b$}
\end{center}

\begin{center}

\vspace{0.25 cm}
\textsl{$^a$Department of Physics, Harvard University, Cambridge, MA, 02138, USA}\\
				\textsl{$^b$Department of Mathematical Sciences, Durham University, Durham, DH1 3LE, UK}\\

\end{center}

\vspace{1cm}

\noindent 
Towers of electrically and magnetically charged states in quantum gravity often exhibit two important properties. First, the ratio of the mass (or tension) of electrically charged states to magnetically charged states is of order $e^2/(4\pi)$, which we refer to as ``co-scaling.'' Second, in theories of multiple gauge fields, the towers of states that exhibit co-scaling have charges that point in approximately the same direction in charge space as measured by the gauge kinetic matrix, which we refer to as ``alignment.'' After motivating these ideas with some heuristic arguments, we examine the spectrum of BPS states in the 5d supergravity landscape arising from M-theory on a Calabi-Yau threefold. In this setting, every tower of magnetically charged strings is paired with a corresponding tower of electrically charged particles that exhibits co-scaling and rapid alignment. In particular, this motivates a sharp mathematical characterization of the magnetic infinity cone in Calabi-Yau geometry. We propose a universal conjecture about quantum gravity: towers of charged states which, in some limit in moduli space, have maximally divergent charge-to-mass ratios always have corresponding magnetic partner states exhibiting co-scaling and alignment. Co-scaling is not a general feature of extremal black hole solutions in theories of gauge fields and scalars, suggesting that it is a principle of UV complete quantum gravity. We briefly remark on possible phenomenological applications, including to axion physics.

 \vspace{1.1cm}

\bigskip
\noindent\today

\end{titlepage}
\setcounter{tocdepth}{2}
\tableofcontents

\section{Introduction}\label{INTRO}

\subsection{Co-scaling and alignment: context and definitions}\label{ss.coscale}

Quantum gravity theories predict {\em towers}, infinite families of particles (or extended objects) of increasing mass and (often) increasing charge or spin. Familiar examples include Kaluza-Klein modes and towers of perturbative string states with oscillator modes excited. Far up in the spectrum, these towers include states corresponding to classical black hole or black brane solutions. 
In recent years, substantial progress has been made toward understanding the nature and spectrum of certain special towers of states that become arbitrarily light (relative to the Planck scale) in asymptotic (infinite-distance) regions of moduli space~\cite{Ooguri:2006in}. In fact, there are strong arguments that the existence of infinite-distance limits is linked to the existence of towers through the phenomenon of emergence: the kinetic terms of moduli fields diverge in regions of moduli space where a tower of states becomes light, and can be thought of as being generated by integrating out the towers~\cite{Grimm:2018ohb,Heidenreich:2018kpg}. The towers that are best understood are those that are lightest, and contain the largest number of states, in some infinite-distance limit~\cite{Ooguri:2006in,Heidenreich:2015nta,Heidenreich:2016aqi,Grimm:2018ohb,Heidenreich:2018kpg,Blumenhagen:2018nts,Grimm:2018cpv,Lee:2018urn,Lee:2018spm,Corvilain:2018lgw,Lee:2019xtm,Marchesano:2019ifh,Font:2019cxq,Lee:2019wij,Gendler:2020dfp,Lanza:2021udy,Lanza:2022zyg,Etheredge:2022opl,vandeHeisteeg:2023ubh,Etheredge:2023odp,Etheredge:2023usk,Rudelius:2023odg,Bedroya:2024ubj,Friedrich:2025gvs}. In the case where a gauge coupling goes to zero in this limit, states in these towers also satisfy the Weak Gravity Conjecture (WGC)~\cite{Arkanihamed:2006dz, Heidenreich:2015nta, Heidenreich:2016aqi, Montero:2016tif, Heidenreich:2017sim, Andriolo:2018lvp, Heidenreich:2024dmr}. An important open problem, with potential phenomenological applications, is to characterize other, subleading towers of states~\cite{Etheredge:2024tok,Grieco:2025bjy,Etheredge:2025ahf}.

In this paper, we explore relationships between states that are electrically and magnetically charged under the same gauge symmetry. We will argue that the charge-to-mass (more generally, charge-to-tension) ratio vectors of an electric tower and a corresponding magnetic brane often exhibit two special relationships that we refer to as {\em co-scaling} and {\em alignment}. The meaning of ``often,'' and more precise conjectures about both quantum gravity and mathematics, will be explained below. To fix notation, consider a theory of U(1) $p$-form gauge fields $A^i$ with field strengths $F^i \equiv \rmd A^i$ and kinetic terms
\begin{equation}
-\frac{1}{2} \int K_{ij}(\phi) F^i \wedge \star F^j,
\end{equation}
where the kinetic matrix $K_{ij}$ is a function of a family of moduli fields $\phi_a$. A $(p-1)$-brane carries electric charges $q_i$ under the gauge fields if it has a worldvolume coupling $\int q_i A^i$. Similarly, we define dual $(d-p-2)$-form gauge fields $\tA_i$ with kinetic matrix $\tK \equiv (2\pi)^{-2} K^{-1}$, i.e., we have $\tK^{ij}K_{jk} = (2\pi)^{-2} \delta^i_k$ (the Dirac quantization condition). Magnetically charged $(d-p-3)$-branes are then labeled by $\tq^i$ and have worldvolume coupling $\int \tq^i \tA_i$. It is often convenient to define the gauge theory in an integral basis where $q_i, \tq^i \in \ZZ$, but we will also find it convenient later to work in a basis in which (for some chosen point in moduli space) $K_{ij}$ is the identity. We define the charge-to-mass ratio vectors $\zvel$ of an electrically charged $(p-1)$-brane of tension $\Tel$ and $\zvmag$ of a magnetically charged $(d-p-3)$ brane of tension $\Tmag$, and their corresponding norms as measured by the kinetic terms:
\begin{align}
({\zel})_i = \frac{1}{\Tel} q_i, & \quad \lVert q\rVert^2 = q_i (K^{-1})^{ij} q_j = (2\pi)^2 q_i \tK^{ij} q_j, \nonumber \\
({\zmag})^i = \frac{1}{\Tmag} \tq^i, & \quad  \lVert \tq \rVert^2 = \tq^i (\tK^{-1})_{ij} \tq^j = (2\pi)^2 \tq^i K_{ij} \tq^j.  \label{eq:zdefinition}
\end{align}
The values of $\zvel$ and $\zvmag$ are basis-dependent but their norms are invariant. (Note that, for convenience, we will often refer to ``charge-to-mass ratio'' even when ``charge-to-tension ratio'' or ``charge-to-action ratio'' is more appropriate.) In the case of a tower of states with increasing charge and (proportional) mass, we define these vectors by their asymptotic values high in the tower. Below, we will often use the notation 
\begin{equation}
\gel = \lVert q \rVert, \quad \gmag = \lVert \tq \rVert
\end{equation}
for electric and magnetic couplings, especially when there is a single U(1) and we can focus on states of unit charge.

We now have the ingredients to define the primary concepts that we study in this paper. First, co-scaling:
\begin{definition}[Co-Scaling] \label{def:coscaling} We say that states exhibit {\em co-scaling} when the norms of their charge-to-mass ratio vectors $\lVert z \rVert$ agree within an $O(1)$ factor, even in a limit where the value of $\lVert z \rVert$ becomes arbitrarily large. 
\end{definition}
In particular, we will be focused on co-scaling of electrically and magnetically charged states, i.e., cases in which
\begin{equation} \label{rr}
   \lVert\zel\rVert= \frac{\gel}{\Tel} =  \frac{\sqrt{q_i \left(K^{-1}\right)^{ij} q_j}}{\Tel} \sim \lVert \zmag \rVert = \frac{\gmag}{\Tmag} = 2\pi  \frac{\sqrt{\tq^i K_{ij} \tq^j}}{\Tmag}.
\end{equation}
Let us give two familiar examples of electric and magnetic states that exhibit co-scaling:
\begin{example}['t~Hooft-Polyakov] When higgsing the group SU(2) to U(1), the $W$ boson has mass $m_W = g v$ while the 't~Hooft-Polyakov monopole has $m_M = 4\pi v/g$~\cite{tHooft:1974kcl,Polyakov:1974ek}. The U(1) gauge coupling (normalized so the smallest charge is 1) is $e = g/2$, while the $W$ boson has charge $q = 2$ and the magnetic monopole has $\tq = 1$. Correspondingly, we have charge-to-mass ratios $\lVert z_W \rVert = \lVert z_M \rVert = 1/v$, independent of $g$ even as $g \to 0$. Note that this is an example of co-scaling for specific particles, rather than towers of states. For a geometrical interpretation of this co-scaling relationship, see~\cite{Halverson:2015vta}.
\end{example}
\begin{example}[Extremal dilatonic black holes] In a gravitational theory in $d$ spacetime dimensions with a canonically normalized scalar field $\phi$ that couples dilatonically to the kinetic term of a $p$-form gauge field, 
\begin{equation}
-\frac{1}{2e^2} \int d^dx\,\sqrt{-g}\,\exp\left[-\sqrt{2\kappa_d^2} \alpha \phi\right] F_{p+1} \wedge \star F_{p+1},
\end{equation}
both electrically and magnetically charged extremal black holes have charge-to-mass ratio vectors of the same length~\cite{Myers:1986un, Horowitz:1991cd, Lu:1993vt, Duff:1993ye, Duff:1996hp},
\begin{equation} 
\lVert z \rVert^2 = \left[\frac{\alpha^2}{2} + \frac{p(d-p-2)}{d-2}\right] \kappa_d^2.   \label{eq:zdilatonic}
\end{equation}
This is an example of perfect co-scaling for an infinite tower of states, even for arbitrarily large $\alpha$, where $\lVert z \rVert \to \infty$. (We are not, however, aware of realizations of large-$\alpha$ dilatonic couplings in string theory.)
\end{example}

The concept of electric-magnetic co-scaling in quantum gravity (but not the term ``co-scaling'' itself) was recently proposed in~\cite{Reece:2024wrn}, primarily in the context of axion physics. In this paper, we will sharpen the hypothesis that towers of electric and magnetic states exhibiting co-scaling exist, finding counter-examples to various candidate sharp conjectures. We will also resolve a puzzle raised in~\cite{Reece:2024wrn} regarding the charges of states exhibiting co-scaling, which were sometimes non-obvious. For example, in one example of a theory with two axions, states of (electric) instanton charge $(0,1)$ and (magnetic) axion string charge $(-1,3)$ exhibited co-scaling. The reason why one should consider this string, rather than simply the $(0,1)$ string, was left unclear. In this paper, we resolve this question by observing that the electric and magnetic charges in question exhibit the phenomenon of {\em alignment}.

\begin{definition}[Alignment and Rapid Alignment] \label{def:alignment} An electric state with charge-to-mass ratio vector $\zel$ and a magnetic state with charge-to-mass ratio vector $\zmag$ that exhibit co-scaling are said to also exhibit {\em alignment} if $\zel$ and $\zmag$ asymptotically point in the same direction (as measured with the metric defined by $K_{ij}$), i.e., if
\begin{equation}
\cos \varphi = \frac{2\pi q_i \tq^i}{\lVert q \rVert \, \lVert \tq \rVert} \to 1 \quad \text{as} \quad \lVert z \rVert \to \infty.   \label{eq:cosphialign}
\end{equation}
We further say that these states exhibit {\em rapid alignment} if the angle between the charge-to-mass ratio vectors vanishes at least as fast as their inverse length:
\begin{equation}
| \varphi | \lesssim {\lVert z \rVert}^{-1} \quad \text{as} \quad \lVert z \rVert \to \infty.   \label{eq:rapidalign}
\end{equation}  
\end{definition}

It is straightforward to check that the example of~\cite{Reece:2024wrn} with a $(0,1)$ instanton and $(-1,3)$ axion string exhibits rapid alignment. In this paper we will see several other examples of alignment, both rapid and not.

Our definition of alignment explicitly refers to a limit in which $\lVert z \rVert$ can be made arbitrarily large. One could instead define alignment at an arbitrary point in moduli space as the condition that $\varphi \lesssim \lVert z \rVert^{-\gamma}$ for some chosen exponent $\gamma > 0$ (equal to $1$ for rapid alignment). This definition, like that of co-scaling, contains an ambiguous $O(1)$ prefactor, which one might hope to make precise in future work. However, because our focus in this paper is on {\em towers} of states, we expect that the restriction to cases where $\lVert z \rVert$ can be taken to infinity can be made without loss of generality. For an individual particle, we could imagine the mass becoming accidentally small (and thus the charge-to-mass vector accidentally long) at an innocuous point in the interior of moduli space, where different contributions happen to cancel. For this to happen to every state in a tower, however, would be a remarkable accident. Our intuition, then, is that the charge-to-mass vector of a tower can become long only when there is a singularity or boundary in moduli space where the entire tower becomes massless.

Towers of both electrically and magnetically charged states are expected to exist in quantum gravity. Aside from black holes themselves, we expect states of smaller mass and charge. The Weak Gravity Conjecture posits the existence of a superextremal state, i.e., a particle (or brane) whose charge to mass (or tension) ratio is greater than or equal to that of a large, extremal black hole~\cite{Arkanihamed:2006dz}:
\begin{equation}
\lVert z \rVert \geq \lVert z_\textsc{BH} \rVert \sim M_{\text{Pl};d}^{\frac{2-d}{2}}
\label{WGC}
\end{equation}
(in $d$ spacetime dimensions), where the $\sim$ estimate holds at least in the absence of scalar forces much stronger than gravity. It is often the case that the particles (or branes) satisfying the WGC come in infinite towers~\cite{Arkanihamed:2006dz, Heidenreich:2015nta, Heidenreich:2016aqi, Montero:2016tif, Andriolo:2018lvp} (but perhaps not always; see~\cite{Cota:2022yjw,FierroCota:2023bsp}). The WGC together with the existence of towers of black hole states leads us to expect that in generic directions in charge space, one finds towers of both electrically and magnetically charged objects with $\lVert z \rVert \sim M_{\text{Pl};d}^{\frac{2-d}{2}}$, while only in certain special directions might one find towers of much larger $\lVert z \rVert$. A generic expectation is that these are directions with strong scalar forces. Extremal black holes are characterized by a no-force condition: the repulsive gauge force between two identically charged black holes is compensated by the attractive gravitational and scalar forces. In order for the gauge force to be much stronger than gravity, it must be balanced by a strong attractive scalar force. Similarly, a variant on the WGC calls for the existence of self-repulsive charged particles~\cite{Arkanihamed:2006dz,Palti:2017elp,Heidenreich:2019zkl}. If a tower nearly saturates this bound and has large $\lVert z \rVert$, the repulsive gauge force is approximately balanced by attractive scalar forces. The study of towers of large $\lVert z \rVert$ is thus closely related to the study of scenarios where scalar forces dominate over gravity. Intuitively, co-scaling and alignment might arise when scalar forces interact strongly with the gauge fields, since the same gauge fields are turned on in the background of both electrically and magnetically charged objects. Indeed, we saw a precise realization of this intuition in the example of dilatonic extremal black holes.

\subsection{Conjectures}\label{ss.introconj}

In this paper we will analyze a number of scenarios and find that co-scaling and alignment (in many cases rapid) of electrically and magnetically charged states are ubiquitous in consistent quantum gravity theories. Formulating a sharp and useful version of this observation is somewhat challenging. For example, it is not the case that {\em any} electrically charged state is necessarily accompanied by a co-scaling magnetically charged state. If this were true, a magnetic monopole with mass of order $m_e / \alpha \sim 50$ MeV would exist in our universe, contrary to observations. This particular example, previously mentioned in~\cite{Reece:2024wrn}, hints that we should focus on towers of states rather than isolated light particles. Before we hazard any precise statement regarding when co-scaling and alignment occur in the quantum gravity landscape, we emphasize that our work is exploratory. The notions of co-scaling and alignment appear to be useful tools for thinking about the spectrum of objects in a theory, and (as we will see below) they lead to new ideas about phenomenology and mathematics. The main message of our work---that these are productive definitions to think about---stands independent of the precise details of the conjectures that follow. 

A general statement consistent with all the evidence that we have is the following. Consider a limit in moduli space in which some towers of charged states have divergent charge-to-mass ratio vectors $\vec z$. We parametrize the limit as $s \to 0$ along some path, and find towers of states with $\lVert z \rVert \propto s^{-\gamma}$ for some $\gamma > 0$. In general, there may be multiple towers with $\lVert z \rVert$ diverging at different rates. We focus on the towers exhibiting the parametrically maximal rate of divergence, i.e., the set of towers that share the largest $\gamma$ (independent of the prefactor). We then conjecture: 
\begin{conjabc}[Co-Scaling and Alignment of Maximally Divergent Electric and Magnetic Towers]\label{ConA}
Consider a limit in moduli space in which there are towers of charged states with divergent charge-to-mass ratio vectors. For any electric (magnetic) tower exhibiting the maximal rate of divergence in this limit, there will be a corresponding magnetic (electric) brane that co-scales and aligns with it.
\end{conjabc}
In fact, the available evidence is consistent with a stronger statement. An electric tower exhibiting the maximal rate of divergence may co-scale not only with a magnetic brane of charge $q$, but also with either (a) a tower of magnetic branes of charge $k q$, $k \in \mathbb{Z}$ or (b) a tower of magnetic branes of charge $kq + q'$, $k \in \mathbb{Z}$ for some charge $q'$. We refer to the latter possibility as a \emph{skew tower}. A simple example of a skew tower arises in the circle reduction of a particle of charge $q$ under a $U(1)$ gauge field; after reduction, this particle gives rise to a skew tower of KK modes of charge $(q, n)$, where $n \in \mathbb{Z}$ is the charge under the KK photon. Geometrically, the convex hull generated by a tower is identical to that of a skew tower.

The restriction to towers exhibiting the \emph{maximal} rate of divergence is important; in~\S\ref{ss.GS}, we will see an example where there are electric towers with $\lVert z \rVert$ diverging as $s^{-1/2}$ and $s^{-3/2}$, and no magnetic tower co-scales with the former.

\autoref{ConA} also makes no reference to rapid alignment. Indeed, we find examples with electric towers exhibiting the maximal rate of divergence that co-scale and align, but do not rapidly align, with a magnetic tower. Interestingly, in such cases we have always found an {\em additional} electric tower that co-scales and aligns (but not rapidly) with the first, and co-scales and {\em rapidly} aligns with a magnetic tower. This shows that rapid alignment is a very common phenomenon, but not in a manner that lends itself to a universal conjecture that is simple to state.

The setting that we study in the most detail is the landscape of 5d supersymmetric quantum gravity theories, where our results are all consistent with a more precise statement: 
\begin{conjabc}[Co-Scaling and Rapid Alignment in the 5d Supergravity Landscape]\label{ConB}
In a 5d supersymmetric quantum gravity theory, every tower of magnetically charged BPS strings exhibits co-scaling and rapid alignment with a tower of electrically charged BPS particles. 
\end{conjabc}
It is not obvious how to extend this statement to a universal conjecture, in part because it is not clear how the asymmetry between electric and magnetic towers in the statement should be generalized. In the 5d case, there are multiple sources of such an asymmetry: for example, electrically and magnetically charged objects have different dimension, and only the electric gauge fields participate in Chern-Simons terms of the form $C_{IJK} A^I \wedge F^J \wedge F^K$.

Within the 5d supergravity landscape, our results motivate two novel, precise conjectures about Calabi-Yau geometry, discussed in detail in~\S\ref{ss.conj}. In M-theory compactifications on Calabi-Yau threefolds, electrically charged particles come from M2 branes wrapping curves and magnetically charged strings come from M5 branes wrapping divisors. Mathematically, directions in electric charge space where infinite families of BPS charged particles arise from holomorphic curves (the electric infinity cone~\cite{Gendler:2022ztv}) are much more well-understood than directions in magnetic charge space where infinite families of BPS strings arise from effective divisors (the magnetic infinity cone). One of our precise mathematical conjectures, \autoref{Con2}, is a characterization of the magnetic infinity cone: namely, it is the dual to the cone consisting of all $\cF_I = C_{IJK} Y^J Y^K$ for $Y \in \cK_{\rm hyp}$. Here, $\cK_{\rm hyp}$ is the hyperextended K\"ahler cone defined in~\cite{Gendler:2022ztv}, which extends K\"ahler moduli space across all flops and stable Weyl reflections. Essentially, $\cK_{\rm hyp}$ is the largest extension of K\"ahler moduli space for which BPS states do not decay.

The other precise mathematical conjecture that we formulate, \autoref{Con1}, claims that the set of all $C_{IJK} Y^K \tq^J$, where $\tq^J$ is in the magnetic infinity cone and $Y^K$ is in $\cK_{\rm hyp}$, is precisely the electric infinity cone. This provides a formulation of the conjecture of co-scaling and rapid alignment in the 5d supergravity landscape that is potentially rigorously provable.

\subsection{The geometry of rapid alignment}\label{ss.geometry}

In the case of a theory with multiple U(1) gauge fields, it is often convenient to consider the {\em convex hull} of the charge-to-mass vectors of objects in the theory. For example, the multi-U(1) WGC can be stated as the condition that the convex hull of charge-to-mass vectors of all particles should enclose the region populated by charge-to-mass vectors of all asymptotically large black holes~\cite{Cheung:2014vva}.

\begin{figure}[!ht]
\centering
\includegraphics[width=\textwidth]{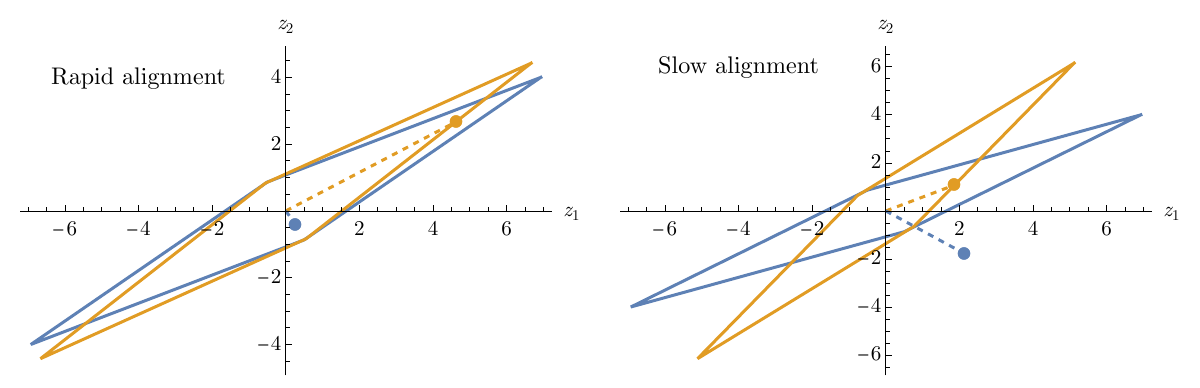}
\caption{Illustration of rapid alignment (left) and slow alignment (right). The electric (magnetic) convex hull is shown in blue (orange). The dashed orange line shows the distance to the magnetic convex hull in the long direction of the electric convex hull. The dashed blue line shows the projection of the long direction of the electric convex hull onto the short direction of the magnetic convex hull. For these figures, we have taken the long axes to have length $z = 8$, short axes perpendicular and of length $1$, $\varphi_\mathrm{rapid} = 1/(2z)$, and $\varphi_\mathrm{slow} = 1/\sqrt{z}$.}
\label{f.rapidalignmentillustration}
\end{figure}

Our definition of alignment singled out {\em rapid alignment} as the case when the angle between charge-to-mass ratio vectors satisfies $|\varphi| \lesssim \lVert z \rVert^{-1}$ at large $\lVert z \rVert$. The reason for this criterion is that, roughly speaking, if we have rapid alignment of all the long charge-to-mass ratio vectors in the theory, the convex hulls of the electric and magnetic charge-to-mass vectors ``look similar'' in all directions. We illustrate two considerations for this notion of similarity in \autoref{f.rapidalignmentillustration}, by plotting the convex hull of charge-to-mass vectors in an orthonormal basis (where the gauge kinetic matrix is the identity), for a case of rapid alignment and a case where alignment is not rapid (labeled ``slow alignment'' in the figure).

The first consideration is whether the projection of long electric and magnetic charge-to-mass ratio vectors onto short directions have similar length. Roughly speaking, this asks whether a given gauge field in the theory has a significant interaction with both light electric and light magnetic degrees of freedom, neither, or one but not the other. Suppose we pick a (unit length) direction $\dv$ in the charge lattice, and wish to find what the projection is of $\zvel$ and $\zvmag$ onto this direction.
We can write
\begin{equation}
    \zvel = \bravv{\zel}\bras{\frac{\zvmag}{\bravv{\zmag}}\cos\varphi + \zvperp \sin\varphi}\,,
\end{equation}
for some unit length $\zvperp$ perpendicular to $\zvel$.
Then
\begin{eqnarray}
    \dv \cdot \zvel
    &=& \bravv{\zel}\bras{\frac{\dv\cdot\zvmag}{\bravv{\zmag}}\cos\varphi + \dv\cdot \zvperp\sin\varphi}\\
    &=& \brap{\frac{\bravv{\zel}}{\bravv{\zmag}}\cos\varphi}\dv \cdot \zvmag + \brap{\bravv{\zel}\sin\varphi}\dv\cdot \zvperp\,.
\end{eqnarray}
If $\lVert \zmag \rVert \sim \lVert\zel \rVert\sim z \gg 1$, then $\dv \cdot \zvel \sim \dv \cdot \zvmag$ for generic $\dv$ if and only if $\varphi\lesssim\frac1z$.
A similar statement applies if we exchange the roles of $\zvel$ and $\zvmag$.

The second consideration is the distance from the origin to the convex hull in a given direction. This is most easily illustrated through an example, rather than in complete generality. Suppose, as in \autoref{f.rapidalignmentillustration}, that we have a long electric charge-to-mass vector $\zvel$ of length $z$ and an orthogonal short electric charge-to-mass vector of length 1, and that the magnetic charge-to-mass vectors are of identical length but rotated with respect to the electric charge-to-mass vectors by an angle $\varphi$. Then, in the direction of $\zvel$, the distance to the magnetic convex hull is
\begin{equation}
d_{\rm mag} = \frac{z}{\left| z \sin \varphi + \cos \varphi \right|}.
\end{equation}
If $\varphi \lesssim 1/z$, we have $d_{\rm mag} \sim \lVert \zvel \rVert = z$, whereas if $\varphi \gtrsim 1/z$ we have $d_{\rm mag} \sim O(1)$.

Thus, we see that rapid alignment is, from multiple viewpoints, the condition that the light electrically and magnetically charged states in the theory behave similarly in all directions, up to $O(1)$ factors. 

\subsection{Outline}

The outline of the remainder of this paper is as follows. In \S\ref{sec:heuristic} we discuss some heuristic arguments for why co-scaling should be common. The core of the paper is \S\ref{sec:5d}, which contains a detailed study of co-scaling and alignment in 5d Calabi-Yau compactifications of M-theory, with a focus on behavior near the boundaries of K\"ahler moduli space. In \S\ref{sec.non.UV}, we consider extremal black hole solutions in theories with scalars interacting with gauge fields in a more general way than the well-understood dilatonic couplings. We find that the classical solutions do not always exhibit co-scaling, suggesting that co-scaling is a swampland criterion~\cite{Vafa:2005ui}, i.e., it holds for consistent quantum gravity theories but not for theories without UV completions. We offer a concluding discussion in \S\ref{s.discussion}. Two appendices give additional results: Appendix~\ref{appendix.6d} on the 6d supergravity landscape, and Appendix~\ref{appendix.black.hole} on technical details of the black hole computations.

\paragraph{Note added:} While we were completing this paper, the work~\cite{Blanco:2025qom} appeared on arXiv, with some overlap in results but a different perspective following on earlier results of the same authors~\cite{Marchesano:2023thx, Marchesano:2024tod, Castellano:2024gwi} regarding rigid field theories (RFTs). In particular, in their Appendix A, they classify allowed scaling behaviors of BPS strings in a manner similar to our classification in~\S\ref{s.coscaling}. They also show (to use our terminology) co-scaling of magnetic strings of charge $\tq^J$ and electric particles of charge $q_I \propto a_{IJ} \tq^J$ with $a_{IJ}$ the gauge kinetic matrix. In~\S\ref{ss.sufficient}, we derive a similar result for $\cF_{IJ}$, the matrix of second derivatives of the prepotential. In the rigid limit, these statements are equivalent at leading order.

\section{Heuristic Arguments}
\label{sec:heuristic}

In subsequent sections of this paper, we will outline several quantitative arguments for co-scaling from string/M-theory compactifications. Before we encounter these more precise arguments, however, we may first sketch several more heuristic arguments based on simple scaling behavior.

\subsection{Monopole mass estimates}

Barring a fine-tuned cancellation, one generically expects that the mass of a monopole will be at least as large as the energy stored in its magnetic field. This can be estimated as (see e.g.~\cite{Arkanihamed:2006dz})
\be
m_{\rm mag} \sim \frac{\Lambda}{\gel^2}\,, 
\ee
where $\gel$ is the electric gauge coupling and $\Lambda$ is a UV cutoff given by the inverse of the semiclassical radius of the monopole.
This expression generalizes straightforwardly to the case of a $(d-p-2)$-dimensional monopole for a $p$-form electric gauge field in $d$ dimensions as \cite{ Hebecker:2017wsu}:
\begin{equation}
T_{\rm mag} \sim \frac{\Lambda^p}{\gel^2}\,.
\label{magest}
\end{equation}

Meanwhile, we also expect this theory to feature an electrically charged $(p-1)$-brane.
Setting the cutoff $\Lambda$ equal to the mass scale of the electrically charged $(p-1)$-brane, $T_{\rm el} \sim \Lambda^p$, \eqref{magest} reduces to the expected relation,
\be
T_{\rm mag} \sim \frac{T_{\rm el}}{\gel^2}\,.
\label{magest2}
\ee

\subsection{Types of boundaries}

In quantum gravity, particle masses and brane tensions are controlled by vacuum expectation values of scalar fields---a consequence of the widely held presumption that there are no free parameters in quantum gravity \cite{Ooguri:2006in}. A light brane or tower of particles signals a breakdown of effective field theory and thus a boundary of the scalar field moduli space, and there are four common patterns for the physics at such boundaries: (1) a fundamental string emerges, (2) a decompactification occurs, (3) a CFT emerges, or (4) a nonabelian enhancement of the U(1) gauge group occurs.

Cases (1) and (2) are predicted by the Emergent String Conjecture \cite{Lee:2019wij}, and they lie at infinite distance in the moduli space metric. Both limits involve dilatonic couplings of the modulus to the gauge kinetic term, resulting in approximate saturation of the both the electric and magnetic WGC by towers of black holes/black branes \cite{Heidenreich:2015wga}. In this case, we expect $|\zel| \sim 1$ and $|\zmag| \sim 1$, so co-scaling follows trivially.

Case (3), in which an interacting CFT emerges at low energies, occurs at finite distance in moduli space. Here, gravity is decoupled, and the scale invariance of the CFT implies that the low-energy physics is independent of the Planck scale $M_{{\rm Pl};d}$ and depends only on the moduli space distance to the CFT locus $|\Delta \phi|$. Thus we expect that $T_{\rm el}$, $T_{\rm mag}$, and $\gel^2$ will scale as
\begin{equation}
T_{\rm el}^{\frac{1}{p}} \sim T_{\rm mag}^{\frac{1}{d-p-2}} \sim \gel^{\frac{2}{2+2p-d}} \sim |\Delta \phi|^{\frac{2}{d-2}} \equiv m_{\rm CFT}\,,
\end{equation}
where $p$ is the worldvolume dimension of the electrically charged brane.
This gives
\begin{align}
\frac{|\zel|}{|\zmag|} \sim \frac{\gel^2 T_{\rm mag}}{T_{\rm el}} \sim \frac{m_{\rm CFT}^{2+2p-d} \cdot m_{\rm CFT}^{d-p-2}}{m_{\rm CFT}^p} \sim 1\,.
\end{align}
in accordance with \eqref{rr}.

Finally, we have case (4), in which a U(1) gauge group enhances to SU(2) at finite distance in moduli space. Again, gravity decouples, and we expect W-bosons and 't~Hooft-Polyakov monopoles whose mass/tension scale respectively as
\begin{equation}
m_W \sim \gel |\Delta \phi|\,,~~~T_{\rm mag} \sim \frac{|\Delta \phi|}{\gel}\,,
\end{equation}
so once again
\be
\frac{|\zel|}{|\zmag|} \sim \frac{\gel^2 T_{\rm mag}}{m_W} \sim 1\,,
\ee
exhibiting co-scaling. Note that in this case, we do not expect a tower associated with the W-boson mass scale $m_W$ or the monopole tension scale $T_{\rm mag}$.\footnote{However, many UV-complete models of SU(2) enhancement feature skew towers of charge $q' + k q_W$, $k \in \mathbb{Z}$, where $q_W$ is the charge of the W-boson and $q'$ is some other electric charge.} This shows that the co-scaling relation \eqref{rr} is sometimes satisfied not only for towers, but also for isolated numbers of light particles. 

\subsection{Emergence Proposal}

Another way to motivate co-scaling in the particle case ($p=1$) is by the emergence arguments of~\cite{Heidenreich:2018kpg, Grimm:2018ohb}. As shown in those works, a tower of light particles that vanish at a point $\phi_0$ in field space will ordinarily renormalize the scalar kinetic term and drive the point to infinite distance. If the tower of light particles is charged under the gauge group with $\zel \lesssim 1$, it will likewise renormalize the gauge kinetic term and drive the gauge field to weak coupling \cite{Heidenreich:2017sim}. These emergence arguments typically assume, for simplicity, that there is a single tower of states that drives both the modulus and gravity to strong coupling at short distances (or, from a different point of view, explain the weakness of their interactions at long distances). It is useful to review this argument and then to relax some of the assumptions to see how other, non-leading towers might behave. (There has been substantial recent work on the emergence proposal, including more concrete string-theoretic computations; see, e.g.,~\cite{Castellano:2021mmx, Castellano:2022bvr, Blumenhagen:2023yws, Blumenhagen:2023tev, Hattab:2023moj}. Here we restrict our remarks to heuristic scaling arguments as in the early papers.)

The first ingredient is the species bound~\cite{Veneziano:2001ah,Dvali:2007hz,Dvali:2009ks}, which posits that if there are $N_\mathrm{tot}(E)$ total weakly coupled degrees of freedom below the energy $E$ in a quantum gravity theory in $d$ spacetime dimensions, then the cutoff scale $\Lambda_\mathrm{QG}$ (often called the ``species scale'') satisfies
\begin{equation} \label{eq:speciesbound}
    N_\mathrm{tot}(\Lambda_\mathrm{QG}) \Lambda_\mathrm{QG}^{d-2} \lesssim M_\mathrm{Pl}^{d-2}.
\end{equation}
One way to understand this bound is to ask that loop corrections to the graviton propagator do not overwhelm the leading-order result. A similar argument for scalar fields leads to the conclusion that, if $N_\phi(E)$ denotes the number of weakly coupled fields with mass proportional to $\phi$ (near some singular point $\phi = 0$ in field space) and smaller than $E$, then requiring corrections to the $\phi$ propagator to be subdominant up to the cutoff $\Lambda_\mathrm{QG}$ implies that~\cite{Heidenreich:2018kpg}
\begin{equation} \label{eq:phirenorm}
    \frac{N_\phi(\Lambda_\mathrm{QG}) \Lambda_\mathrm{QG}^{d-2}}{\phi^2} \lesssim K(\phi),
\end{equation}
where $K(\phi)$ is the factor preceding $(\partial \phi)^2$ in the kinetic term for $\phi$. 

Familiar asymptotic infinite-distance limits correspond to cases where $N_\phi(\Lambda_\mathrm{QG}) \sim N_\mathrm{tot}(\Lambda_\mathrm{QG})$, i.e., the states that become massless as $\phi \to 0$ dominate the total counting of degrees of freedom and hence the relationship between the Planck scale and the species scale. In that case, we expect the inequalities to be parametrically saturated, with $K(\phi) \sim M_\mathrm{Pl}^{d-2}/\phi^2$. This gives the familiar logarithmic kinetic term, and implies that infinite-distance moduli typically couple with gravitational strength. In the case that the tower coupled to such an infinite-distance modulus is charged, gauge, scalar, and gravitational forces are of comparable strength, and the black hole spectrum will differ from the Reissner-Nordstr\"om case only by $O(1)$ factors. Thus, we expect that the co-scaling of electric and magnetic black holes will hold. Indeed, these cases are described by the extremal dilatonic black hole co-scaling~\eqref{eq:zdilatonic}, with $O(1)$ values of $\alpha$.

The cases of most interest to us in this paper are those that do not correspond to infinite-distance limits, so that we cannot assume logarithmic kinetic terms and dilatonic black hole solutions. For example, the renormalization of the modulus and gauge kinetic terms can be avoided if the UV cutoff $\Lambda_{\rm UV}$ scales linearly with the characteristic mass scale of the charged particles, $\Lambda_{\rm UV} \sim m_{\rm el}$. Since the tower mass $m_{\rm el}$ never drops parametrically below the UV cutoff, there are only an order-one number of light fields that contribute to the renormalization of the kinetic terms, so the point $\phi_0$ remains at finite distance and the gauge coupling $\gel$ is not significantly renormalized away from its UV value, which we assume to be of order $\Lambda_{\rm UV}^{(d-4)/2}$. One simple way to enforce this is to add a monopole at the mass scale $m_{\rm mag} \sim m_{\rm el}$, which acts as a UV cutoff on the EFT, ensuring that the boundary lies at finite distance and the gauge coupling is order-one in units of $m_{\rm mag}$. Thus we have co-scaling: $\gel^2 m_{\rm mag} / m_{\rm el} \sim 1$.

Alternatively, we can apply~\eqref{eq:phirenorm} in an asymptotic infinite-distance limit but to subdominant moduli that do not control the limit. We measure this subdominance by $\epsilon_\phi \equiv N_\phi(\Lambda_\mathrm{QG}) / N_\mathrm{tot}(\Lambda_\mathrm{QG})$. The kinetic term of $\phi$ is correspondingly smaller by a factor of $\epsilon_\phi$, and we expect such a modulus to have couplings suppressed by $\sqrt{\epsilon_\phi} M_\mathrm{Pl}$, i.e., to interact much more strongly than gravity. As an example where this logic applies, consider the case of a Calabi-Yau with volume depending on big and small divisor volumes $\tau_{b,s}$ as ${\cal V} \sim \tau_b^{3/2} - \tau_s^{3/2}$ (as in the well-studied LVS scenario~\cite{Balasubramanian:2005zx}). Here $\tau_b \to \infty$ is an infinite-distance limit in which a tower of KK modes becomes light, but $\tau_s \ll \tau_b$ is a subdominant modulus. From the volume formula, it is apparent that $\tau_s^{3/2}$ measures the volume of a six-dimensional hole in the manifold, and we might estimate that the counting of KK modes sensitive to $\tau_s$ is smaller than the total number by $\epsilon_s = (\tau_s/\tau_b)^{3/2}$. The mass of a KK mode on the small cycle scales as $\Lambda_\mathrm{QG} \tau_s^{-1/4}$, which we identify as $\phi$ in~\eqref{eq:phirenorm}. This leads to the estimate
\begin{equation}
    K(\phi_s)(\partial \phi_s)^2 \sim \left(\frac{\tau_s}{\tau_b}\right)^{3/2}\frac{M_\mathrm{Pl}^2}{\phi_s^2} (\partial \phi_s)^2 \sim  \left(\frac{\tau_s}{\tau_b}\right)^{3/2}\frac{M_\mathrm{Pl}^2}{\tau_s^2} (\partial \tau_s)^2 \sim \frac{M_\mathrm{Pl}^2}{\tau_b^{3/2} \tau_s^{1/2}} (\partial \tau_s)^2,
\end{equation}
which is indeed the correct scaling of the $\tau_s$ kinetic term (as we can readily read off from the second derivative of the K\"ahler potential). In this case, the volume formula was particularly simple. In a general setting, the KK spectrum depends in an intricate way on many moduli, and it is not straightforward to extract the full structure of the kinetic terms from an emergence argument. Nonetheless, the message that we take away for this paper is that moduli that control subleading towers in infinite-distance limits will typically couple much more strongly than gravity. This means that the black hole solutions can be radically different from Reissner-Nordstr\"om solutions, as gauge forces primarily balance scalar rather than gravitational forces. We will see that in such cases co-scaling often, but not always, holds.

\subsection{Classical solutions and Bogomolnyi bounds}\label{ss.bogomolnyi}

Here we generalize a heuristic argument for electric-magnetic co-scaling made only in the case of axions (0-form gauge fields) in~\cite{Reece:2024wrn} to the general case of higher-form gauge fields. In the limit that scalar (modulus-mediated) and gauge forces are both much stronger than gravity, we can find classical solutions for both electrically and magnetically charged objects by solving the coupled modulus-gauge field equations of motion. For the case of a single scalar, the Bogomolnyi trick can be used to understand these solutions. Specifically, we consider the action
\begin{equation}
I = \int \left[-\frac{1}{2}f(\phi)^2 F^{(p)} \wedge \star F^{(p)} - \frac{1}{2} h(\phi)^2 \rmd\phi \wedge \star\rmd \phi\right],
\end{equation} 
with $F^{(p)}$ the $p$-form field strength of a $(p-1)$-form gauge field. The electric gauge coupling is $e = 1/f(\phi)$ and the magnetic gauge coupling is $\tilde e = 2\pi f(\phi)$. We focus on a magnetically charged solution for which $\int_{S^p} F^{(p)} = 2\pi$. In this case, we can separate our spatial coordinates into a radius $r$ away from the object, angular directions $\Omega_p$ surrounding the object, and $t$, $y_i$ ($i \in \{1, \ldots d - p - 2\}$) along the worldvolume of the object. Then we define $\ast$ as the Hodge star for the $r, \Omega_p$ directions only; the other directions are spectators, and the action density of the magnetically charged object is
\begin{equation}
{\cal I} = -\left[\frac{1}{2} \int \left|h(\phi) \rmd \phi \mp f(\phi) \ast F^{(p)}\right|^2 \pm \int h(\phi)f(\phi)\rmd \phi \int_{S^p} F^{(p)}\right].
\end{equation}
In the last term, the flux of $F^{(p)}$ is a topological invariant, whereas the prefactor is an integral over $\phi$ that is fixed at $r \to \infty$ by a choice of asymptotic value (or vacuum) $\phi_*$, whereas at $r \to 0$ a boundary condition often selects a special value of $\phi$ (typically $0$, $+\infty$, or $-\infty$), so that the integral is fixed. When this holds (as can be checked for explicit examples of functions $h(\phi)$ and $f(\phi)$), the Bogomolnyi trick applies: the first term is a perfect square and is extremized when it is zero, while the latter term is invariant under small deformations of the fields obeying fixed boundary conditions. Thus, we can read off the tension of the magnetically charged object from the second term. For electrically charged objects, the same argument applies, taking $p \mapsto d - p - 2$ and $f(\phi) \mapsto 1/[2\pi f(\phi)]$. Thus, we expect the electric and magnetic tensions to take the form
\begin{align}
T_\mathrm{el} &= \left| \int_{\phi^{\rm el}_0}^{\phi_*} \frac{h(\phi)}{f(\phi)} \rmd \phi \right| \nonumber \\
T_\mathrm{mag} &= 2\pi \left| \int_{\phi^{\rm mag}_0}^{\phi_*} h(\phi) f(\phi) \rmd \phi \right|.
\end{align}
(We have allowed for the possibility that the boundary conditions select different values of $\phi_0 = \lim_{r\to 0} \phi(r)$ in the electric and magnetic cases, disambiguated with superscripts.) Now, provided that these integrals are dominated near $\phi_*$ (as is the case in exponential or power-law examples discussed in~\cite{Reece:2024wrn}), we will have
\begin{equation}
\frac{T_\mathrm{el}}{T_\mathrm{mag}} \sim \frac{1}{2\pi f(\phi_*)^2} = \frac{e^2}{2\pi},
\end{equation}
which is the statement of electric-magnetic co-scaling.

The most unsatisfying aspect of this argument is that it is not straightforward to generalize to the case of multiple moduli fields. We have also omitted gravity. In \S\ref{sec.non.UV}, we will return to the study of classical solutions but include gravity by examining extremal black hole solutions for different forms of scalar couplings to gauge fields.

\subsection{Dimensional reduction}\label{ss.dimred}

Suppose that we have a theory in $D$ dimensions which exhibits co-scaling and alignment. In this section, we show that it still exhibits co-scaling and alignment after dimensional reduction to $d=D-1$ dimensions.

We begin by verifying that co-scaling is preserved. There are, in fact, two different relations to check: one in which the dimensionality of the electric objects is preserved, and one in which it is reduced. These correspond, respectively, to the cases where the dimensionality of the magnetic monopoles are reduced and preserved.

Let us then begin by assuming the co-scaling relation is satisfied for an electrically charged brane of worldvolume dimension $P$. That is, we assume
\be
{\gel^{(D)}} T^{(D)}_{\text{mag},D-P-2} \sim {\gmag^{(D)}}T^{(D)}_{\text{el},P}\,.
\ee
After reduction, preserving $P$, we have \cite{Heidenreich:2019zkl}
\begin{align}
T^{(d)}_{\text{el},P} &= \exp\left[-\frac{P}{\sqrt{(d-1)(d-2)}} \rho\right]  T^{(D)}_{\text{el},P} \\
T^{(d)}_{\text{mag},d-P-2} &= (2 \pi R) \exp\left[-\frac{P}{\sqrt{(d-1)(d-2)}} \rho\right]  T^{(D)}_{\text{mag},D-P-2} \\
(\gel^{(d)})^2 &= \frac{1}{2\pi R}  \exp\left[-\frac{2P}{\sqrt{(d-1)(d-2)}} \rho\right] (\gel^{(D)})^2 \\
(\gmag^{(d)})^2 &= (2 \pi R)  \exp\left[+\frac{2P}{\sqrt{(d-1)(d-2)}} \rho\right] (\gmag^{(D)})^2\,,
\end{align}
where $R$ is the radius of the compactification circle and $\rho$ is the canonically normalized radion. Thus, we have
\begin{equation}
\frac{\gel^{(d)} T^{(d)}_{\text{mag},d-P-2}}{\gmag^{(d)} T^{(d)}_{\text{el},P}} =
\frac{\gel^{(D)} T^{(D)}_{\text{mag},D-P-2}}{\gmag^{(D)} T^{(D)}_{\text{el},D}}
\sim 1\,.
\end{equation}

Similarly, reducing the dimension of the electrically charged object from $P$ to $p=P-1$, we have \cite{Heidenreich:2019zkl}:
\begin{align}
T^{(d)}_{\text{el},p} &= (2 \pi R) \exp\left[\frac{d-p-2}{\sqrt{(d-1)(d-2)}} \rho\right]  T^{(D)}_{\text{el},P} \\
T^{(d)}_{\text{mag},d-p-2} &=  \exp\left[-\frac{d-p-2}{\sqrt{(d-1)(d-2)}} \rho\right]  T^{(D)}_{\text{mag},D-P-2} \\
(\gel^{(d)})^2 &= (2 \pi R)  \exp\left[\frac{2(d-p-2)}{\sqrt{(d-1)(d-2)}} \rho\right] (\gel^{(D)})^2 \\
(\gmag^{(d)})^2 &= \frac{1}{2 \pi R}  \exp\left[-\frac{2(d-p-2)}{\sqrt{(d-1)(d-2)}} \rho\right] (\gmag^{(D)})^2 \,.
\end{align}
So, once again, we have
\begin{equation}
\frac{\gel^{(d)} T^{(d)}_{\text{mag},d-p-2}}{\gmag^{(d)} T^{(d)}_{\text{el},p}} =
\frac{\gel^{(D)} T^{(D)}_{\text{mag},D-P-2}}{\gmag^{(D)} T^{(D)}_{\text{el},D}}
\sim 1\,.
\end{equation}

Next, let us verify that alignment is preserved. In $D$ dimensions, we have
\be
\cos \varphi_D = \frac{2\pi q_i \tilde q^{i}}{\lVert q \rVert_D  \lVert \tilde q \rVert_D }\,,
\ee
with
\be
\lVert q \rVert_D^2 = (2\pi)^2 q_i \tilde K_D^{ ij}q_j\,,~~~~\lVert \tilde q \rVert_D^2 = (2\pi)^2 \tilde q^{i}K_{D, ij} \tilde q^{j}\,.
\ee
Under dimensional reduction, $q_i$ and $\tilde q^i$ are preserved, while $K_{d,ij}$ is related to $K_{D,ij}$ by an overall factor:
\be
K_{d,ij} =2 \pi R \exp\left[\frac{2P}{\sqrt{(d-1)(d-2)}} \rho\right] K_{D,ij}\,,
\ee
if the dimensionality of the electric $P$-brane is preserved, while
\be
K_{d,ij} = \frac{1}{2 \pi R} \exp\left[\frac{-2(d-p-2)}{\sqrt{(d-1)(d-2)}} \rho\right] K_{D,ij}\,,
\ee
for $P$ reduced to $p=P-1$ by wrapping the electric brane on the circle. In either case, this overall factor $C=C(R, \rho, P)$ cancels out in the computation of the angle, since
\begin{align}
\lVert q \rVert_d \lVert\tilde q\rVert_d  = \sqrt{q_i \tilde K_d^{ ij}q_j \tilde q^{k}K_{d, kl} \tilde q^{l}} = \sqrt{q_i C^{-1} \tilde K_D^{ ij}q_j \tilde q^{k} C K_{D, kl} \tilde q^{l}} = \lVert q \rVert_D \lVert \tilde q \rVert_D\,,
\end{align}
where we have used the fact that $\tilde K_d^{ij} K_{d,jk}=\tilde K_D^{ij} K_{D,jk} = (2 \pi)^{-2} \delta^i_k$. Thus, we find
\be
\cos \varphi_D = \cos \varphi_d\,,
\ee
so alignment is preserved under dimensional reduction.

In the case of particles, $p=1$, we could also consider Kaluza-Klein (KK) modes, which carry charge under a KK photon. However, KK modes of the graviton always have $\zel \sim 1$, and KK monopoles have $\zmag \sim 1$, so they are uninteresting from the perspective of co-scaling and alignment. More generally, a tower of particles carrying charge under the KK photon has a characteristic mass $m$ at least as large as the KK gauge coupling, $m \gtrsim e_{\rm KK}$. Thus, in order to have
$\zel \gg 1$, the higher-dimensional charge of the tower must dominate its KK charge, in which case the KK charge plays a negligible role in co-scaling and alignment.\footnote{We have verified this heuristic argument in decompactification limits of 4d supergravity theories.} Similarly, for a monopole to have $\zel \gg 1$, its higher-dimensional charge must dominate its KK magnetic charge, in which case the KK magnetic charge plays a negligible role in co-scaling and alignment.

\section{Five Dimensions}
\label{sec:5d}

\subsection{Basics of 5d supergravity}

At a generic point in vector multiplet moduli space, the action for the massless bosonic fields in a gauge theory with $n_v$ vector multiplets is given by
\begin{align}
  S &= \frac{1}{2 \kappa_5^2}  \int d^5 x \sqrt{- g}  \left( \mathcal{R} -
  \frac{1}{2} \mathfrak{g}_{i j} (\phi) \partial \phi^i \cdot \partial \phi^j
  \right) - \frac{1}{2 g_5^2} \int a_{I J} (\phi) F^I \wedge
  \star F^J \nonumber \\
  &+ \frac{1}{6(2\pi)^2} \int C_{I J K} A^I \wedge F^J \wedge F^K,
  \label{eqn:5dsugra}
\end{align}
where $I = 0, \ldots, n_v$, $i = 1, \ldots, n_v$, and $g_5^2 = (2\pi)^{4/3} (2\kappa_5^2)^{1/3}$. 

Many of the relevant features of the Coulomb branch of a 5d supergravity theory are captured by its prepotential, a cubic homogeneous polynomial:
\begin{equation}
\mathcal{F} = \frac{1}{6} C_{I J K} Y^I Y^J Y^K.
\end{equation}
Here, $Y^I$ is real-valued, and $I$ runs from $0$ to $n_v$.

The gauge kinetic matrix is related to the prepotential $\cF$ by 
\begin{equation}
  a_{I J} = \frac{\mathcal{F}_I \mathcal{F}_J}{\cF^{4/3}} - \frac{\mathcal{F}_{I J}}{\cF^{1/3}} ,
  \label{eq:gaugekinetic}
\end{equation}
with
\begin{equation}
   \mathcal{F}_I = \partial_I \cF= \frac{1}{2} C_{I J K} Y^J Y^K , 
   \qquad \mathcal{F}_{I
   J} = \partial_I \partial_J \cF  = C_{I J K} Y^K .
   \label{eq:prepottrip}
   \end{equation}
The 5d vector multiplet moduli space (also known as the Coulomb branch) is the slice of moduli space $\cF = 1$, and the metric on moduli space is the pullback of $a_{IJ}$ to this slice. However, it is often easier to think of the $Y^I$'s as homogenous coordinates, which are identified under simultaneous rescaling of the $Y^I$ by a positive real number $\lambda$.

The metric on the $n_v+1$-dimensional space of homogeneous coordinates is then given by
\begin{equation}
\mathfrak{g}_{IJ} = \frac{2}{3} \frac{\cF_I \cF_J}{\mathcal{F}^2} - \frac{\cF_{IJ}}{\cF} .
\label{hgmet}
\end{equation}
This metric is positive-semidefinite: it has a single null eigenvalue corresponding to the rescaling $Y^I \rightarrow \lambda Y^I$.

5d supergravity theories have both BPS particles (which carry electric charge under the vector gauge fields) and BPS strings (which are magnetically charged). The mass of a BPS particle of charge $q_I$ is given by
\be
m = \frac{g_5}{\sqrt{2} \kappa_5} \frac{|q_I Y^I|}{\cF^{1/3}}\,.
\label{5dBPS}
\ee
The tension of a BPS string of charge $\tilde q^I$ is given by
\be
T = \frac{2 \pi}{\sqrt{2} g_5 \kappa_5} \frac{|\tilde q^I \cF_I|}{\cF^{2/3}}\,.
\label{5dBPSstring}
\ee 
In the remainder of this section, we will often set $M_{{\rm Pl};d}^3 = \kappa_5^{-2} = 1$. For ease of notation, we further define
\be
\lVert q \rVert^2  = q_I a^{IJ} q_J\,,~~~~\lVert\tilde q\rVert^2 = \tilde q^I a_{IJ} \tilde q^J\,,
\ee
which differ from the definitions in \S\ref{INTRO} by $O(1)$ factors of $g_5^2$ and $2 \pi/g_5^2$, respectively. With these definitions, the charge-to-mass ratios of particles and strings are given by
\be\label{eqn.def.z.5d}
\zel = \frac{g_5 \lVert q \rVert}{m} \,,~~~~ \zmag = \frac{2 \pi}{g_5} \frac{\lVert \tilde q \rVert}{ T}\,,
\ee
and the angle between an electric charge $q_I$ and a magnetic charge $\tilde q^I$ is given by
\be
\cos \varphi = \frac{q_I \tilde q^I}{\lVert q \rVert \lVert \tq \rVert}\,.
\ee

\subsubsection{Calabi-Yau Geometry}\label{ss.CY3}

5d supergravity theories arise from M-theory compactified on Calabi-Yau threefolds. The vector multiplet moduli space of the supergravity theory is identified with the K\"ahler moduli space of the Calabi-Yau threefold, which is divided into phases known as K\"ahler cones. At the boundary of a K\"ahler cone $\mathcal{K}_X$, one of four things happens \cite{Witten:1996qb}:
\begin{enumerate}
\item The entire Calabi-Yau collapses to a manifold of lower dimension.
\item A divisor collapses to a point.
    \item A curve collapses to a point.
    \item A divisor collapses to a curve.
\end{enumerate}
The first case corresponds to an asymptotic boundary, which lies at infinite distance in moduli space. The second corresponds to an SCFT boundary, where a tower of BPS particles become massless and a BPS string becomes tensionless. We will discuss these boundaries in greater detail in what follows.

The third case corresponds to a conifold locus, where a conifold singularity develops. Via a flop transition, one may continue through the boundary into a distinct phase of the moduli space. The union of all K\"ahler cones related by flop transitions form the \emph{extended} K\"ahler cone, $\cK$:
\be
\cK = \bigcup_X \cK_X\,.
\ee
The final case---a divisor collapsing to a curve---corresponds to an SU(2) enhancement. Crossing a boundary of this type corresponds to a \emph{Weyl reflection}, which is an isomorphism associated with the $\mathbb{Z}_2$ Weyl group of SU(2). Such Weyl reflections, also known as Weyl flops, come in two types \cite{Gendler:2022ztv}: \emph{stable} and \emph{unstable}, depending on whether the BPS particles are stable or unstable under wall-crossing. The \emph{hyperextended} K\"ahler cone is defined to be the union of the extended K\"ahler cones across all stable Weyl reflections:
\be
\cK_{\rm hyp} = \bigcup_{w \in \mathcal{W}_{\rm stable}} \bigcup_X \cK_X\,.
\ee
BPS strings arise from M5-branes wrapped on effective divisors. These necessarily live inside the effective cone $\mathcal{E}$, which contains the hyperextended K\"ahler cone. Thus, we have a set of containment relations
\be
\mathcal{K}_X \subseteq \mathcal{K} \subseteq \mathcal{K}_{\rm hyp} \subseteq \mathcal{E} \subseteq H_4(X, \mathbb{R})\,.
\ee
There is a similar series of cones associated with electrically charged particles, which are related to these cones by cone dualities. To begin, the K\"ahler cone $\mathcal{K}_X$ is dual to the Mori cone $\mathcal{M}_X$ of the Calabi-Yau $X$, which is generated by effective curves. Just as BPS strings must reside in the effective cone of divisors, BPS particles live only in the Mori cone. The number of BPS particles of a given charge inside the Mori cone can be counted using Gopakumar-Vafa (GV) invariants \cite{Gopakumar:1998ii,Gopakumar:1998jq}.\footnote{More precisely, GV invariants compute an index, so they count BPS particles up to signs. In special cases, a vanishing GV invariant may represent a cancellation between BPS particles of different spin rather than the absence of BPS particles altogether.}

Infinite towers of BPS particles do not necessarily exist along all rays of the Mori cone. Instead, they form a smaller cone, called the infinity cone, denoted $\mathcal{M}_\infty$. In \cite{Gendler:2022ztv}, it was argued that the infinity cone is dual to the hypereffective cone, $\cM_\infty = \cK_{\rm hyp}^\vee$.

Given a coordinate $Y^I \in \mathcal{K}$, we may associate a \emph{dual coordinate} $\mathcal{F}_I = C_{IJK} Y^J Y^K$. Such dual coordinates parametrize the \emph{cone of dual coordinates} $\mathcal{T}$, which according to \cite{Alim:2021vhs} is equal to the \emph{movable} cone of the Calabi-Yau manifold and is dual to the effective cone, $\mathcal T = \mathcal E^\vee$. Analogously, the hyperextended cone of dual coordinates $\mathcal{T}_{\rm hyp}$ can be determined by extending the map $Y^I \mapsto F_I$ to the hyperextended K\"ahler cone. The region in charge space with BPS black holes $\mathcal{C}_{\rm BH}$ is at least as large as $\mathcal{T}_{\rm hyp}$, and in some cases it is strictly larger \cite{Gendler:2022ztv}. Thus, we have a chain of containment relations,
\be
\mathcal{T}  \subseteq \mathcal{T}_{\rm hyp} \subseteq \mathcal{C}_{\rm BH} \subseteq \cM_\infty \subseteq \cM_{X} \subseteq H_2(X, \mathbb{R})\,.
\label{econtain}
\ee

\subsection{Co-scaling at moduli space boundaries}\label{s.coscaling}

In this subsection, we classify the allowed scaling behaviors for towers of BPS particles and BPS strings at both infinite-distance boundaries and finite-distance boundaries of vector multiplet moduli space.
A similar analysis can be found in Appendix A of \cite{Blanco:2025qom}.

By a convenient choice of basis, we set the boundary to be at $Y^0=1$, $Y^i=0$, where lowercase indices run from $1$ to $n_v$.
From \eqref{5dBPS}, we see that if a tower of BPS particles of charge $k q_I$, $k \in \mathbb{Z}$ becomes massless at a locus of moduli space, then we must have $q_I Y^I\rightarrow 0$ at this point. Hence, these particles are precisely those with $q_0=0$.

We then consider linear paths of the form $Y^i=s n^i$, $s\rightarrow 0$,\footnote{One could, in principle, also consider the limit $s \rightarrow \infty$, where some $Y^I$ diverge. However, such paths are related via homogeneous rescaling to paths where some $Y^I$ vanish, and the rest remain finite. Additional paths in moduli space can be constructed by dropping the assumption of linear dependence on $s$ (see, e.g.,~\cite{Rudelius:2023odg}), but we will not consider such paths in this work.} where the unit vector $n^i$ is restricted by the condition that $Y^I$ must lie in the K\"ahler cone. Along such a path, the prepotential and its first and second derivatives take the form
\begin{align}
\cF  &=\frac{1}{6} C_{000} (Y^0)^3 +\frac{1}{2} C_{00i} (Y^0)^2 (Y^i) + \frac{1}{2} C_{0ij} (Y^0) (Y^i) (Y^j) + \frac{1}{6} C_{ijk} (Y^i)(Y^j)(Y^k) \nonumber \\
&=\frac{1}{6} C_{000}+ \frac{1}{2}n^p C_{00p} s +  \frac{1}{2}n^pn^q C_{0pq} s^2 + \frac{1}{6} n^pn^qn^rC_{pqr} s^3\
\label{prepot5d}\\
\cF_I &= \frac12 C_{00I} + n^pC_{0pI}s + \frac12n^pn^qC_{pqI}s^2\label{eqn.5d.limit.cFI}\\
\cF_{IJ} &= C_{0IJ} + n^pC_{pIJ}s\label{eqn.5d.limit.cFIJ}\,.
\end{align}
Note that in this basis, the $C_{IJK}$'s are not necessarily integers. Furthermore, the electric charges $q_I$ and the magnetic string charges $\tilde q^I$ are not necessarily integers (or even rational numbers), and it is not clear which directions in the charge lattice support BPS particles/strings.
Such issues of charge quantization cannot be addressed at the level of supergravity, so we will ignore them in this subsection and revisit them in UV-complete examples below.

For later notational convenience, we define the following six spaces; we will see that the scaling behavior of particle and string masses, tensions and couplings is determined by which of these spaces the electric/magnetic charges reside in:
\begin{align}
    \cMlight &:= \brac{q:q_0=0}
    \,,&
    \cElimdir &:={\cMlight}^\orthogcomplement = \brac{{\tilde q}: {\tilde q}^I \propto \delta^I_0}
    \,,\\
    \cEperp &:= \brac{{\tilde q}: C_{00I}{\tilde q}^I=0}
    \,,&
    \cMtau &:= {\cEperp}^\orthogcomplement = \brac{q: q_I \propto C_{00I}}
    \,,\\
    \cEker &:= \brac{{\tilde q}: {\tilde q}^I \in \ker\brap{C_{0IJ}}}
    \,,&
    \cMkerorth &:= {\cEker}^\orthogcomplement
    \,,
\end{align}
where orthogonal complement is taken with respect to the inner product ${\tilde q}\cdot{q}={\tilde q}^I{q}_I$.

We now specialize to the case of an infinite-distance boundary.
Limits of this type were studied previously in \cite{Etheredge:2022opl, Rudelius:2023odg}. They fall into two classes, depending on whether the prepotential vanishes linearly ($\mathcal{F} \sim s \Rightarrow C_{000}=0, C_{00i} \neq \mathbf{0}$) or quadratically ($\mathcal{F} \sim s^2 \Rightarrow C_{000} = 0$, $C_{00i}=\mathbf{0}$) in the limit $s \rightarrow 0$.\footnote{For a 5d supergravity arising from a Calabi-Yau compactification of M-theory, these two limits correspond to genus-one fibrations with collapsing genus-one fiber and K3/$T^4$ fibrations with collapsing surface fiber, respectively, as shown in the seminal work \cite{Lee:2019wij}.}

In the latter case,
\begin{equation}
 \cF_I = n^pC_{0pI}s + \frac12n^pn^qC_{nnI}s^2\,.
\end{equation}
Hence, a string with charge ${\tilde q}$ has tension that scales like
\begin{equation}
 T \sim \frac{\brav{{\tilde q}^I\cF_I}}{\cF^{2/3}} \sim
 \begin{cases}
  \frac{s^2}{s^{4/3}} \sim
  s^{2/3} & \text{if $\tilde q \in \cEker$,}\\
  \frac{s^1}{s^{4/3}} \sim
  s^{-1/3} & \text{otherwise.}
 \end{cases}
\end{equation}
Note that ${\tilde q}^I\propto \delta^I_0$ is always in $\cEker$.

In this limit, the gauge kinetic matrix is given by
\begin{align}
 a_{IJ} = \cF^{-4/3}\bigg\{
    n^pn^q&\bras{C_{0pI}C_{0qJ} - \frac12C_{0pq}C_{0IJ}}s^2
    \nonumber\\+
    n^pn^qn^r&\bras{
        \brap{\frac12C_{pqI}C_{0rJ} - \frac16C_{pqr}{C_{0IJ}}}
        +
        \brap{\frac12C_{0rI}C_{pqJ} - \frac12C_{0pq}C_{rIJ}}
    }s^3
    \nonumber\\+
    n^pn^qn^rn^t&\bras{\frac14C_{pqI}C_{rtJ} - \frac16C_{pqr}C_{tIJ}}s^4
 \bigg\}\,.
\end{align}
Assuming no accidental cancellations at higher orders in $s$, one can show that its eigenvectors and eigenvalues are as follows
\begin{itemize}
 \item The basis element $\tilde q^I = (1,0,0,\dots,0) \in \cElimdir$, is an eigenvector of eigenvalue $\sim s^{4/3}$.
 \item There are a set of eigenvectors $\{\tilde q^I\}$ with eigenvalues $\sim s^{1/3}$ that extend $(1,0,0,\dots,0)$ to a basis of $\cEker$.
 \item All other eigenvectors are not in $\cEker$ and have eigenvalues $\sim s^{-2/3}$.
\end{itemize}
Using this, we find that the scaling behavior of $\gmag^2$ is given by that of the least suppressed eigenvalue above to which ${\tilde q}^I$ is not orthogonal.
Similarly, barring cancellations, we expect that the scaling behavior of $\gel^{-2}$ is given by that of the \textit{most} suppressed eigenvalue above to which $q_I$ is not orthogonal. (Note the asymmetry of least vs.~most suppressed between magnetic and electric couplings.)

Hence we find
\begin{equation}
 \gmag \sim \sqrt{{\tilde q}^I a_{IJ} {\tilde q}^J} \sim
 \begin{cases}
    s^{2/3} & \text{if } {\tilde q} \in \cElimdir\\
    s^{1/6} & \text{if ${\tilde q}\in\cEker$, but ${\tilde q} \notin \cElimdir$}\\
    s^{-1/3} & \text{otherwise (i.e. if ${\tilde q}\notin\cEker$),}
 \end{cases}
\end{equation}
and similarly for the electric case,
\begin{equation}
 \gel \sim \sqrt{{q}_I a^{IJ} {q}_J} \sim
 \begin{cases}
    s^{1/3} & \text{if $q\in \cMkerorth$}\\
    s^{-1/6} & \text{if $q \in \cMlight$, but $q \notin \cMkerorth$}\\
    s^{-2/3} & \text{otherwise (i.e. if $q \notin \cMlight$).}
 \end{cases}
\end{equation}
Lastly,
\begin{equation}
 m \sim
 \begin{cases}
  s^{1/3} & \text{if $q \in \cMlight$}\\
  s^{-2/3} & \text{otherwise.}
 \end{cases}
\end{equation}

Combining this, we find
\begin{align}
 \zel \sim
 &\begin{cases}
    \frac{s^{-2/3}}{s^{-2/3}}\sim 1 & \text{if $q \notin \cMlight$}\\
    \frac{s^{-1/6}}{s^{1/3}}\sim s^{-1/2} & \text{if $q \in \cMlight$, but $q \notin \cMkerorth$}\\
    \frac{s^{1/3}}{s^{1/3}}\sim 1 & \text{if $q\in \cMkerorth$ (and so also $q \in \cMlight$)}
 \end{cases}
\\
 \zmag \sim
 &\begin{cases}
    \frac{s^{2/3}}{s^{2/3}}\sim 1 & \text{if ${\tilde q} \in \cElimdir$}\\
    \frac{s^{1/6}}{s^{2/3}}\sim s^{-1/2} & \text{if ${\tilde q}\in\cEker$, but ${\tilde q} \notin \cElimdir$}\\
    \frac{s^{-1/3}}{s^{-1/3}}\sim 1 & \text{if ${\tilde q}\notin\cEker$.}
 \end{cases}
\end{align}

\begin{table}[h]\centering

\begin{subtable}{1.0\textwidth}
\centering
    \begin{tabular}{|c|ccc|}
    \hline
    \multirow{2}{*}{${q}$ for this behavior} & \multicolumn{3}{c|}{Scaling with $s$}\\
    & $\gel$ & $m$ & $\zel$\\
    \hline
    $q \notin \cMlight$
        & $s^{-2/3}$    & $s^{-2/3}$    & $1$\\
    $q \in \cMlight$, but $q \notin \cMkerorth$
        & $s^{-1/6}$    & $s^{ 1/3}$    & $s^{-1/2}$\\
    $q\in \cMkerorth$ (so also $q \in \cMlight$)
        & $s^{ 1/3}$    & $s^{ 1/3}$    & $1$\\
    \hline
    \end{tabular}
\subcaption{Electric particles}
\label{table.5d.scaling.cF.sim.s.pow.2.electric}
    \end{subtable}
    
\vspace{.2cm}

\begin{subtable}{1.0\textwidth}
\centering
    \begin{tabular}{|c|ccc|}
    \hline
    \multirow{2}{*}{${\tilde q}$ for this behavior} & \multicolumn{3}{c|}{Scaling with $s$}\\
    & $\gmag$ & $T$ & $\zmag$\\
    \hline
    ${\tilde q} \in \cElimdir$
        & $s^{ 2/3}$    & $s^{ 2/3}$    & $1$\\
    ${\tilde q}\in\cEker$, but ${\tilde q} \notin \cElimdir$
        & $s^{ 1/6}$    & $s^{ 2/3}$    & $s^{-1/2}$\\
    ${\tilde q}\notin\cEker$
        & $s^{-1/3}$    & $s^{-1/3}$    & $1$\\
    \hline
    \end{tabular}
    \subcaption{Magnetic strings}
\label{table.5d.scaling.cF.sim.s.pow.2.magnetic}
    \end{subtable}

    \caption{The scaling behaviors of electric coupling, mass and charge-to-mass ratio of particles and magnetic coupling, tension, and charge-to-tension ratio of strings, in infinite-distance limits in which $\cF\sim s^2$.}
    \label{table.5d.scaling.cF.sim.s.pow.2}
\end{table}

The various $m$, $T$, $\gel$, $\gmag$, $\zel$ and $\zmag$ scaling behaviors are shown in \autoref{table.5d.scaling.cF.sim.s.pow.2}. One can show that a particle in the $n$th row of electric portion of this table has non-zero Dirac pairing with a string in $n$th row of the magnetic portion, and vice versa. Said differently, in limits of this type, we find particles and strings come in pairs such that $\zel\sim\zmag$. Note that in these limits, $\cElimdir \subseteq \cEker$ and $\cMkerorth \subseteq \cMlight$, which shows that any electric/magnetic charge can be sorted uniquely into one of the rows of \autoref{table.5d.scaling.cF.sim.s.pow.2}.

The rows of \autoref{table.5d.scaling.cF.sim.s.pow.2} can be given simple physical interpretations within the context of an emergent string limit \cite{Etheredge:2022opl}. The electrically charged particles with $\gel \sim m \sim s^{1/3}$ correspond (in an appropriate duality frame) to KK modes, while the magnetic strings with $\gmag \sim T \sim s^{-1/3}$ correspond to KK monopoles. The magnetic strings with $\gmag \sim T \sim s^{2/3}$ correspond to fundamental strings charged under a 2-form $B_2$, while the electric particles with $\gel \sim m \sim s^{-2/3}$ correspond to wrapped NS5-branes, which carry magnetic charge under $B_2$. Finally, the particles with $\gel \sim s^{-1/6}$, $m \sim s^{1/3}$ and the strings with $\gmag \sim s^{1/6}$, $T \sim s^{2/3}$ form part of a rigid field theory \cite{Blanco:2025qom} that decouples in the limit.

\begin{table}[h]\centering

   \begin{subtable}{1.0\textwidth}
\centering
    \begin{tabular}{|c|ccc|}
    \hline
    \multirow{2}{*}{${q}$ for this behavior} & \multicolumn{3}{c|}{Scaling with $s$}\\
    & $\gel$ & $m$ & $\zel$\\
    \hline
    $q \notin \cMkerorth \cap \cMlight$ but $q \in \cMlight$ & $s^{-1/3}$ & $s^{2/3}$ & $s^{-1}$\\
    $q \notin \cMlight$ & $s^{-1/3}$ & $s^{-1/3}$ & $1$\\
    $q \in \cMkerorth \cap \cMlight$, but $q \notin \cMtau$ & $s^{1/6}$ & $s^{2/3}$ & $s^{-1/2}$\\
    $q \in \cMtau$ & $s^{2/3}$ & $s^{2/3}$ & $1$\\
    \hline
    \end{tabular}
    \subcaption{Electric particles}
\label{table.5d.scaling.cF.sim.s.pow.1.electric}
\end{subtable}
    
\vspace{.2cm}

\begin{subtable}{1.0\textwidth}
\centering
    \begin{tabular}{|c|ccc|}
    \hline
    \multirow{2}{*}{${\tilde q}$ for this behavior} & \multicolumn{3}{c|}{Scaling with $s$}\\
    & $\gmag$ & $T$ & $\zmag$\\
    \hline
    ${\tilde q} \in \cEker$ & $s^{1/3}$ & $s^{4/3}$ & $s^{-1}$\\
    ${\tilde q} \in \cEker\oplus\cElimdir$, but ${\tilde q} \notin\cEker$ & $s^{1/3}$ & $s^{1/3}$ & $1$\\
    ${\tilde q} \in \cEperp$, but ${\tilde q} \notin \cEker\oplus\cElimdir$ & $s^{-1/6}$ & $s^{1/3}$ & $s^{-1/2}$\\
    ${\tilde q} \notin \cEperp$ & $s^{-2/3}$ & $s^{-2/3}$ & $1$\\
    \hline
    \end{tabular}
\subcaption{Magnetic strings}
\label{table.5d.scaling.cF.sim.s.pow.1.magnetic}
\end{subtable}

    \caption{The scaling behaviors of electric coupling, mass and charge-to-mass ratio of particles and magnetic coupling, tension, and charge-to-tension ratio of strings, in infinite-distance limits in which $\cF\sim s$.}
    \label{table.5d.scaling.cF.sim.s.pow.1}
\end{table}

In the former case of a decompactification limit (i.e., a limit with $\cF\sim s$), we can perform a similar analysis. For brevity, we omit the details, and instead provide the results of the possible scaling behaviors in \autoref{table.5d.scaling.cF.sim.s.pow.1}.
We note that in limits of this type, $\cEker \subsetneq \cEker \oplus \cElimdir \subseteq \cEperp$ and $\cMtau \subseteq \cMkerorth \cap \cMlight \subseteq \cMlight$, which shows that any electric/magnetic charge can be sorted uniquely into one of the rows of \autoref{table.5d.scaling.cF.sim.s.pow.1}.

While it is not immediate, one can again show that each particle in the $n$th row of the electric table has nonzero Dirac-pairing with a string in the $n$th row of the magnetic string. Since the $n$th row of the electric table has the same parametric scaling behavior for $\zel$ as the $n$th row of the magnetic table for $\zmag$, those particles and strings have $\zel\sim\zmag$.

 The rows of \autoref{table.5d.scaling.cF.sim.s.pow.1} again admit physical interpretations within the context of a decompactification limit. The electrically charged particles with $\gel \sim m \sim s^{2/3}$ correspond to KK modes in a decompactification limit to six dimensions, while the magnetic strings with $\gmag \sim T \sim s^{-2/3}$ correspond to KK monopoles. The magnetic strings with $\gmag \sim T \sim s^{1/3}$ correspond to fundamental strings charged under a 2-form $B_2$, while the electric particles with $\gel \sim m \sim s^{-1/3}$ correspond to wrapped strings of the 6d theory. These particles and strings exist in every decompactification limit.
 
 On the other hand, the particles with $\gel \sim s^{-1/3}$, $m \sim s^{2/3}$ and the strings with $\gmag \sim s^{1/3}$, $T \sim s^{4/3}$ exist only if $\cE_{\rm ker}$ is nontrivial. These particles and strings form part of a rigid field theory sector in six dimensions.

Let us now turn our attention to the case of a finite-distance limit. The requirement of a finite-distance boundary implies that the prepotential remains finite and nonzero in the $s\rightarrow 0$ limit, hence $C_{000} \neq 0$.

\begin{table}[h]\centering

\begin{subtable}{1.0\textwidth}
\centering
    \begin{tabular}{|c|ccc|}
    \hline
    \multirow{2}{*}{${q}$ for this behavior} & \multicolumn{3}{c|}{Scaling with $s$}\\
    & $\gel$ & $m$ & $\zel$\\
    \hline
    $q\in \cMkerorth$ and $q \notin \cMlight$
        & $1$           & $1$       & $1$\\
    $q\in \cMkerorth$ and $q \in \cMlight$
        & $1$           & $s$       & $s^{-1}$\\
    $q\notin \cMkerorth$ and $q \in \cMlight$
        & $s^{-1/2}$    & $s$       & $s^{-3/2}$\\
    $q\notin \cMkerorth$ and $q \notin \cMlight$
        & $s^{-1/2}$    & $1$       & $s^{-1/2}$\\
    \hline
    \end{tabular}
    \subcaption{Electric particles}
\label{table.5d.scaling.cF.sim.s.pow.0.electric}
\end{subtable}

    \vspace{.2cm}

   \begin{subtable}{1.0\textwidth}
\centering
    \begin{tabular}{|c|ccc|}
    \hline
    \multirow{2}{*}{${\tilde q}$ for this behavior} & \multicolumn{3}{c|}{Scaling with $s$}\\
    & $\gmag$ & $T$ & $\zmag$\\
    \hline
    ${\tilde q}\notin \cEperp$
        & $1$           & $1$       & $1$\\
    ${\tilde q}\in \cEperp$, but ${\tilde q} \notin \cEker$
        & $1$           & $s$       & $s^{-1}$\\
    ${\tilde q} \in \cEker$
        & $s^{1/2}$     & $s^{2}$   & $s^{-3/2}$\\
    \hline
    \end{tabular}
    \subcaption{Magnetic strings}
\label{table.5d.scaling.cF.sim.s.pow.0.magnetic}
\end{subtable}

    \caption{The scaling behaviors of electric coupling, mass and charge-to-mass ratio of particles and magnetic coupling, tension, and charge-to-tension ratio of strings, in finite-distance limits.}
    \label{table.5d.scaling.cF.sim.s.pow.0}
\end{table}

Following a similar analysis, one finds the scaling behaviors given in \autoref{table.5d.scaling.cF.sim.s.pow.0}.
Note that $\cEker \subseteq \cEperp$, which shows that any electric/magnetic charge can be sorted uniquely into one of the rows of \autoref{table.5d.scaling.cF.sim.s.pow.0}.

Neglecting the fourth row of \autoref{table.5d.scaling.cF.sim.s.pow.0.electric}, one can again show that each particle in the $n$th row of \autoref{table.5d.scaling.cF.sim.s.pow.0.electric} has nonzero Dirac-pairing with a string in the $n$th row of \autoref{table.5d.scaling.cF.sim.s.pow.0.magnetic}, so those particles have $\zel\sim\zmag$. The same applies swapping electric and magnetic.

However, there is no fourth row of \autoref{table.5d.scaling.cF.sim.s.pow.0.magnetic}, and indeed no row in that magnetic table has the $\zmag \sim s^{-1/2}$ scaling behavior required by the fourth row of \autoref{table.5d.scaling.cF.sim.s.pow.0.electric}. Hence any particle in the fourth row has no magnetic partner such that $\zel\sim\zmag$.
However, the fourth row of \autoref{table.5d.scaling.cF.sim.s.pow.0.electric} is non-empty if and only if the kernel of $C_{0IJ}$ is non-trivial. The third row of \autoref{table.5d.scaling.cF.sim.s.pow.0.magnetic}, and hence also of \autoref{table.5d.scaling.cF.sim.s.pow.0.electric}, is also non-empty if and only if the kernel of $C_{0IJ}$ is non-trivial. Hence, any theory with electric particles with $\zel \sim s^{-1/2}$ also has electric particles with the parametrically more divergent $\zel \sim s^{-2/3}$, the latter of which co-scales with a magnetic string.

Thus, the fourth row does not provide a counterexample to \autoref{ConA}: whenever such particles are present, they do not have maximally divergent $\zel$.

In some cases, these electric particles with $\zel \sim z^{-1/2}$ align with one of the particles with $\zel \sim z^{-3/2}$, and hence they are not even the maximally divergent electric particle aligning with that given direction. In fact, as we show momentarily, it is always possible to find a $q^\brap{L} \in \cMlight \setminus \cMkerorth$ that aligns with $q$. However, as we shall see in \S\ref{sec.sub.sub.KMV.intersection.of.2.SCFTs}, this charge $q^\brap{L}$ is not always in $\cM_\infty$, which means that there do not necessarily exist towers of particles of this charge.

We now explain the process of finding such a $q^\brap{L} \in \cMlight \setminus \cMkerorth$.
We begin with a charge $q$ with $\zel \sim z^{-1/2}$, i.e., $q\notin \cMkerorth$ and $q \notin \cMlight$.
We can uniquely write
\begin{equation}
    q = q^\brap{{\text{ker}}} + q^\brap{\text{ker}^\bot}\, \text{, where}~
    q^\brap{{\text{ker}}} \in \ker\brap{\delta^{IK}C_{0KL} \delta^{LJ}}\,,~
    q^\brap{\text{ker}^\bot} \in \bras{\ker\brap{\delta^{IK}C_{0KL} \delta^{LJ}}}^\orthogcomplement\,,
\end{equation}
and the orthogonal complement is taken with respect to the $\delta^{ij}$ metric.

We know that $q^\brap{{\text{ker}}}\notin\cMkerorth$ and $q^\brap{\text{ker}^\bot}\in\cMkerorth$.
We also know $q^\brap{{\text{ker}}}\neq0$, since $q\notin \cMkerorth$.
Hence in the limit $s\rightarrow\infty$, $\frac{q}{\bravv{q}} \rightarrow \frac{q^\brap{{\text{ker}}}}{\bravv{q^\brap{{\text{ker}}}}}$, so in particular $q$ and $q^\brap{{\text{ker}}}$ align.

Suppose we could find a $q^\brap{L}\in\cMlight$ such that we can uniquely write $q^\brap{L} = q^\brap{{\text{ker}}} + q^\brap{\text{ker}^\bot;L}$, for some $q^\brap{\text{ker}^\bot;L}\in \bras{\ker\brap{\delta^{IK}C_{0KL} \delta^{LJ}}}^\orthogcomplement$.
Then by the same arguments as for $q$, $q^\brap{L}$ would align with $q^\brap{{\text{ker}}}$ and hence also with $q$.

We now show that we can find such a $q^\brap{L}$ via proof by contradiction. Suppose that we cannot find such a $q^\brap{L}$. $\cMlight$ is codimension-1, so for us to not be able to intersect it by starting at $q^\brap{{\text{ker}}}$ and adding on elements in $\bras{\ker\brap{\delta^{IK}C_{0KL} \delta^{LJ}}}^\orthogcomplement$ we would have to have
$\bras{\ker\brap{\delta^{IK}C_{0KL} \delta^{LJ}}}^\orthogcomplement \subseteq \cMlight$ (i.e., changing $q^\brap{\text{ker}^\bot;L}$ is moving parallel to $\cMlight$).
Taking the orthogonal complement, we find
$
    {\ker\brap{\delta^{IK}C_{0KL} \delta^{LJ}}} \supseteq \cMlight^\orthogcomplement = \brac{q:q_I\propto\delta_I^0}
$,
so in particular the charge $p_I=\delta_I^0$ is in ${\ker\brap{\delta^{IK}C_{0KL} \delta^{LJ}}}$, and so $C_{00I}=0$ for all $I$.
But this limit is at finite distance, so $C_{000}\neq0$, and hence we have reached a contradiction.

If $q^\brap{L}$ lies in the infinity cone $\cM_\infty$, then there exists of tower of particles of charge proportional to $q^\brap{L}$ that (a) aligns with $q$, (b) has maximally divergent $\zel \sim s^{-3/2}$, and (c) aligns and co-scales. Once again, however, we stress that this $q^\brap{L}$ need not reside in $\cM_\infty$.

In M-theory compactifications on Calabi-Yau threefolds, the scaling behavior in the second rows (both electric and magnetic) of \autoref{table.5d.scaling.cF.sim.s.pow.0}, i.e.,
\begin{equation}
 m \sim s\,,~~~T \sim s\,,~~~\frac{1}{\gel^2 }\sim \gmag^2 \sim s^0\,,
 \label{5dSU2}
\end{equation}
is realized at SU(2) boundaries. Here, the magnetic string of charge $\tilde q \in \cEperp \setminus \cEker$ corresponds to a divisor that collapses to a curve. Physically, the light particles of mass $m$ are W-bosons, while the light string of tension $T$ is an 't~Hooft-Polyakov monopole. We will discuss such SU(2) boundaries further in \S\ref{ss.infinitetowers} below.

Meanwhile, the scaling behavior in the third rows (both electric and magnetic) of \autoref{table.5d.scaling.cF.sim.s.pow.0}, i.e.,
\begin{equation}
 m \sim s\,,~~~T \sim s^2\,,~~~\frac{1}{\gel^2 } \sim \gmag^2 \sim s \,,
 \label{5dreal}
\end{equation}
is realized at SCFT boundaries of the moduli space. Here, the magnetic string of charge $\tilde q \in \cEker$ corresponds to divisor of the Calabi-Yau manifold that shrinks to a point.

That scaling behavior \eqref{5dreal} is also special in that it involves a decoupling of the Planck scale. Introducing the canonically normalized scalar field
\be
\phi \sim \frac{1}{\sqrt{2} \kappa_5} \frac{2}{3} s^{3/2}
 \,,
\ee
we have
\begin{equation}
m \sim \sqrt{2 \pi T} \sim \frac{1}{g^2} \sim \phi^{2/3}\,.
\end{equation}
In other words, the mass scale $m$, the string scale $\sqrt{2 \pi T}$, and the gauge coupling scale $1/g^2$---all of which have dimension one in 5d---depend only on the vev of the scalar field $\phi$ and not on the Planck scale, vanishing in the SCFT limit $\phi \rightarrow 0$. This is precisely what we would expect for an SCFT sector that is decoupling from gravity in this limit.

We will discuss such SCFT boundaries further in \S\ref{ss.SCFT} below.

\subsection{A sufficient condition for co-scaling and alignment}\label{ss.sufficient}

In this subsection, we present a sufficient condition for co-scaling and (rapid) alignment in 5d supergravity. In particular, let us suppose that there exists a BPS particle of  electric charge $q_I$ and a BPS string of magnetic charge $\tilde q^I$ which are related by%
\footnote{The authors of \cite{Blanco:2025qom} showed that under the similar condition $q_I=\lambda a_{IJ}{\tilde q}^J$, the electric particles and magnetic monopoles exactly co-scale, i.e., $\zel=\zmag$.
It is also immediate that they exactly align: $\varphi=0$.
This condition reduces to \eqref{eqn.relation.between.electric.and.magnetic.charge.for.theta.going.like.1.over.z.linear} in a rigid limit at leading order.
}
\begin{equation}
 q_I=\lambda \cF_{IJ}{\tilde q}^J\label{eqn.relation.between.electric.and.magnetic.charge.for.theta.going.like.1.over.z.linear}\,,
\end{equation}
for some scalar $\lambda$.
We then claim that these objects exhibit co-scaling and rapid alignment.

From the gauge kinetic matrix \eqref{eq:gaugekinetic},
\begin{equation}
  a_{I J} = \frac{\mathcal{F}_I \mathcal{F}_J}{\cF^{4/3}} - \frac{\mathcal{F}_{I J}}{\cF^{1/3}} ,
\end{equation}
one can show \cite{Reece:2024wrn} (using $Y^J\cF_{IJ}=2\cF_I$ and $Y^I\cF_I=3\cF$) that the inverse gauge kinetic matrix is given by
\begin{equation}
 a^{IJ} = \frac{Y^I Y^J}{2\cF^{2/3}} - {\cF^{1/3}}{\mathcal{F}^{IJ}}\,,
\end{equation}
where $\cF^{IJ}$ is the inverse of $\cF_{IJ}$, i.e., $\cF_{IJ} \cF^{JK} = \delta_I^K$.
Hence, a magnetic string of charge $\tilde q$ has tension and coupling
\begin{equation}\label{eqn.magnetic.string.tension.and.coupling}
    T = T_\ast \brap{\frac{{\tilde q}^I\cF_I}{\cF^{2/3}}}
    \quad,\quad
    \bravv{\tilde q}^2 = {-\frac{{\tilde q}^I\cF_{IJ}{\tilde q}^J}{\cF^{1/3}}+\brap{\frac{T}{T_\ast}}^2}\,,
\end{equation}
and an electric particle of charge $q$ has mass and coupling
\begin{equation}
    m = \frac{m_\ast}{\sqrt2}\brap{\frac{q_IY^I}{\cF^{1/3}}}
    \quad,\quad
   \lVert q \rVert^2 = -\cF^{1/3}q_I\cF^{IJ}q_J + \brap{\frac{m}{m_\ast}}^2\,,
\end{equation}
where $T_\ast=\frac{2\pi}{\sqrt{2}g_5}=\brap{\frac\pi2}^{1/3}$ and $m_\ast=g_5 = 2^{\frac56}\pi^{\frac23}$, where here and throughout this subsection have set $M_{\text{Pl};5}^3 = \kappa_5^{-2} =1$.

For later convenience, we define the constants ${\zmag}_\ast=\sqrt2$, ${\zel}_\ast=1$ so that $\zel$ and $\zmag$ in \eqref{eqn.def.z.5d} are given by
\begin{equation}
    \frac{\zmag}{{\zmag}_\ast} = \bravv{\tilde q} \frac{T_\ast}{T}
    \quad,\quad
    \frac{\zel}{{\zel}_\ast} = \bravv{q}\frac{m_\ast}{m}\,.
\end{equation}

Suppose now that $q$ and $\tilde q$ are related by \eqref{eqn.relation.between.electric.and.magnetic.charge.for.theta.going.like.1.over.z.linear}.
We thus have
\begin{eqnarray}
 \frac{m}{m_\ast} &=& \sqrt{2}\lambda\cF^{1/3} \brap{\frac{T}{T_\ast}}\label{eqn.m.in.terms.of.T.for.q.related.to.qtilde}\,,\\
 \cF^{1/3}\bras{q_I\cF^{IJ}q_J} &=& \lambda^2\cF^{2/3}\bras{\frac{{\tilde q}^I\cF_{IJ}{\tilde q}^J}{\cF^{1/3}}}\,,
\end{eqnarray}
and so
\begin{eqnarray}
 \bravv{q}^2
 &=& -\cF^{1/3}\bras{q_I\cF^{IJ}q_J} + \brap{\frac{m}{m_\ast}}^2\\
 &=& \lambda^2\cF^{2/3}\brac{ -{\frac{{\tilde q}^I\cF_{IJ}{\tilde q}^J}{\cF^{1/3}}} + 2\brap{\frac{T}{T_\ast}}^2 }\\
 &=& \lambda^2\cF^{2/3}{\bravv{\tilde q}^2}\bras{1+\brap{\frac{{\zmag}_\ast}{\zmag}}^2}\,.
\end{eqnarray}
Using \eqref{eqn.m.in.terms.of.T.for.q.related.to.qtilde} for $m/m_\ast$, we then have
\begin{equation}
\frac{\zel}{\zmag} =
 \frac
 {\bravv{q}}{\bravv{\tilde q}} \cdot \frac{T}{T_\ast} \cdot \frac{m_\ast}{m} \cdot \frac{{\zel}_\ast}
 {{\zmag}_\ast}
 =\frac12\sqrt{1+\brap{\frac{{\zmag}_\ast}{\zmag}}^2}
 \label{eqn.coscaling.q.related.to.qtilde}\,,
\end{equation}
which is order-one, since from the enumeration of $\zmag$ scaling behaviors in \S\ref{s.coscaling} we have $\zmag \gtrsim 1$. Thus, the particle of charge $q_I$ and the string of charge $\tilde q^I$ exhibit co-scaling.

We now show that these also demonstrate (rapid) alignment.
We have
\begin{eqnarray}
    \brap{\frac{T}{T_\ast}}^2 &=& \brap{\frac{{\zmag}_\ast}{\zmag}}^2\bravv{\tilde q}^2\\
    &=&  \brap{\frac{{\zmag}_\ast}{\zmag}}^2\brap{-\frac{{\tilde q}^I \cF_{IJ} {\tilde q}^J}{\cF^{1/3}} + \brap{\frac{T}{T_\ast}}^2}\,.
\end{eqnarray}
Rearranging, we obtain
\begin{equation}
 \brap{\frac{T}{T_\ast}}^2 = \brap{-\frac{{\tilde q}^I \cF_{IJ} {\tilde q}^J}{\cF^{1/3}}}\bras{\frac1{\brap{\frac{\zmag}{{\zmag}_\ast}}^2-1}}\,,
\end{equation}
and so substituting this back into expression \eqref{eqn.magnetic.string.tension.and.coupling} for $\bravv{\tilde q}^2$, we find
\begin{equation}
 \bravv{\tilde q}^2 = -\frac{{\tilde q}^I \cF_{IJ} {\tilde q}^J}{\cF^{1/3}}\bras{\frac{\brap{\frac{\zmag}{{\zmag}_\ast}}^2}{\brap{\frac{\zmag}{{\zmag}_\ast}}^2-1}}\,.
\end{equation}
Following similar reasoning to the magnetic case, we find
\begin{equation}
 \bravv{q}^2 = -{\cF^{1/3}}{{q}_I \cF^{IJ} {q}_J}\brap{\frac{\brap{\frac{\zel}{{\zel}_\ast}}^2}{\brap{\frac{\zel}{{\zel}_\ast}}^2-1}}\,.
\end{equation}

Combining the electric and magnetic result, we find that the angle $\varphi$ between the electric and magnetic charges is given by
\begin{eqnarray}
 \cos\varphi &=& \frac{q_I {\tilde q}^I}{\bravv{q} \bravv{\tilde q}}\\
 &=&
 \frac{q_I \tilde q^I}{\sqrt{ {({q}_I \cF^{IJ} {q}_J})( {{\tilde q}^I \cF_{IJ} {\tilde q}^J)} }}
 \times
 \brac{\bras{1-\brap{\frac{{\zel}_\ast}{{\zel}}}^2}\bras{1-\brap{\frac{{\zmag}_\ast}{{\zmag}}}^2}}^{\frac12}\,.
\end{eqnarray}

Under the assumption \eqref{eqn.relation.between.electric.and.magnetic.charge.for.theta.going.like.1.over.z.linear}, we have
\begin{eqnarray}
 {{q}_I \cF^{IJ} {q}_J} &=& \lambda {q_I {\tilde q}^I}\,,\\
 {{\tilde q}^I \cF_{IJ} {\tilde q}^J} &=& \frac1\lambda {q_I {\tilde q}^I}\,,
\end{eqnarray}
yielding $\frac{q_I {\tilde q}^I}{\sqrt{ {({q}_I \cF^{IJ} {q}_J})( {{\tilde q}^I \cF_{IJ} {\tilde q}^J)} }}=1$. Substituting this back into the previous expression, we find
\begin{equation}
    \cos\varphi = \brac{\bras{1-\brap{\frac{{\zel}_\ast}{{\zel}}}^2}\bras{1-\brap{\frac{{\zmag}_\ast}{{\zmag}}}^2}}^{\frac12}\,.
\end{equation}

This is order 1 since we do not have $\zmag \ll 1$.
If further $\zmag\rightarrow\infty$ (and so also $\zel\rightarrow\infty$)
\begin{equation}
 \cos\varphi = 
 \bras{1-\frac12\brap{\frac{{\zel}_\ast}{{\zel}}}^2-\frac12\brap{\frac{{\zmag}_\ast}{{\zmag}}}^2+\dots}\,.
\end{equation}
Hence
\begin{equation}
    \varphi \sim \frac1z\,,
    \label{5drapid}
\end{equation}
so by the definition in \eqref{eq:rapidalign}, the particle and the string exhibit rapid alignment.

\subsection{SCFT boundaries and rapid alignment}\label{ss.SCFT}

We have seen in \S\ref{s.coscaling} that light particles and strings at SCFT boundaries follow the scaling behavior \eqref{5dreal}, which implies that their charge-to-mass vectors co-scale as
\be
z \equiv |\vec{z}_{\rm el}| \sim |\vec{z}_{\rm mag}|  \sim s^{-3/2}  \,.
\ee

We will now argue that these electric and magnetic charge-to-mass vectors also exhibit rapid alignment.
By the argument of \S\ref{ss.sufficient}, it suffices to show that the light particles and strings satisfy  \eqref{eqn.relation.between.electric.and.magnetic.charge.for.theta.going.like.1.over.z.linear}.

For this, let us first suppose that our SCFT boundary represents a codimension-1 facet of the moduli space in homogeneous coordinates (i.e., the K\"ahler cone) at $Y^1 = 0$.
More precisely, we assume that there exists a tensionless string and a light tower of particles with the scaling of \eqref{5dreal} in the limit $Y^1 = s \rightarrow 0$ for any fixed values of $Y^I = Y^I_0$, $I \neq 1$.
In other words, we drop the previous restriction $Y^0 = 1$, $Y^I  = sn^I$ and instead set
\begin{align}
Y^I = Y^I_0 + s Y^I_1\,,~~~Y^I_1 = \delta^I_1\,,~~~Y^1_0 = 0\,.
\end{align}
In the limit $s \rightarrow 0$, the tensionless string charge $\tilde q^I$ then lies in the kernel of the symmetric matrix $C_{IJK} Y_0^K$, so that
\be
 \tilde q^I C_{IJK} Y_0^K = 0 ~~\text{for all } J \,.
\ee
But now, allowing $Y_0^K$ to vary along the codimension-1 boundary with $Y^1 = s = 0$, we conclude that
\be
 \tilde q^I C_{IJK}  = 0 ~~\text{for all } (J, K) \neq (1, 1)   \,. \label{cijkid}
\ee
Since $q_I\propto \delta_I^1$, we can rewrite this as
\be
 \tilde q^I C_{IJK} = \tilde q^I C_{I11} \delta_J^1 \delta_K^1 = cq_Jq_K\,,
\ee
for some constant $c$.
Contracting with $Y^K$ and rearranging, we find \eqref{eqn.relation.between.electric.and.magnetic.charge.for.theta.going.like.1.over.z.linear} with $\lambda=\frac{2}{c\brap{q_IY^I}}$. By the argument in \S\ref{ss.sufficient}, this establishes the claim that $\varphi \sim 1/z$, i.e., the light particles and the light string exhibit rapid alignment.

We may extend this result to the case of a codimension-$p$ boundary as follows. Suppose that our boundary locus lies at $Y^1 = Y^2 = \cdots = Y^p = 0$. Then, by an analogous argument, we have 
\be
 \tilde q^I C_{IJK}  = 0 ~~\text{unless } J, K \in \{1, ..., p \}   \,. \label{cijkid2}
\ee
Contracting with $Y^K$, we have
\be
\tilde q^I C_{IJK} Y^K = c_{jk}Y^k\,, 
\ee
where now $j, k \in \{1, ..., p\}$ and $c_{jk} = \tilde q^I C_{Ijk}$ are constants. This implies that \eqref{eqn.relation.between.electric.and.magnetic.charge.for.theta.going.like.1.over.z.linear} is satisfied for a particle of charge
\be
q_I \propto \delta_I^j c_{jk} Y^k \,,
\ee
which notably becomes massless at the codimension-$p$ boundary, where $Y^k = 0$ for all $k$.

A priori, it is unclear that BPS particles of this charge exist (or even that this charge must lie in the Mori cone of the compactification manifold), since (in an appropriate basis) the effectiveness of $q_I$ requires non-negativity of all $c_{jk} Y^k$. Nonetheless, we shall verify this result in examples below.

\subsubsection{Chern-Simons terms and anomaly inflow}

We have seen that SCFT boundaries feature light particles and light strings. In the basis introduced in \S\ref{ss.SCFT}, where the SCFT boundary lies at $Y^1 = 0$, the light particles have charge $q_I \propto \delta_I^1$. However, the magnetic charge $\tilde q^I$ is not simply a multiple of $\delta^I_1$ in this basis. This raises the question: is there a simple way to understand the relationship between the magnetic and electric charges? As we now explain, the two are indeed related via anomaly inflow on the magnetic string worldvolume.

To begin, we once again assume that the SCFT boundary is a codimension-1 facet of the extended K\"ahler cone with $Y^1 = 0$. 
This implies \eqref{cijkid}:
\be
 \tilde q^I C_{IJK}  = 0 ~~\text{for all } (J, K) \neq (1, 1)   \,. \label{cijkid3}
\ee

Next, we set $B_1$ to be the 1-form gauge field to which the tensionless string magnetically couples. The Chern-Simons coupling of $B_1$ in the action is given by \cite{Alim:2021vhs}
\be
S \supset \frac{1}{6(2 \pi)^2} \int \tilde q^I C_{IJK} B_1 \wedge F^J \wedge F^K\,.
\ee
From \eqref{cijkid3}, this becomes
\be
\frac{1}{6(2 \pi)^2} \int B_1 \wedge F^1 \wedge F^1 \,.
\ee
As discussed in \cite{Callan:1984sa, Naculich:1987ci} (see reviews in \cite{Harvey:2005it, Heidenreich:2021yda}), in the presence of this Chern-Simons coupling, a string charged magnetically under $B_1$ will admit zero-mode excitations charged electrically under the gauge field $A^1$, due to anomaly inflow. The mass scale of these excitations is given simply by the string scale, $m \sim \sqrt{2 \pi T}$. This agrees precisely with the scaling of the light particles in \eqref{5dreal}, which indeed carry charge only under $A^1$.

We expect that these string excitations will come in infinite towers; this expectation is borne out by GV invariants of the charged particles. In contrast, however, we expect that the tensionless strings will not come in infinite towers, as the corresponding collapsing divisors are rigid. Intuitively, a stack of M5-branes wrapping such a divisor cannot be deformed to give a single multi-wrapped M5-brane, so a charge $k \tilde q$ multi-string state cannot be deformed to single string of that charge.\footnote{We thank Timo Weigand and Max Wiesner for explaining this to us.} However, it is plausible that there may nonetheless exist a skew tower of BPS strings of charge $\tilde q' + k \tilde q$ for some $\tilde q'$.

\subsubsection{The species scale}\label{ss.5dspecies}

As a final comment on SCFT boundaries, let us consider the behavior of the species scale. In \cite{vandeHeisteeg:2022btw, vandeHeisteeg:2023dlw, Castellano:2023aum} it was argued that gravitational higher-derivative corrections should take the schematic form
\begin{equation}
    S = \frac{M_{\text{Pl};d}^{d-2}}{2} \int d^dx \sqrt{-g} \left( R  
 + \frac{\tilde c_2}{\Lambda_{\rm QG}^2 } R^2 + \frac{\tilde c_3}{\Lambda_{\rm QG}^4 } R^r + ...\right)\,,
 \label{speciescoef}
\end{equation}
where $\Lambda_{\rm QG}$ is the species scale and the $\tilde c_i$'s are order-one numbers. In other words, the $R^2$ terms in the low-energy effective action are suppressed by a factor that is no larger than $\LQG^2$ in Planck units.\footnote{An argument for this must deal with scale-dependence in a careful way, or assume sufficient supersymmetry; in the real world, at low energies, $R^2$ and higher terms are generated suppressed by neutrino masses. See~\cite{Bedroya:2024ubj} for some discussion along these lines, and~\cite{Calderon-Infante:2025ldq} for discussion of how the mass scale of towers of states and the species scale both appear in a double EFT expansion.}
By computing this coefficient, we may therefore place a lower bound on the species scale (up to order-one factors).

Via \eqref{speciescoef}, we expect that the $R^2$ coefficient will take the form
\be
\frac{\tilde c_2 M_{\text{Pl};5}^3}{\Lambda_{\rm QG}^2} R^2\,,
\ee
in five dimensions, where $\tilde c_2$ is an order-one number.

As shown in \cite{Hanaki:2006pj}, the $R^2$ coefficient of a 5d supergravity theory takes the schematic form
\begin{align}
    \mathcal{L}  \supset \frac{C_I Y^I}{\cF^{1/3}} R^2\,,
\end{align}
where the $C_I$ are constants and the $\cF^{1/3}$ scaling with the prepotential is fixed by invariance under homogeneous scaling $Y^I \rightarrow \lambda Y^I$. In a Calabi-Yau compactification, the coefficients $C_I$ are determined by the second Chern class $c_2$ \cite{Grimm:2017okk}:
\be
C_I Y^I \propto \int  c_2 \wedge J = Y^I \int c_2 \wedge \omega_I \,,
\ee
where $J= \omega_I Y^I$ is the K\"ahler form and $\omega_I$ represent a basis of 2-cycles.

Let us specialize to a limit of the form studied above, where $Y^0 =1$, $Y^1=s$, $Y^I=0$, $I>1$. If $C_0 = 0$, then the $R^2$ coefficient vanishes in the limit $s \rightarrow 0$, indicating a breakdown of the relationship between this coefficient and the species scale.\footnote{A vanishing $R^2$ coefficient would naively imply a diverging species scale $\LQG \gg M_{\rm Pl;5
}$, which is clearly unphysical.}
Assuming instead that $C_0 > 0$,\footnote{Miyaoka has proven that $\int c_2 \wedge J$ is non-negative \cite{Miyaoka} (see also \cite{kanazawawilson}). However, this establishes only the weaker constraint $C_0 \geq 0$.} which is true in all examples of SCFT boundaries we have studied, we have
\be
\Lambda_{\rm QG}^2 \sim \cF^{1/3}\,.
\ee

For the case of an infinite-distance limit, we have that $\cF \rightarrow 0$ as $s \rightarrow 0$. In the case where $\cF$ vanishes linearly, $\cF \sim s$, a string of charge $\tilde q^I = \delta_0^I$ is asymptotically tensionless, with scaling behavior given by the second row of \autoref{table.5d.scaling.cF.sim.s.pow.1.magnetic}:
\be
T \sim s^{1/3}\,.
\ee
Thus we have
\be
\Lambda_{\rm QG} \sim \sqrt{2 \pi T} \sim s^{1/6} \,,
\ee
so the species scale may be identified with the string scale, up to an order-one constant. 

Similarly, when $\cF$ vanishes quadratically, $\cF \sim s^2$, we find that a string of charge $\tilde q^I = \delta_0^I$ is again asymptotically tensionless, with
\be
\Lambda_{\rm QG} \sim \sqrt{2 \pi T} \sim s^{1/3} \,,
\ee
so again the species scale may be identified with the string scale, up to an order-one constant.

In the case of a finite-distance SCFT boundary, on the other hand, the prepotential $\cF$ remains finite in the limit $s \rightarrow 0$, and therefore the species scale remains near the Planck scale even as the SCFT string becomes tensionless. This SCFT string decouples from the gravitational dynamics and does not produce a breakdown of non-gravitational EFT below the Planck scale.

\subsection{Infinite towers vs.~isolated states and the magnetic infinity cone}\label{ss.infinitetowers}

In the previous subsection, we focused on SCFT boundaries, which feature a tower of light, electrically charged BPS particles. We argued that these boundaries also feature magnetically charged BPS strings, which together exhibit co-scaling and alignment.

However, as discussed in \S\ref{ss.CY3}, there are two other types of finite-distance boundaries of the K\"ahler cone: conifold loci and SU(2) boundaries. These boundaries may be traversed by flops and Weyl flops, respectively, and relatedly these boundaries feature only finitely many light BPS particles.

In this subsection, we explore the phenomenon of co-scaling and alignment at such boundaries. We argue that co-scaling need not occur in the absence of infinite towers of light particles, and we discuss the mathematical implications of this fact for curves as well as divisors.

\subsubsection{Flops and Weyl flops}\label{ss.flopWeylflop}

Let us begin by considering the case of a flop transition, whereby a finite number of BPS particles become massless. In accordance with the general argument of \S\ref{s.coscaling}, there exists a particle of charge $q_I$ whose mass and charge follow the scaling behavior of \eqref{5dSU2}.
However, as shown explicitly via an example in \S\ref{s.exinf}, there is no BPS string that becomes light at a flop transition. In particular, the magnetic charge $\tilde q^I$ specified in the second row of \autoref{table.5d.scaling.cF.sim.s.pow.0.magnetic} lies outside the effective cone, so no BPS strings of this charge can possibly exist.

This strongly suggests that co-scaling does not apply to isolated numbers of light particles but only to infinite towers. Indeed, this should come as little surprise: the string landscape is full of examples of massless charged particles, which need not be accompanied by tensionless monopoles. In our own universe, there is no evidence for a monopole whose mass is within a few orders of the electron mass.

This distinction between towers and isolated particles immediately raises questions regarding the magnetic strings. If co-scaling applies only to towers of electrically charged particles, we expect that it similarly should apply only to towers of magnetic monopoles. In general, the distinction between single-string BPS states and multi-string BPS states is subtle and poorly understood due to the vanishing binding energy between BPS strings. From a mathematical perspective, the distinction between infinite towers of divisors and multiply wound divisors is similarly subtle and poorly understood: it is at present impossible to compute the analog of GV invariants for divisors, which one might use to distinguish isolated light BPS string states from infinite towers of light BPS strings. Said differently, there has so far been no discussion in the literature (to our knowledge) of what we might call the \emph{infinity cone of divisors} $\mathcal{K}_\infty$, namely, the cone in $H_4(X, \mathbb{R})$ populated by infinite towers of BPS divisors. 

This distinction between towers and isolated states is especially important when it comes to SU(2) boundaries. At a codimension-1 SU(2) boundary, there exist both massless W-bosons and tensionless 't~Hooft-Polyakov monopole strings, which obey the co-scaling relation \eqref{5dSU2}. These W-bosons are not part of an infinite tower of BPS particles, and by S-duality, we expect that the 't~Hooft-Polyakov monopoles similarly will not come in infinite towers.\footnote{We thank Ben Heidenreich for discussions on this point.} 

However, there is an important distinction here between SU(2) boundaries associated with \emph{stable} Weyl reflections and those associated with \emph{unstable} Weyl reflections. The charge $q_W$ of a W-boson that becomes light at a stable Weyl reflection boundary lies strictly outside the infinity cone $\mathcal{M}_\infty$, whereas at an unstable Weyl reflection boundary $q_W$ lies on the boundary of the infinity cone. This means that at an unstable Weyl flop boundary, there exist a skew tower of BPS particles of charge $q'+k q_W$, $k\in \mathbb{Z}$, for some $q'$. The charge-to-mass vectors of the particles in the skew tower lie arbitrarily close to the charge-to-mass vector of the W-bosons in both magnitude and direction. This skew tower co-scales and aligns with the 't Hooft-Polyakov monopole string.

There is reason to suspect that the 't Hooft-Polyakov monopole associated with a stable Weyl reflection lies strictly outside $\mathcal{K}_\infty$. As discussed in \cite{wilson-92, Gendler:2022ztv}, at a generic point in complex structure moduli space, a genus $g > 1$ Weyl reflection is replaced by an ordinary flop, while a genus $g=1$ Weyl reflection is replaced by a smooth point. These Weyl flops are both stable, so the W-boson charge lies strictly outside the infinity cone $\cM_\infty$.\footnote{While all genus $g \geq 1$ Weyl flops are stable, it remains an open question whether or not all stable flops have genus $g \geq 1$. We thank Jakob Moritz for explaining this to us.}
Physically, this complex structure deformation corresponds to giving a vev to an adjoint hypermultiplet, which breaks SU(2) to U(1). Consequently, (as we will see in the example in \S\ref{ss.GHMMR}) the divisor associated with the light 't~Hooft-Polyakov monopole is no longer effective, as there are no tensionless BPS strings associated with an ordinary flop transition. Barring wall-crossing for BPS strings, there must not exist towers of effective strings outside the effective cone of the deformed Calabi-Yau, so the 't~Hooft-Polyakov string charge must lie strictly outside the magnetic infinity cone.

\subsubsection{Bounds on the magnetic infinity cone}

The preceding discussion highlights the importance of the magnetic infinity cone. In what follows, we use a combination of physics and geometry to place bounds on this cone.

We begin with the results of \cite{BPSstrings} (based on the earlier works \cite{Alim:2021vhs, Gendler:2022ztv}), which show that BPS black strings exist for all charge directions inside the hyperextended K\"ahler cone $\mathcal{K}_{\rm hyp}$. We may reasonably assume that these BPS black strings come in infinite towers,\footnote{We thank Ben Heidenreich for this observation.} since after dimensional reduction to four dimensions, these monopole strings become ordinary monopoles, which (by the Tower WGC \cite{Heidenreich:2016aqi, Andriolo:2018lvp}) should come in infinite towers. Thus, by the same logic that produced the containment relations \eqref{econtain} for the electric charge lattice, we may similarly conclude that 
\be
\cK_{\rm hyp} \subseteq \cC_{\rm BStr} \subseteq \cK_{\infty}
\ee
within the magnetic charge lattice.

Conversely, we also know that BPS strings do not exist outside the effective cone $\mathcal{E}$. Thus, the magnetic infinity cone $\mathcal{K}_{\infty}$ must lie inside the effective cone, $\cK_{\infty
} \subseteq \mathcal{E}$.

Furthermore, by the argument of \S\ref{ss.flopWeylflop}, we know that a genus $g \geq 1$ Weyl flop can be deformed to a smooth point or an ordinary flop. This deformation has the effect of shrinking the effective cone, so in the absence of wall-crossing for BPS strings, the magnetic infinity cone must be contained in the effective cone of the deformed Calabi-Yau.

The effective cone $\cE$ is dual to the cone of dual coordinates $\mathcal{T}$, which is parametrized by all possible values of $\cF_I = C_{IJK}Y^J Y^K$ for $Y^I \in \cK$. After the aforementioned complex structure deformation, the extended K\"ahler cone grows to encompass the entire hyperextended K\"ahler cone, i.e., $\cK' = \cK_{{\rm hyp}}$. Analogously, the cone of dual coordinates also grows to encompass the hyperextended cone of dual coordinates,  $\mathcal{T}' = \mathcal{T}_{\rm hyp}$, which is parametrized by all possible values of $\cF_I = C_{IJK}Y^J Y^K$ for $Y^I \in \cK_{{\rm hyp}}$.

We refer to the dual of the hyperextended cone of dual coordinates as the \emph{hypereffective cone}, $\cE_{{\rm hyp}} = \mathcal{T}_{\rm hyp}^\vee = \cE' $. Thus, the absence of wall-crossing implies that the magnetic infinity cone is contained in the hypereffective cone:
\be
\cK_{\infty} \subseteq \cE_{\rm hyp}\,.
\label{Khyp}
\ee
So, in summary, we have
\be
\mathcal{K} \subseteq \mathcal{K}_{\rm hyp} \subseteq  \cC_{\rm BStr} \subseteq \cK_{\infty} \subseteq \cE_{\rm hyp} \subseteq \cE \subseteq H_4(X, \mathbb{R})\,,
\ee
which parallels \eqref{econtain} in the electric case.

\subsubsection{Two novel conjectures}\label{ss.conj}

Thus far, we have placed upper and lower bounds on the size of the magnetic infinity cone. Unless these bounds coincide, however, we are unable to compute the magnetic infinity cone.

Faced with this obstacle, we shall turn the logic around: rather than using properties of 5d supergravity and Calabi-Yau geometry to argue for co-scaling and alignment, we will instead 
assume alignment and co-scaling and use it to justify a pair of novel mathematical conjectures.

We begin by defining a map from the magnetic charge space $H_4(X, \mathbb{R})$ to the electric charge $H_2(X, \mathbb{R})$ at a given point in the moduli space:
\begin{align}
\cS: ~&\cK_{\rm hyp} \times H_4(X, \mathbb{R}) \rightarrow H_2(X, \mathbb{R}) \\ \nonumber
& (Y^I, \tilde q^J) \mapsto \cF_{IJ}(Y) \tilde q^J .
\end{align}
Recall that $\cF_{IJ} \equiv C_{IJK} Y^K$, where the triple intersection numbers $C_{IJK}$ are constant inside the K\"ahler cone $\mathcal{K}_X$ but shift in a simple way under a flop or Weyl flop, and thus they depend weakly on the position $Y^I \in \cK_{\rm hyp}$.

With this definition in place, we may introduce our first conjecture:
\begin{conj}
    The image of $\cK_{\rm hyp} \times \cE_{\rm hyp}$ under the $\cS$-map is the electric infinity cone. That is, $\cS( \cK_{\rm hyp} \times \cE_{\rm hyp}) = \cM_{\infty}$.
    \label{Con1}
\end{conj}
The primary evidence for this conjecture is phenomenological: we have verified that it holds in the three examples in Section 7 of \cite{Alim:2021vhs}, the three examples in Appendix C of \cite{Gendler:2022ztv}, and the Calabi-Yau given by five bidegree $(1,1)$ hypersurfaces inside $\mathbb{P}^4 \times \mathbb{P}^4$.\footnote{We thank Callum Brodie, Naomi Gendler, and Ben Heidenreich for discussions and (unpublished) computations on this last geometry.} It is also a rather elegant result for two reasons. First, upon setting $\tilde q^I = Y^I$, the $\cS$-map reduces to the $\mathscr{T}$-map studied in \cite{Gendler:2022ztv}, i.e.,
\be
\cS(Y^I, Y^J) \equiv \mathscr{T}(Y^I)\,.
\ee
The image of $\mathscr{T}$ is (by definition) equal to the hyperextended cone of dual coordinates, which is known to be a subset of the electric infinity cone $\cM_\infty$. Here, by allowing $\tilde q^I$ to vary over the larger hypereffective cone, our map $\cS$ maps to another noteworthy cone, $\cM_\infty$.

More importantly, the result is elegant in that it is closely related to the phenomenon of co-scaling and alignment. 
We have seen in \S\ref{ss.sufficient} that a BPS string of charge $\tilde q^I$ and a BPS particle of charge $q_I \propto \cS(Y^I, \tilde q^J) = \cF_{IJ} \tilde q^J$ exhibit co-scaling and rapid alignment. Since, $\cK_\infty \subseteq \cE_{\rm hyp}$ by \eqref{Khyp}, \autoref{Con1} ensures that every BPS string tower co-scales and rapidly aligns with a corresponding tower of electrically charged particles. In other words, \autoref{Con1} implies \autoref{ConB}.

Note, however, that the converse is not true: for a fixed $Y^I \in \cK_{\rm hyp}$, $\cS(Y^I, \cE_{\rm hyp})$ may be a proper subset of $\cM_{\infty}$. This means that the inverse map $q_I \mapsto \cF^{IJ} q_J = \tilde q^I$ will take some charges $q_I \in \cM_\infty$ to charges outside $\cE_{\rm hyp}$. Therefore an electric tower with charge proportional to $q_I$ is not guaranteed to co-scale and rapidly align with any magnetic tower. Indeed, in \S\ref{ss.KMV}, we shall see an example of this where such an electric tower does co-scale and align with a BPS magnetic string, but it does not rapidly align.

Nonetheless, since \autoref{Con1} ensures that every $\tilde q^I \in \cE_{\rm hyp}$ maps to a charge in $\cM_\infty$, it is natural to expect that all such $\tilde q^I$ should correspond to infinite towers of strings. With this, we posit a more speculative conjecture:
\begin{conj}
    $\cE_{\rm hyp} = \cK_{\infty}$.
    \label{Con2}
\end{conj}
Here, let us emphasize that $\cK_\infty$ is defined to be the closure of the cone generated by infinite towers of BPS strings. As explained above in \S\ref{ss.SCFT}, we expect that strings which become light at SCFT boundaries do not come in infinite towers. However, such strings live on the boundary of the effective cone, which means that \autoref{Con2} may still be satisfied if there is a skew tower of charge $\tilde q' + k \tilde q_{\rm SCFT}$, $k \in \mathbb{Z}$, where $q_{\rm SCFT}$ is the charge of the SCFT string. Further analysis of Calabi-Yau geometry is needed to verify or disprove this hypothesis.

\subsection{Example: The GMSV geometry}\label{s.exinf}

In this section, we explore co-scaling and alignment in the UV-complete example of M-theory compactified on the GMSV geometry \cite{Greene:1995hu,Greene:1996dh} (see also \cite{Alim:2021vhs} for further details). The vector multiplet moduli space is one-dimensional and has two phases, which are related geometrically by a flop transition. The first phase features an asymptotic boundary corresponding to an emergent string limit, while the second features an SCFT boundary. In what follows, we will analyze (i) the asymptotic boundary, (ii) the SCFT boundary, and (iii) the flop transition. In the process, we will demonstrate that \autoref{Con1} is satisfied by explicit computation.

\subsubsection{Asymptotic boundary}

The prepotential in the first phase of the GMSV geometry takes the form
\be
\cF = \frac56 X^3+ 2 X^2 Y\,,
\label{GMSV.prepotential.phaseI}
\ee
where we have set $Y^0 = X$, $Y^1=Y$ for ease of notation. With this, the limit $X \rightarrow 0$ represents an infinite-distance, emergent string limit. Setting $X=s$, $Y=1$ in homogeneous coordinates, we find a tower of light particles of charge $q_I = k (1, 0)$ with mass 
\be
m = \frac{g_5}{\sqrt{2} \kappa_5}\frac{k s}{\cF^{1/3}} =  k \pi^{2/3}  s^{1/3} M_{{\rm Pl}; 5}+O(s^{4/3})\,.
\ee
The charge of these particles is given by 
\be
\lVert q \rVert^2 =  q_I a^{IJ} q_J =  2^{-4/3} k^2  s^{2/3} + O(s^{5/3})\,,
\ee
so $\zel \sim \lVert q \rVert/m \sim 1$ in the limit $s \rightarrow 0$.

Meanwhile, a string of charge $\tilde q^I = (\tilde q^0, \tilde q^1)$ has a tension of
\be
T = \frac{2 \pi}{\sqrt{2} g_5 \kappa_5} \frac{|\tilde q^0 \cF_X + \tilde q^1 \cF_Y |}{\cF^{2/3}} = M_{{\rm Pl}; 5}^2 \left( 2 \pi^{1/3} \tilde q^0 s^{-1/3} + \frac{\pi^{1/3}}{36} (25 \tilde q^0 + 36 \tilde q^1) s^{2/3} + O(s^{5/3}) \right)\,.
\ee
The limit $s \rightarrow 0$ is an emergent string limit, as a string of charge $\tilde q^I = (0, 1)$ becomes tensionless. There are also BPS strings of charge $\tilde q^0 \neq 0$ that become heavy in the limit $s \rightarrow 0$, with $T \sim s^{-1/3}$. All of these strings have $\zmag \sim 1$, so the co-scaling relation \eqref{rr} is satisfied trivially.

\subsubsection{SCFT boundary}\label{ss.GS}

A more sophisticated example of co-scaling and alignment occurs in the second phase of the GMSV geometry. Here, the prepotential is given by
\be
\cF = \frac{5}{6} X^3 + 8 X^2 Y + 24 X Y^2 + 24 Y^3\,,
\ee
where the phase II coordinates are related to the coordinates in phase I via $X^{(II)} = X^{(I)} + 4 Y^{(I)}$ and $Y^{(II)} = Y^{(I)}$. We will work in the phase II basis for the remainder of \S\ref{s.exinf} (except in \S\ref{ss.GMSVverify}, where we verify that \autoref{Con1} is satisfied in phase I).

The boundary $X = 0$ of this geometry represents an SCFT boundary. Setting $X = s$, $Y=1$, we find a tower of light particles of charge $q_I = (k, 0)$ and mass
\be
m = \frac{g_5 }{\sqrt{2} \kappa_5} \frac{k |X|}{\cF^{1/3}}= \left(\frac{\pi^2}{12}\right)^{1/3} s k M_{{\rm Pl}; 5}+O(s^2)\,.
\ee
The charge of these particles is given by
\be
\lVert q \rVert^2 =  q_I a^{IJ} q_J = 2 \cdot 3 ^{4/3} k^2 s^{-1}  + O(s^0)\,.
\ee
Towers of BPS particles exist everywhere inside the electric infinity cone $\cM_\infty$ \cite{Gendler:2022ztv}, which is generated by $q_I = (1, 0)$ and $q_I = (1, 4)$, so particles of charge $q_I=(k,0)$ do exist.
Meanwhile, the effective cone of divisors $\cE$ is generated by $\tilde q^I = (-3, 1)$ and $\tilde q^I = (4, -1)$ \cite{Alim:2021vhs}.

The string of charge $\tilde q^I = (-3,1)$ is tensionless with tension
\be
T = \frac{2 \pi}{\sqrt{2}g_5 \kappa_5} \frac{|3 \cF_{X} - \cF_{Y}|}{\cF^{2/3}} = \frac{1}{8}\left(\frac{\pi}{18}\right)^{1/3} s^2 M_{{\rm Pl}; 5}^2 + O(s^3)\,.
\ee
and charge
\begin{equation}
    \bravv{\tilde q}^2 = {\tilde q}^Ia_{IJ}{\tilde q}^J = \frac12 \cdot 3^{2/3} s + O\brap{s^2}\,.
\end{equation}
Together, these obey the scaling behavior \eqref{5dreal}.

Furthermore, the light particle and string charges $q_I = (1, 0)$, $\tilde q^I = (-3, 1)$ satisfy the relation \eqref{eqn.relation.between.electric.and.magnetic.charge.for.theta.going.like.1.over.z.linear}, since
\be
\cF_{IJ} \tilde q^I = (X, 0) = X q_I\,,
\ee
so we further expect that they exhibit rapid alignment near the SCFT boundary. We may verify this explicitly by computing the angle between the electric charge vector $q_I = (1, 0)$ and the magnetic charge vector $\tilde q^I = (3, -1)$ in an orthonormal basis, which is given by
\be
\cos \varphi = \frac{q_I \tilde q^I}{\lVert q \rVert \cdot \lVert \tilde q \rVert }\,.
\ee
By explicit computation, we find
\be
\cos \varphi = 1 - \frac{1}{192} s^3 + O(S^4) ~~~ \Rightarrow ~~~ \varphi = \frac{1}{96} s^{3/2} + O(s^{5/2}) \sim \frac{1}{z} \,,
\label{GMSValign}
\ee
which agrees with \eqref{5drapid} and ensures rapid alignment.

In addition to these light particles, there are also some heavy particles in $\cM_\infty$, namely $q_I=k\brap{1,0}+l\brap{1,4}$ for $l>0$, $k\geq0$.
Those with $l=3k$ have $\zel\sim s^0$., whilst those with $l\neq3k$ have
\begin{equation}
    m=\brap{\frac{16\pi^2}{3}}^{1/3}l + O\brap{s}\,,~~~
    \bravv{q}^2=\frac{2\brap{3k-l}^2}{3^{2/3}}s^{-1} + O\brap{s^0}\,,
\end{equation}
and so $\zel\sim s^{-1/2}$.
All other strings in $\cE$ have $\zmag\sim s^0$, so as found in \S\ref{s.coscaling}, we have $\zel\sim s^{-1/2}$ particles with no partner-string such that $\zel\sim\zmag$.
We now demonstrate how the method in \S\ref{s.coscaling} for finding another tower aligning with the $\zel\sim s^{-1/2}$ tower that has parametrically larger $\zel\sim s^{-3/2}$, and co-scales and aligns with a $\zmag\sim s^{-3/2}$ string works.
From the analysis above, we know that this method should produce a $q_I=(k',0)$ for some $k'\neq0$ (we allow also negative $k$, since this gives an anti-BPS electric tower aligning with the original electric tower, or we can negate the charge and find a BPS electric tower that anti-aligns with the original electric tower).

Following the notation in \S\ref{s.coscaling}, starting with $q_I=k\brap{1,0}+l\brap{1,4}$ for $l>0$, $k\geq0$, $l\neq3k$, we find
$q^\brap{{\text{ker}}}=\frac{l-3k}{10}\brap{-3,1}$
$q^\brap{\text{ker}^\bot}=\frac{k+13l}{10}\brap{1,3}$.
Taking $q^\brap{\text{ker}^\bot;L}=-\frac{\brap{l-3k}}{30}\brap{1,3}$ we find
$q^\brap{L}=-\frac{\brap{l-3k}}{3}\brap{1,0}=\brap{k',0}$ for $k'=-\frac{\brap{l-3k}}{3}$.
This $k'$ is nonzero since $l\neq3k$.

So here we find that while there are partner-less particles with $\zel\sim s^{-1/2}$, they all align with an electric tower with parametrically larger $\zel \sim s^{-3/2}$, the latter of which co-scales and aligns with a magnetic tower.

\begin{figure}[h]
\centering
\begin{subfigure}{0.49\textwidth}
\centering
\includegraphics[width=60mm]{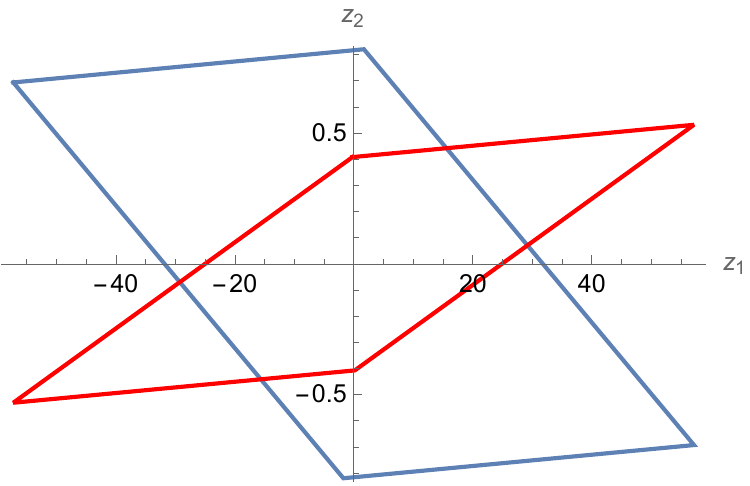}
\caption{$X=0.4$\label{sf40}}
\end{subfigure}
\begin{subfigure}{0.49\textwidth}
\centering
\includegraphics[width=60mm]{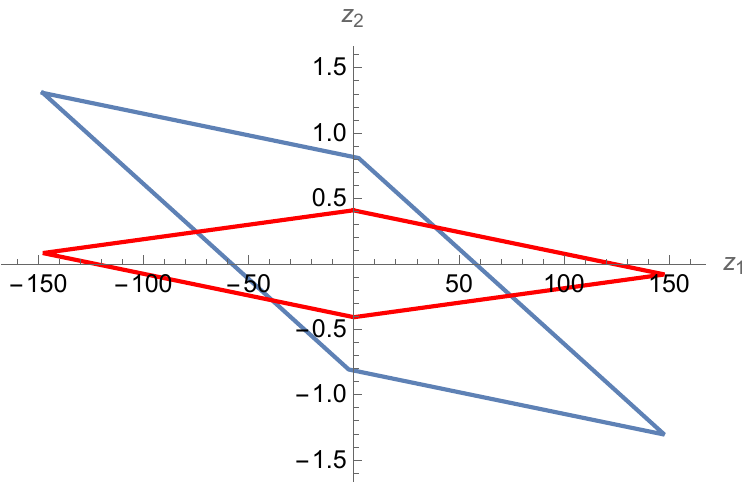}
\caption{$X=0.2$\label{sf20}}
\end{subfigure}
\begin{subfigure}{0.49\textwidth}
\centering
\includegraphics[width=60mm]{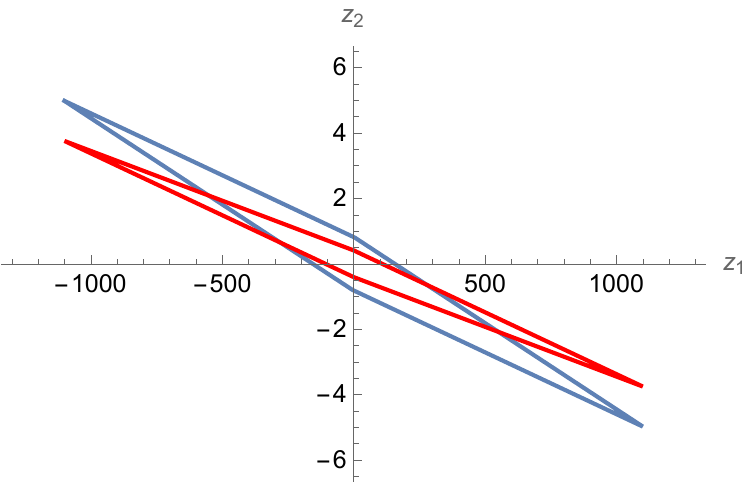}
\caption{$X=0.05$\label{sf05}}
\end{subfigure}
\begin{subfigure}{0.49\textwidth}
\centering
\includegraphics[width=60mm]{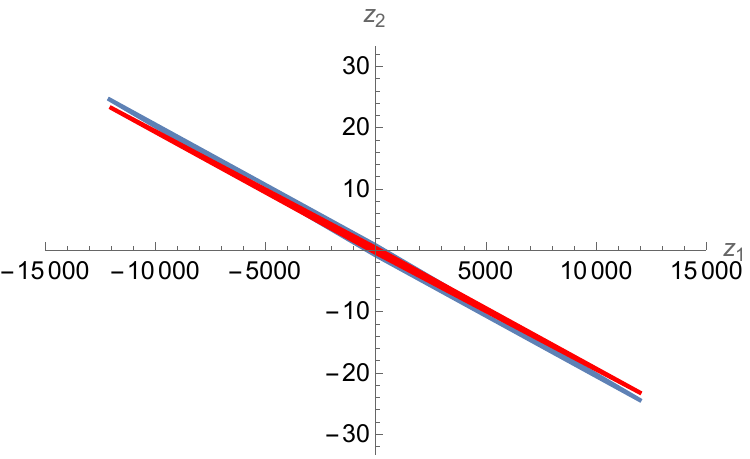}
\caption{$X=0.01$\label{sf01}}
\end{subfigure}
\caption{Electric and magnetic convex hulls near the SCFT boundary of the GMSV geometry. For $Y=1$, $X =s   \in \{0.4, 0.2, 0.05, 0.01\}$, the convex hull generated by towers of electric BPS particles is shown in blue, while the convex hull generated by magnetic strings (rescaled by an overall factor of $\pi$ for ease of comparison) is shown in red. As one approaches the SCFT boundary at $X \rightarrow 0$, one diagonal of each convex hull diverges in length, and the two convex hulls rotate relative to each other so that the diverging diagonals align.}
\label{GMSVCFTfig}
\end{figure}

The co-scaling and alignment of towers of BPS particles with magnetic strings is illustrated in \autoref{GMSVCFTfig}, which compares the convex hull generated by the infinite towers of BPS particles (blue) with the convex hull generated by BPS strings (red) in an orthonormal charge basis in a neighborhood of the SCFT boundary. As one approaches this boundary, the charge-to-mass ratio of the light particles of charge $q_I \propto (1, 0)$ diverges, as does the the charge-to-tension ratio of the light string of charge $\tilde q^I = (3, -1)$. Thus, the electric and magnetic convex hulls grow arbitrarily large along one diagonal. Notably, these two convex hulls grow at the same rate, and in accordance with \eqref{GMSValign}, they rapidly align in the limit $s \rightarrow 0$.

Next, let us consider the Chern-Simons couplings in phase II of the GMSV geometry, which take the form:
\begin{align}
C_{IJK} A^I \wedge F^J \wedge F^K &= 5 A^0  \wedge F^0 \wedge F^0 + 16 A^1 \wedge F^0 \wedge F^0 + 32 A^0 \wedge F^0 \wedge F^1 \nonumber \\
&+ 96 A^1 \wedge F^0 \wedge F^1 + 48 A^0 \wedge F^1 \wedge F^1 +  144 A^1 \wedge F^1 \wedge F^1\,.
\end{align}
This can be rewritten as
\be
C_{IJK} A^I \wedge F^J \wedge F^K = A^1 \wedge F^0 \wedge F^0 +  (A^0 + 3 A^1) \wedge (5 F^0 \wedge F^0 + 32 F^0 \wedge F^1 + 48 F^1 \wedge F^1)\,.
\ee
The tensionless string has charge $\tilde q^I = (3 , -1)$, so its magnetic coupling to the gauge field $A^0 + 3 A^1$ vanishes. The only Chern-Simons coupling that induces an inflow on the string worldvolume, therefore, is the coupling to $F^0 \wedge F^0$, which induces $A^0$ charge on the string worldsheet and implies that the string-scale oscillation modes of the string carry charge $q_I \propto (1, 0)$. It is unsurprising, therefore, that the tower of electrically charged particles has precisely this charge, and its mass scales with the string tension as $m^2 \sim T$.

Finally, let us estimate the species scale near the SCFT boundary using the arguments of \S\ref{ss.5dspecies}. As explained there, the $R^2$ coefficient in the low-energy effective action is determined by second Chern class $c_2$:
\be
\mathcal{L} \supset \frac{C_I Y^I}{\cF^{1/3}} R^2\,,~~~~ C_I Y^I \propto \int c_2 \wedge J\,.
\ee
Since the phase I GMSV geometry is simply the intersection of a bidgree $(4, 1)$ hypersurface a bidgree $(1, 1)$ hypersurface inside $\mathbb{P}^4 \times \mathbb P^1$, its second Chern class can be computed using (2.4) of \cite{Hosono:1994ax}. Passing through the flop transition to phase II, the second Chern class is modified according to (2.15) of \cite{Gendler:2022ztv}. Putting this all together, we find that in the phase II basis,
\be
C_I Y^I = 50 X + 144 Y\,,
\ee
which implies that $C_I Y^I / \cF^{1/3}$ remains order-one near the SCFT boundary at $X = 0$:
\be
\lim_{X \rightarrow 0} \frac{C_I Y^I}{\cF^{1/3}} = \frac{144 Y}{(24 Y^3)^{1/3}} = \frac{72}{\sqrt{3}}\,.
\ee
Thus, in accordance with \S\ref{ss.5dspecies}, we see that the $R^2$ coefficient remains near the Planck scale, suggesting that $\LQG \sim M_{\rm Pl;5}$ even as the SCFT string tension vanishes and a tower of BPS particles become massless.

\subsubsection{Flop transition}

\begin{figure}
\centering
\begin{subfigure}{0.49\textwidth}
\centering
\includegraphics[width=60mm]{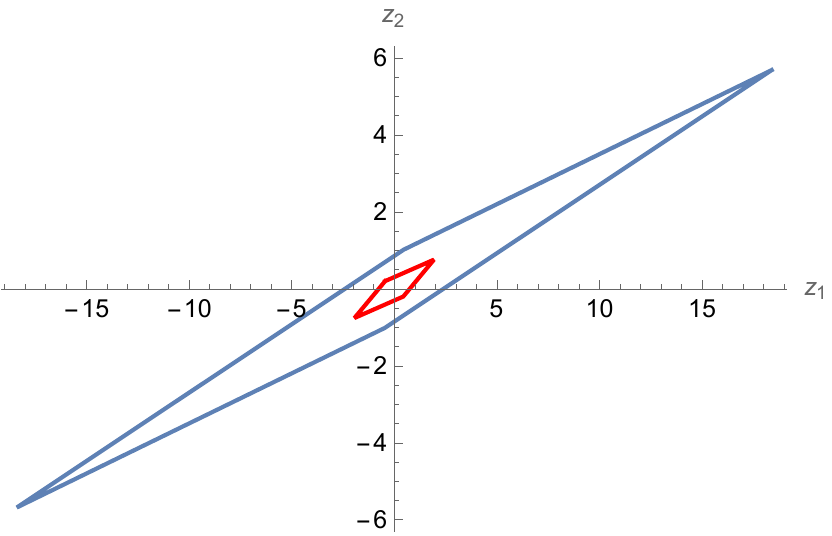}
\caption{$X=0.4$\label{sf40con}}
\end{subfigure}
\begin{subfigure}{0.49\textwidth}
\centering
\includegraphics[width=60mm]{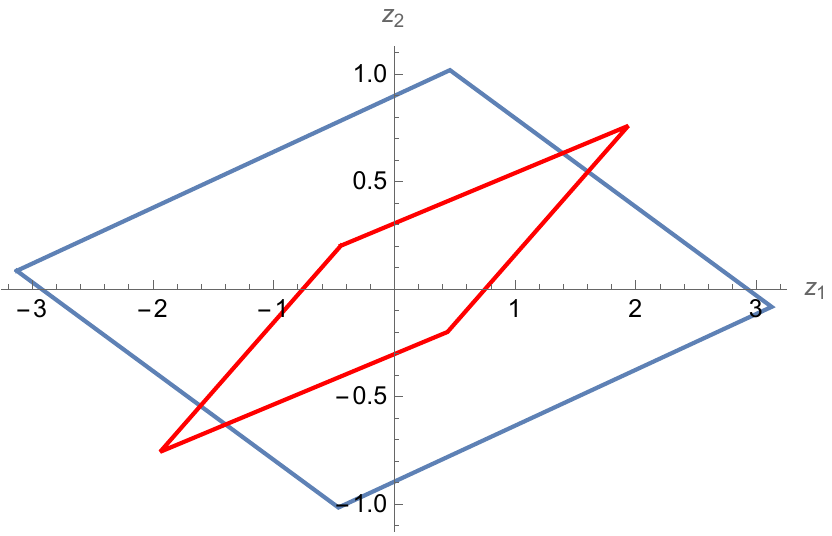}
\caption{$X=0.2$\label{sf20con}}
\end{subfigure}
\caption{Electric and magnetic convex hulls near the conifold locus of the GMSV geometry (a) including the finite number of light particles, (b) not including the light particles. For $Y=0.05$, $X   =1$, the convex hull generated by electric BPS particles is shown in blue, while the convex hull generated by BPS magnetic strings (rescaled by an overall factor of $\pi$ for ease of comparison) is shown in red. The co-scaling relation is violated in the left figure due to the finite number of light particles whose mass vanishes at the conifold locus, but it is satisfied in the right figure, where we restrict our attention to towers of BPS particles.}
\label{f.GMSVcon}
\end{figure}

Recall that the prepotential in the second phase of the GMSV geometry takes the form
\be
\cF = \frac{5}{6} X^3 + 8 X^2 Y + 24 X Y^2 + 24 Y^3\,,
\ee
where again we have set $Y^0 = X$, $Y^1=Y$ for ease of notation. Working in homogeneous coordinates and setting $X = 1$, $Y=s$, a conifold singularity develops when $s \rightarrow 0 $. Here, a finite number of particles of charge $q_I = (0, 1)$ become massless, with
\be
m_{0,1} \sim s^1\,.
\ee
Meanwhile, the charge of these particles remains finite in Planck units,
\be
\lVert q \rVert^2 \sim q_I a^{IJ} q_J \sim s^0\,.
\ee
Co-scaling then requires a magnetic string whose tension vanishes at the conifold limit as $T \sim s^1$. Naively, this is provided by a string of charge $\tilde q^I = (16,-5)$, since 
\be
\frac{|16 \cF_0  -5 \cF_1|}{\cF^{2/3}} \sim s\,.
\ee
However, the effective cone of divisors in the second phase of the GMSV geometry is given by:
\be
\cE = \{\tilde q^0 + 4 \tilde q^1 \geq 0\} \cap \{\tilde q^0 + 3 \tilde q^1 \geq 0 \}\,.
\ee
The charge in question, $\tilde q^I = (16,-5)$, lies outside this cone (as does its negative, $\tilde q^I = (-16,5)$). That means that this divisor is not effective, so no BPS string (or anti-BPS string) of this charge can possibly exist. Thus, barring the existence of a tensionless non-supersymmetric string of charge $\tilde q^I = (16,-5)$, the co-scaling relation is violated.

This is depicted visually in \autoref{f.GMSVcon}. Here, we see that in the limit $Y = s \rightarrow 0$, the electric convex hull grows arbitrarily large if we include the finite number of particles of charge $q_I = (1, 0)$ and thus violates the co-scaling relation. However, if we restrict our attention to towers of electrically charged BPS particles, then both the electric and magnetic convex hulls are order-one in size, and thus they trivially exhibit co-scaling. This demonstrates that a finite number of light particles does not necessarily co-scale or align with any light string.

\subsubsection{Verifying Conjecture 1}\label{ss.GMSVverify}

According to \autoref{Con1}, the $\cS$-map $\cS(Y^I, \tilde q^J) = \cF_{IJ} \tilde q^J$ should map the hypereffective cone to the electric infinity cone:
\be
\cM_\infty = \langle (1,0) ; (1, 4) \rangle\,.
\ee
Since there are no stable Weyl flops, the hypereffective cone is equal to the effective cone, 
\be
\cE_{\rm hyp} = \cE = \langle (-3, 1) ; (4, -1) \rangle \,.
\ee
Applying the $\cS$-map to the generators of the hypereffective cone in phase II, we have
\begin{align}
\cS(Y^I, (-3, 1)) = X \cdot (1,0)\,,~~~\cS(Y^I, (4, -1)) = (4X +12 Y) \cdot (1, 4) + 4 Y \cdot (1, 0)\,.
\end{align}
As expected, the cone generated by these charges is always contained in $\cM_\infty$, and the two cones coincide at the flop boundary $Y \rightarrow 0$.

Finally, let us check that \autoref{Con1} is also satisfied in phase I. For this, we return to the phase I basis, where the prepotential takes the form \eqref{GMSV.prepotential.phaseI}. In the phase I basis, the (hyper)effective cone is given by
\be
\cE_{\rm hyp} = \cE = \langle (1, -1) ; (0, 1) \rangle \,.
\ee
The electric infinity cone is again given by
\be
\cM_\infty = \langle (1,0) ; (1, 4) \rangle\,.
\ee
Applying the $\cS$-map to the generators of the hypereffective cone in phase I, we have
\begin{align}
\cS(Y^I, (1, -1)) = X\cdot (1,4) + 2 Y \cdot(1, 0)\,,~~~\cS(Y^I, (0, 1)) =  5X \cdot (1, 0)\,.
\end{align}
Once again, we find that the cone generated by these charges is contained in $\cM_\infty$ for all $Y^I \in \cK_{\rm phase\,I}$, and the two cones coincide at the flop boundary $Y \rightarrow 0$.

Thus, taking the union over all $Y^I \in \cK = \cK_{\rm hyp}$, we see that \autoref{Con1} is satisfied: $\cS(\cK_{\rm hyp} \times \cE_{\rm hyp}) = \cM_\infty$.

\subsection{Example: The KMV geometry}\label{ss.KMV}

We next examine co-scaling and alignment in the $h^{1,1}=3$ KMV geometry, introduced in \cite{Klemm:1996hh} and further studied in \cite{Alim:2021vhs}. This geometry features two phases connected by a flop transition. In the second phase, the prepotential is given by
\begin{align}
\cF^{(II)} = \frac{4}{3} X^3 &+ \frac{3}{2} X^2 Y + \frac{1}{2}X Y^2 + \frac{9}{2}X^2 Z + \frac{9}{2}X Z^2 + \frac{1}{2}Y^2 Z \nonumber \\ &+ \frac{3}{2}Y Z^2 + \frac{3}{2} Z^3 + 3 X Y Z \,.
\end{align}
The K\"ahler cone of this phase is given by $X, Y, Z, \geq 0$. The boundaries at $X=0$ and $Y=0$ each represent SCFT boundaries. Geometrically, the former corresponds to the collapse of a dP$_8$ surface, while the latter corresponds to the collapse of a $\mathbb{P}^2$.

The electric infinity cone $\cM_\infty$ is non-simplicial and is generated by the electric charges
\be
\cM_\infty \langle (0,1,1) ; (1,0,1); (1,0,0); (0,1,0) \rangle \,.
\label{KMVinfty}
\ee
The cone of effective divisors is simplicial and is generated by magnetic charges
\be
\cE = \langle (-1,0,1);(0,-3,1);(1,1,-1)\rangle\,.
\label{KMVeffcone}
\ee

In the limit $X \rightarrow 0$, a tower of BPS particles with $q_I = k (1,0,0)$ becomes light, as does a BPS string of charge $\tilde q^I = (-1,0,1)$.
The gauge coupling of this string vanishes in the limit $X \rightarrow 0$, and together these light particles and string follow the scaling behavior \eqref{5dreal}. Furthermore, these particle and string charges satisfy \eqref{eqn.relation.between.electric.and.magnetic.charge.for.theta.going.like.1.over.z.linear}, as expected for an SCFT boundary.

Analogously, at the $Y=0$ boundary, electric particles of charge $q_I = k(0,1,0)$ and a string of charge $\tilde q^I = (0,-3,1)$ become light; these also scale as \eqref{5dreal} and satisfy the relation \eqref{eqn.relation.between.electric.and.magnetic.charge.for.theta.going.like.1.over.z.linear}.

\subsubsection{Intersection of two SCFT boundaries}\label{sec.sub.sub.KMV.intersection.of.2.SCFTs}

As a novel test of co-scaling and alignment, we restrict our attention to a neighborhood of the codimension-2 boundary at $X=Y=0$, where the two codimension-1 SCFT boundaries intersect.

\begin{figure}[h]
\centering
\begin{subfigure}{0.49\textwidth}
\centering
\includegraphics[width=60mm]{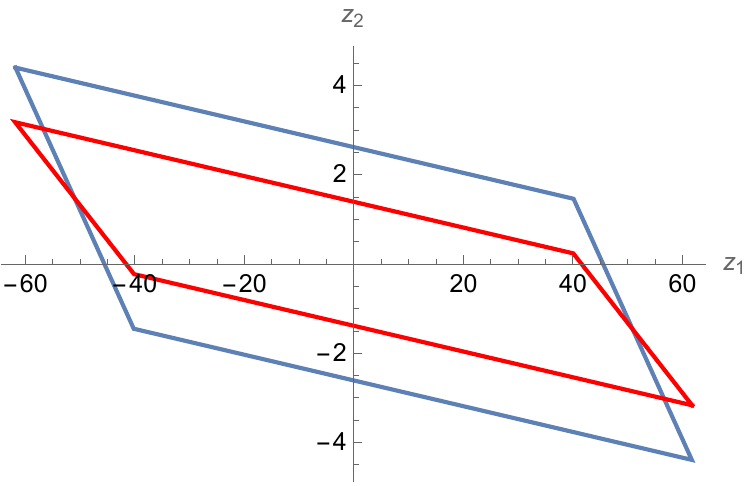}
\caption{$s=0.1$\label{sf10corn}}
\end{subfigure}
\begin{subfigure}{0.49\textwidth}
\centering
\includegraphics[width=60mm]{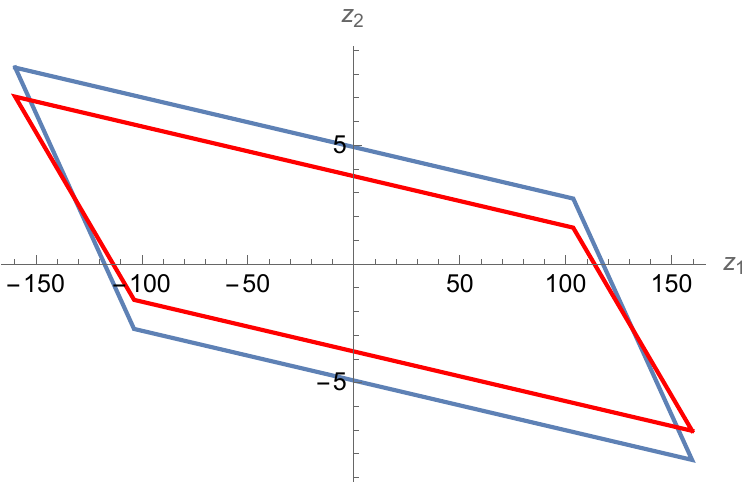}
\caption{$s=0.05$\label{sf05corn}}
\end{subfigure}
\begin{subfigure}{0.49\textwidth}
\centering
\includegraphics[width=60mm]{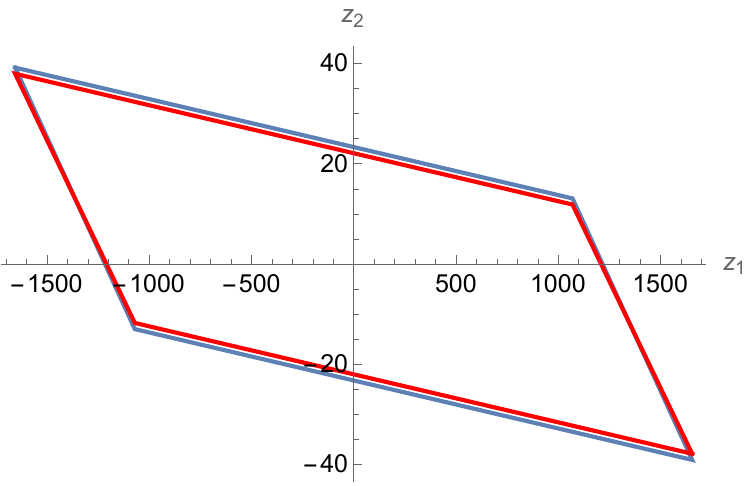}
\caption{$s=0.01$\label{sf01corn}}
\end{subfigure}
\caption{ Electric and magnetic convex hulls for the KMV geometry. For $Z=1$, $X  =Y \in \{0.1,  0.05, 0.01\}$, the convex hull generated by towers of electric BPS particles is shown in blue, while the convex hull generated by BPS magnetic strings (rescaled by an overall factor of $\pi$ for ease of comparison) is shown in red. As one approaches the codimension-2 intersection of two SCFT boundaries at $s \rightarrow 0$, the size of the convex hulls diverge, and the two convex hulls rotate relative to each other so that the figures agree.}
\label{figKMVONB}
\end{figure}

To study this, we consider the limit $X=s$, $Y=s$, $Z=1$. As in the case of the GMV geometry above, we construct the convex hull of the charge-to-mass vectors of the BPS towers in \eqref{KMVinfty} and the convex hull of the charge-to-tension vectors of the BPS strings in \eqref{KMVeffcone}. One 2d projection of the convex hulls in an orthonormal basis is shown for decreasing values of $s$ in \autoref{figKMVONB}. In the limit $s \rightarrow 0$, the diameters of these convex hulls diverge at the same rate, and simultaneously they rotate so as to align with each other. We have verified numerically that these convex hulls differ in size by no more than an order-one factor in any U(1) direction.

Since this boundary is codimension-2, we can also consider the more general limit
$X=s\cos\theta$, $Y=s\sin\theta$, $Z=1$, for any fixed $\theta\in\bras{0,\frac\pi2}$.
In this limit, the tensionless strings inside $\cE$ are ${\tilde q}^I=a\brap{-1,0,1}+b\brap{0,-3,1}$, for $a,b\geq0$ (and not both zero).
The corresponding $q$ given by \eqref{eqn.relation.between.electric.and.magnetic.charge.for.theta.going.like.1.over.z.linear}, which co-scales and aligns with this string, is given by $q_I=\brap{a\cos\theta,b\sin\theta,0}$, which is both in $\cM_\infty$ and a light electric particle. Indeed, if $\theta\notin\brac{0,\frac\pi2}$ -- i.e., if the perturbation takes us into the bulk -- then any light electron in $\cM$ can be written as $q_I=\brap{a\cos\theta,b\sin\theta,0}$ for some $a,b\geq0$ (and not both zero).
Hence, for all perturbations, each light electron in $\cM_\infty$ co-scales and aligns with a tensionless string in $\cE$, but which electron aligns with which string depends upon the perturbation direction.

Because there are multiple towers of electric particles with $\zel\sim s^{-3/2}$, the test of the procedure to ignore the $\zel \sim s^{-1/2}$ electrons in \S\ref{s.coscaling} is less trivial, so we again demonstrate it here.
In this case, the heavy electrons in $\cM_\infty$ are those with $q_I=\brap{a,b,c}$, $a, b, c \geq0$, $c \leq a+b$.
Those with $\zel \sim s^{-1/2}$ are those that do not have $c=a=3b$.
Following the notation of \S\ref{s.coscaling}, this $q$ gives
\begin{align}
    q^\brap{{\text{ker}}} &=
    \frac{1}{19}\brap{
         10 a-3 b-9 c,
         -3 a+18 b-3 c,
         -9 a-3 b+10 c
        }
\\
    q^\brap{\text{ker}^\bot}  &=
    \frac{{3 a+b+3 c}}{19}
    \brap{3,1,3}\,.
\end{align}
Then taking
$
    q^\brap{\text{ker}^\bot;L} = 
    \frac{9  a + 3  b - 10  c}{57} \brap{3,1,3}
$
we find
$
q^\brap{L} = 
\brap{a-c,b-\frac{c}{3},0}
$.
This is non-zero since we do not have $c=a=3b$.
However, this is only sometimes in $\cM_\infty$, because $\brap{a-c}$ and $\brap{b-\frac{c}{3}}$ may or may not have opposite sign.
For example, if we started with $q_I=\brap{2,2,1}$, we find $q^\brap{L} =\frac13\brap{3,5,0}$, so we find that $q_I=\brap{2,2,1}$ aligns with $\brap{3,5,0}\in\cM_\infty$, the latter of which co-scales and aligns with a magnetic string in $\cE$.
But if instead we started with $q_I=\brap{0,2,1}$, we find $q^\brap{L} =\frac13\brap{-3,5,0}$, which is not in $\cM_\infty$ and neither is $-q^\brap{L}$.
Hence in this example there are electric towers with $\zel\sim s^{-1/2}$ that do not co-scale or align with any magnetic tower, and furthermore they do not align with any electric tower that does co-scale and align.

\subsubsection{Intersection of a flop and an SCFT boundary}

As a second illustrative case, let us instead consider the limit $X=1$, $Y=Z=s$. This limit approaches a codimension-2 SCFT boundary at the intersection of a codimension-1 SCFT boundary (at $Y=0$) and a codimension-1 conifold locus (at $Z=0$). At this boundary, the string of charge $\tilde q^I = (0,-3,1)$ and the BPS particles of charge $q_0 = 0$ satisfy the co-scaling relation \eqref{5dreal}:
\be
m \sim s\,,~~~T \sim s^2 \,,~~~g^2 \sim s^{-1}\,.
\ee
Furthermore, for $\tilde q^I = (0,-3,1)$, we have
\be
\cF_{IJ} \tilde q^J = (0, Y, 0)\,,
\ee
so \eqref{eqn.relation.between.electric.and.magnetic.charge.for.theta.going.like.1.over.z.linear} is obeyed for $q_I \propto (0, 1, 0)$, and correspondingly the angle between $-\tilde q^I = (0, 3,-1)$ and $q_I = (0, 1, 0)$ is given by
\be
1-\cos \varphi = \frac{3}{32}s^3 + O(s^4) \sim \zel^{-2}  ~~~\Rightarrow ~~~ \varphi \sim \frac{1}{\zel}\,.
\ee
The other generator of the infinity cone that corresponds to a light tower in this limit is given by $q_I = (0, 1, 1)$. This does not satisfy \eqref{eqn.relation.between.electric.and.magnetic.charge.for.theta.going.like.1.over.z.linear}, and as a result its charge-to-mass vector aligns with that of the light magnetic string at a slower rate:
\be
1-\cos \varphi' = \frac{s}{3} + O(s^2) \sim |\zel|^{-2/3} \,.
\ee
Said differently, these particles and strings exhibit alignment in the limit $s \rightarrow 0$ but not rapid alignment.

\begin{figure}
\centering
\begin{subfigure}{0.49\textwidth}
\centering
\includegraphics[width=60mm]{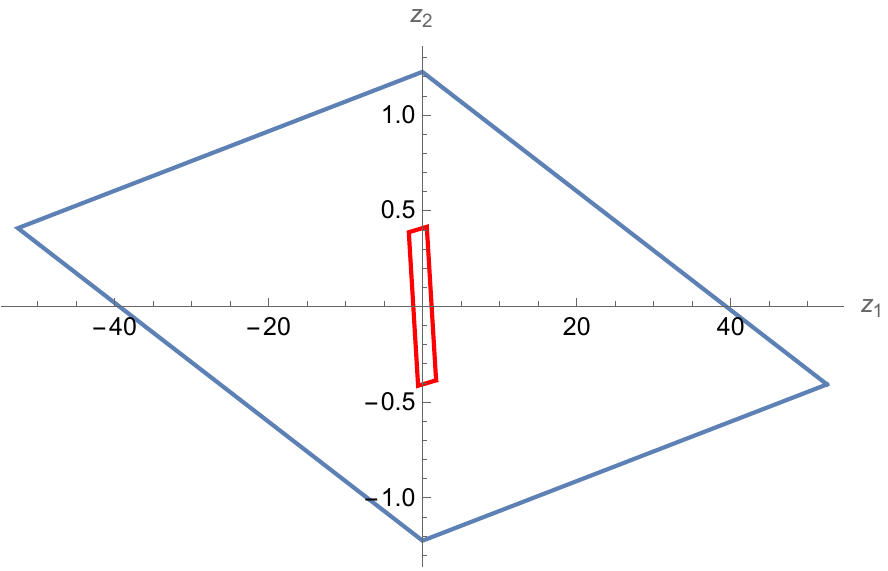}
\caption{$s=0.01$\label{sfp01}}
\end{subfigure}
\begin{subfigure}{0.49\textwidth}
\centering
\includegraphics[width=60mm]{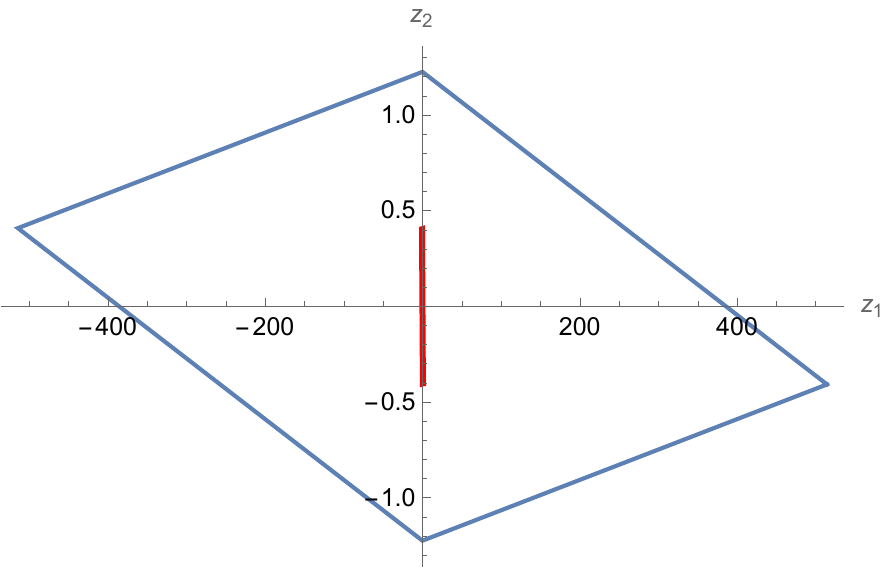}
\caption{$s=0.001$\label{sfp001}}
\end{subfigure}
\caption{ Electric and magnetic convex hulls for the KMV geometry, projected on the plane orthogonal to the magnetic string charge $\tilde q^I = (0, -3, 1)$. For $X=1$, $Y  =Z \in \{0.01, 0.001\}$, the convex hull generated by towers of electric BPS particles is shown in blue, while the convex hull generated by magnetic strings (rescaled by an overall factor of $\pi$) is shown in red.
}
\label{f.nonequal}
\end{figure}

As explained in \S\ref{ss.geometry}, this rate of alignment is too slow to ensure that the electric and magnetic convex hulls differ by an order-one amount in every direction. In particular, \autoref{f.nonequal} shows the projection of the electric and magnetic convex hulls onto the plane orthogonal to the string charge $\tilde q^I = (0, -3, 1)$. One can see that, in this plane, the electric convex hull grows without bound in the limit $s \rightarrow 0$, whereas the magnetic convex hull remains bounded. This demonstrates that the phenomenon of electric and magnetic co-scaling does not always apply in every direction in charge space. Nonetheless, the co-scaling and alignment of the maximally divergent electric towers and magnetic strings persists in this example.

\subsubsection{Verifying Conjecture 1}

To conclude our analysis of the KMV geometry, let us verify \autoref{Con1}, which holds that the $\cS$-map should map the $\cK_{\rm hyp}\times\cE_{\rm hyp}$ to the electric infinity cone $\cM_\infty$.

To check this,
we will find it useful to work in phase I coordinates rather than phase II coordinates, which are related via
\be
X^{(II)} = X^{(I)} + Z^{(I)} ~,~~~Y^{(II)} = Y^{(I)} + Z^{(I)}~,~~~ Z^{(II)} = - Z^{(I)}\,.
\ee
In phase I coordinates, the K\"ahler cone of the first phase $\cK_{\text{phase I}}$ is generated simply by
\begin{equation}
 \cK_{\rm phase \, I}   = \langle (1,0,0); (0,1,0);(0,0,1) \rangle \,,
\end{equation}
and the prepotential in phase I is given by
\be
\cF^{(I)} = \frac{8}{3} X^3 + \frac{3}{2} X^2 Y + X^2 Z + \frac{1}{2} X Y^2 + XYZ\,.
\ee
Meanwhile, in phase I coordinates, the K\"ahler cone of phase II is generated by
 \begin{equation}
     \cK_{\rm phase \, II} = \langle (1,0,0); (0,1,0);(1,1,-1) \rangle\,,
 \end{equation} 
 and the prepotential is given by 
 \be
\cF^{(II)} = \frac{8}{3} X^3 + \frac{3}{2} X^2 Y + X^2 Z + \frac{1}{2} X Y^2 + XYZ - \frac{1}{6} Z^3\,.
\ee

Since there are no stable Weyl flops in this geometry, the hypereffective cone is the same as the effective cone, $\cE_{\rm hyp}=\cE$. In phase I coordinates, this cone is given by the generators
\be
 \cE  =  \langle (0,1,-1);(1,-2,-1);(0,0,1) \rangle \,. 
\ee
Meanwhile, the electric infinity cone $\cM_{\infty}$ in phase I coordinates is given by
\be
\cM_{\infty} = \langle (0,1,1);(1,0,1);(1,0,0);(0,1,0) \rangle \,.
\ee

It is then straightforward to check that
\be
\cS(\cK_{\text{phase I}}, \cE_{\rm hyp}) = \cM_\infty\,,
\ee
that is, when $Y^I$ is restricted to lie in phase I, the image of the hypereffective cone under the $\cS$-map is precisely the electric infinity cone.

The same is not true of phase II. The image of $\cK_{\text{phase II}}\times\cE_{\rm hyp}$ under the $\cS$ map is the strictly smaller cone given (in phase I coordinates) by
\begin{equation}
 \cS(\cK_{\text{phase II}}, \cE_{\rm hyp})  = \langle (2,1,0) ; (1,0,0); (1,0,1); (0,1,1) \rangle \,.
\end{equation}
In particular, this cone does not include the generator $(0,1,0)$, but instead it has the generator $(2,1,0) = 2(1,0,0)+(0,1,0)$. Of course, this is perfectly consistent with \autoref{Con1}, since together we have
\be
\cS(\cK_{\rm hyp},\cE_{\rm hyp}) = \cS(\cK_{\rm phase \,I} \cup \cK_{\rm phase \, II},\cE_{\rm hyp}) = \cS(\cK_{\rm phase \,I},\cE_{\rm hyp}) \cup \cS(\cK_{\rm phase \, II},\cE_{\rm hyp}) = \cM_\infty\,.
\ee

\subsection{Example: The GHMMR Geometry}\label{ss.GHMMR}

In studying the conifold locus of the GMSV geometry, we saw that the co-scaling and alignment relations do not necessarily apply to isolated light particles. Instead, they seem to deal with light towers of charged particles.

As we saw in \S\ref{ss.infinitetowers}, the distinction between isolated states vs. towers of states can be quite subtle on the magnetic string side of the spectrum. With this in mind, we turn our attention to the geometry studied in Appendix C.4 of \cite{Gendler:2022ztv}. This geometry features a simplicial K\"ahler cone with three boundaries, one of which represents an asymptotic boundary, and two of which represent SU(2) boundaries. The prepotential is given by
\be
\cF= X Y Z + Y^2 Z + X Z^2 +2 Y Z^2 + \frac43 Z^3 \,.
\ee
Since there are no flop boundaries, the extended K\"ahler cone is simply the K\"ahler cone, which is given by
\be
\mathcal{K}  =  \{ X, Y, Z \geq  0\}\,. 
\ee
The boundary $Z = 0$ is an asymptotic boundary, which corresponds to a decompactification limit. The boundary $X=0$ represents a genus 1 Weyl flop: at this boundary, a divisor collapses to a curve of genus 1. This Weyl flop is stable, which in particular means that the collapsing curve of charge $q_I = (1, 0, 0)$ lies strictly outside the infinity cone. Conversely, the Weyl flop at the boundary $Y=0$ is a genus 0, unstable Weyl flop: the charge $q_I = (0, 1, 0)$ lies on the boundary of the infinity cone. The codimension-2 point of intersection between these two SU(2) boundaries at $X=Y=0$ represents an SCFT boundary.

The infinity cone is simplicial and is generated by the curve classes
\be
\cM_\infty =  \langle (1, 1, 0) ; (0,0,1); (0,1,0) \rangle\,.
\ee
This is dual to the hyperextended K\"ahler cone, which includes not only the extended K\"ahler cone, but also its reflection under the stable Weyl flop at $X=0$. The hyperextended K\"ahler cone is generated by
\be
\cK_{\rm hyp} = \langle (-1,1,0) ; (1,0,0) ; (0,0,1) \rangle \,.
\label{hyper}
\ee
Finally, the effective cone is generated by 
\be
\cE = \langle   (1, 0,0) ; (-2, 1, 0) ; (0, -2, 1)  \rangle \,.
\ee
These three generators represent the charges of the strings that become light at the asymptotic boundary $Z=0$, the SU(2) boundary $X=0$, and the SU(2) boundary $Y=0$, respectively.

By our previous general analysis, we know that at a codimension-1 SU(2) boundary, the mass of the light particles (i.e., the W-bosons), the tension of the light strings, and the associated electric gauge coupling scale as 
\be
m \sim s\,,~~~~T \sim s\,,~~~~ g^2 \sim s^0\,,
\label{su2scaleG}
\ee
ensuring that the co-scaling relation is satisfied.

The W-bosons for the stable Weyl flop, by definition, do not lie on the boundary of the electric infinity cone $\cM_\infty$. The co-scaling relation therefore suggests that the 't~Hooft-Polyakov monopole of charge $\tilde q^I = (-2, 1, 0)$ should similarly lie strictly outside the magnetic infinity cone $\cK_\infty$. We cannot prove this supposition due to the previously discussed limitations, but we can provide some evidence for it as follows: as discussed in \cite{Gendler:2022ztv}, at a generic point in complex structure moduli space, the genus $1$ Weyl flop disappears. Physically, this deformation corresponds to giving a vev to an adjoint hypermultiplet, which breaks SU(2) to U(1) at the boundary $X=0$. Consequently, the hyperextended K\"ahler cone in \eqref{hyper} becomes the extended K\"ahler cone, and the effective cone shrinks. At generic complex structure, the effective cone is then given by the hypereffective cone,
\be
\cE_{\rm hyp} =  \langle (1, 0,0) ; (-1, 1, 0) ; (0, -2, 1)  \rangle \,,
\ee
so $\tilde q^I = (-2, 1, 0)$ is no longer effective. In the absence of wall-crossing for BPS strings, then, we conclude that there must not exist a tower of BPS strings of charge $\tilde q^I = ( -2, 1, 0)$. Thus, it is plausible that just as the charge $q_I  = (1, 0, 0)$ lies strictly outside the infinity cone $\mathcal{M}_\infty$ for towers of BPS particles, so too should the charge $\tilde q^I = (-2, 1, 0)$ lie strictly outside the analogous infinity cone of divisors $\mathcal{K}_\infty$ for towers of BPS strings.

On the other hand, the Weyl flop across the SU(2) boundary at $Y=0$ is unstable, so the charge $q_I = (0,1, 0)$ lies on the boundary of $\mathcal{M}_\infty$, and there is a skew tower of BPS particles of charge $q_I = (1,k,0)$, $k \in \mathbb{Z}$.
The corresponding tensionless BPS monopole string has charge $\tilde q^I = (0, -2, 1)$. This charge lies strictly outside the hyperextended cone \eqref{hyper}, so there is no guarantee that BPS black strings of this charge exist. However, $\tilde q^I = (0, -2, 1)$ lies on the boundary of the hypereffective cone, so \autoref{Con2} implies that there must exist a tower or skew tower of this charge. At present, we can neither confirm nor deny this, and we leave this as a nontrivial prediction of co-scaling and alignment.

\section{Extremal Black Holes}\label{sec.non.UV}

Above, we have encountered a number of general heuristic arguments, semi-analytic arguments and explicit computations suggesting that co-scaling and (rapid) alignment are ubiquitous in the landscape. One may wonder to what extent these phenomena follow from purely low-energy constraints.

In this section, we address this question by computing the electric and magnetic extremal black hole spectrum for several (UV-incomplete) examples of EFTs. We will see that some of these examples violate the co-scaling relation, indicating that the presence of co-scaling provides nontrivial constraints not only on the spectrum of charged matter, but also on the 2-derivative effective action.

Consider the low-energy Einstein frame effective action
\begin{equation}
 S=\int d^dx\sqrt{-g}\brac{\frac1{2\kappa^2}\bras{\mathcal{R}-\frac12\nabla\phi\cdot\nabla\phi} - \frac12t\brap{\phi}F_{p+1} \cdot F_{p+1}}
\end{equation}
for a singular modulus field $\phi$, and single U(1) $\brap{p-1}$-form vector field $F_{p+1}=dA_p$.
Let $\phi_\infty$ be the value of $\phi$ at $r=\infty$.
We shall compute the charge-to-mass ratios of electric and magnetic extremal black branes in this background. Since there is only one U(1) field, there is effectively only one electric and one magnetic tower, so while alignment is trivial, this theory exhibits co-scaling if and only if these towers satisfy $\zel\sim\zmag$.
We shall find that for some (assumed to be non-UV-complete) choices of $t\brap{\phi}$, there exist values of $\phi_\infty$ such that $\zel \nsim \zmag$.

In the rest of this section, we examine co-scaling behavior for three choices of $t\brap{\phi}$.
Details of using the attractor mechanism to calculate $\zel$ and $\zmag$ can be found in Appendix \ref{appendix.black.hole}.
For brevity, in this section we just provide only the resulting $\zel$ and $\zmag$.
In our first example we provide an analytic expression, while in the other cases there is no known analytic expression, so we instead turn to numerics.\footnote{
When we rely upon numerics, we specialize to the case $d=4$, $p=1$. Indeed, this is without loss of generality provided $1 \leq p \leq d-3$, as the differential equations for generic $p$ can be reduced to the equations for $p=1$ by a suitable rescaling of $z$, $\alpha$ and $\phi$.
}

For reference throughout this section, we define the quantity $\zeta$ to be mass-to-charge ratio (i.e., the inverse of the charge-to-mass ratio):
\be
\zeta_e:=\frac1\zel \,,~~~~\zeta_m:=\frac1\zmag\,.
\ee
We further define the expression
\begin{align}
\tilde\zeta\brap{\phi}:=\bras{\frac{\bras{\alpha\brap{\phi}}^2}2 + \xi}^{-\frac12}\,,
\end{align}
where 
\begin{align}
\alpha\brap{\phi}:=\frac{\partial\bras{\log\brap{t^{-1}\brap{\phi}}}}{\partial \phi}   \,,~~~~\xi:=\frac{p\brap{d-p-2}}{d-2}\,.
\label{alphaxidef}
\end{align}

\subsection{Example: Constant $\alpha$}\label{sec.sub.non.UV.eg.1}

As a warm up, we begin with the simple example of a dilatonic coupling to the Maxwell term, which was briefly discussed in \S\ref{ss.coscale} above and previously analyzed in, e.g.,~\cite{Horowitz:1991cd, Heidenreich:2015nta, Harlow:2022ich}. We set $t= \frac{1}{\gel^2}= \frac1{e_p^2}\mathrm{e}^{-\alpha\phi}$, with $\alpha$ and $e_p$ constant.
With this choice, $\alpha(\phi)$ and $\tilde \zeta(\phi)$ are both constants:
\begin{equation}
    \alpha\brap{\phi}\equiv\alpha \,,\quad\tilde\zeta\brap{\phi}\equiv\tilde \zeta=\bras{\xi+\frac{\alpha^2}2}^{-\frac12} \,.
\end{equation}

The unique positive solution to the governing equations over all $\phi_\infty \in \mathbb R$ is simply 
\cite{Harlow:2022ich}
\be
\zeta_e=\zeta_m=\tilde\zeta\,.
\ee

With this, 
\begin{equation}
\frac{|z_{\rm el}|}{|z_{\rm mag}|} = \frac{\zeta_{m}}{\zeta_e} = 1\,,
\end{equation}
which is notably independent of both $\alpha$ and $\phi_\infty$, and so we observe co-scaling.
Indeed, here we have \textit{exact} co-scaling.

\subsection{Example: Quadratic $\alpha$}\label{sec.sub.non.UV.eg.2}

We now consider a theory with
\begin{equation}
    t\brap{\phi} = c\exp\bras{-\lambda\brap{\frac13\phi^3-\phi}}\,,
\end{equation}
for some constant $c$.
The resulting $t^{-1}\brap{\phi}$ is plotted in \autoref{fig.black.hole.Qe.squared.quadratic}.

This choice of $t\brap{\phi}$ leads to a quadratic function for $\alpha\brap{\phi}$, as defined in \eqref{alphaxidef}:
\be
\alpha\brap{\phi}=  \lambda\brap{\phi-1}\brap{\phi+1} \,.
\ee

Let us remark that we do not expect this sort of behavior to arise in the string landscape; in all known theories, charge functions decay or diverge exponentially with geodesic distance in infinite-distance limits, so $\alpha$ approaches a constant as $\phi \rightarrow \pm \infty$.
It shall turn out that we do not always have co-scaling in this theory, demonstrating that co-scaling can be absent in theories in the swampland.

\begin{figure}[h]
    \centering
    \includegraphics[width=0.45\linewidth]{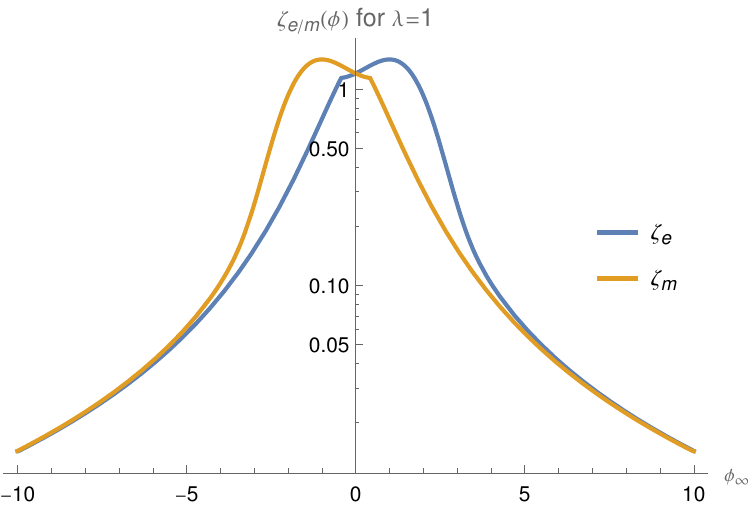}
    \quad
    \includegraphics[width=0.45\linewidth]{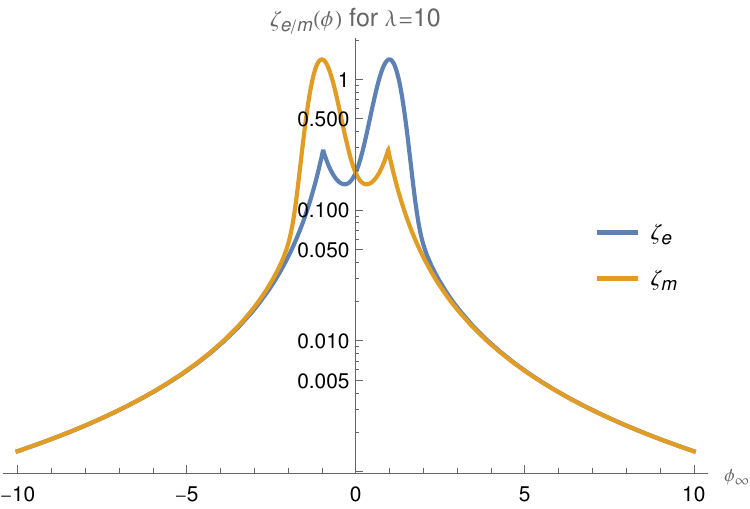}

    \includegraphics[width=0.45\linewidth]{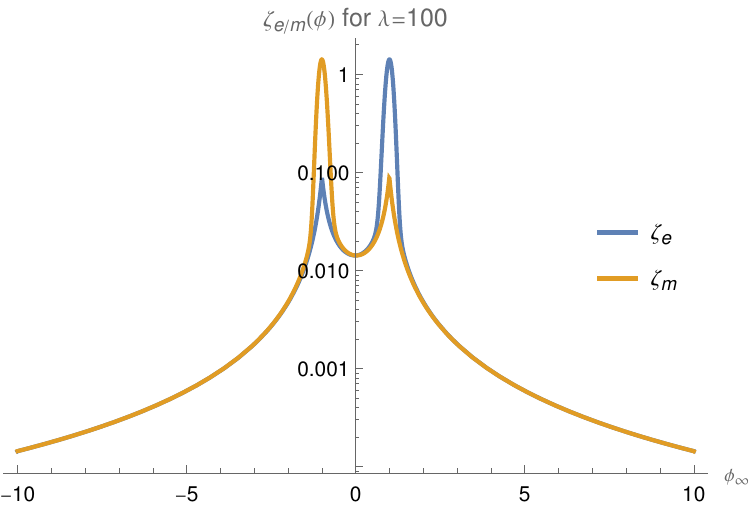}
    \quad
    \includegraphics[width=0.45\linewidth]{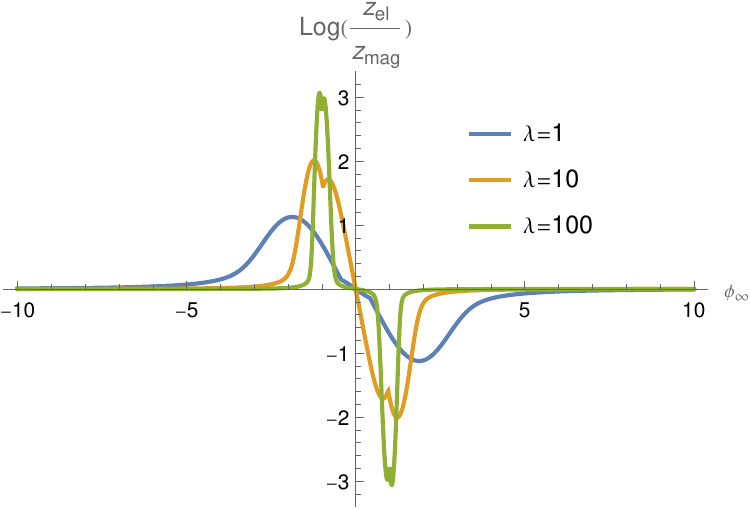}
    \caption{Plots of $\zeta_e$, $\zeta_m$, and the log of their ratio for the example in \S\ref{sec.sub.non.UV.eg.2} with quadratic $\alpha$.
    The first three sub-figures each depict $\zeta_e$, $\zeta_m$ for a given value of $\lambda$; note the logarithmic scale in the y-axis. The final sub-figure plots the natural $\log$ of the ratio $\frac{\zeta_m}{\zeta_e}=\frac{|z_{\rm el}|}{|z_{\rm mag}|}$ for the $\lambda$ values from the other sub-figures.
   }
    \label{fig.black.holes.quadratic.zetas}
\end{figure}

Unlike the simple case of constant $\alpha$ considered above, this choice of $t\brap{\phi}$ does not admit analytic solutions.
\autoref{fig.black.holes.quadratic.zetas} shows some numerically obtained solutions for different values of $\lambda$.

The charge-to-mass ratio $z_{\rm el}$ for a black hole solution in this theory depends on both the parameter $\lambda$ and the boundary condition $\phi_\infty$.
It will prove useful for us to consider two types of limits:
\begin{enumerate}
\item $\lambda\rightarrow\infty$. This limit corresponds to sending the value of $t^{-1}(\phi)$ at its local minimum and maximum to zero and infinity, respectively.
\item $\phi_\infty\rightarrow\pm\infty$. This corresponds to a limit in which the electric black brane charge $Q^2(\phi_\infty)$ vanishes (for $\phi \rightarrow - \infty$) or diverges (for $\phi \rightarrow + \infty$). Conversely, the magnetic black brane charge diverges and vanishes in these respective limits.
\end{enumerate}
From \autoref{fig.black.holes.quadratic.zetas}, we see that limits of type (2) exhibit co-scaling. That is, for fixed $\lambda$, as $\phi_\infty\rightarrow\pm\infty$, we find 
$
 \frac{|z_{\rm el}|}{|z_{\rm mag}|} \rightarrow \text{const}
$.

\begin{figure}[h]
    \centering
    \includegraphics[width=\linewidth]{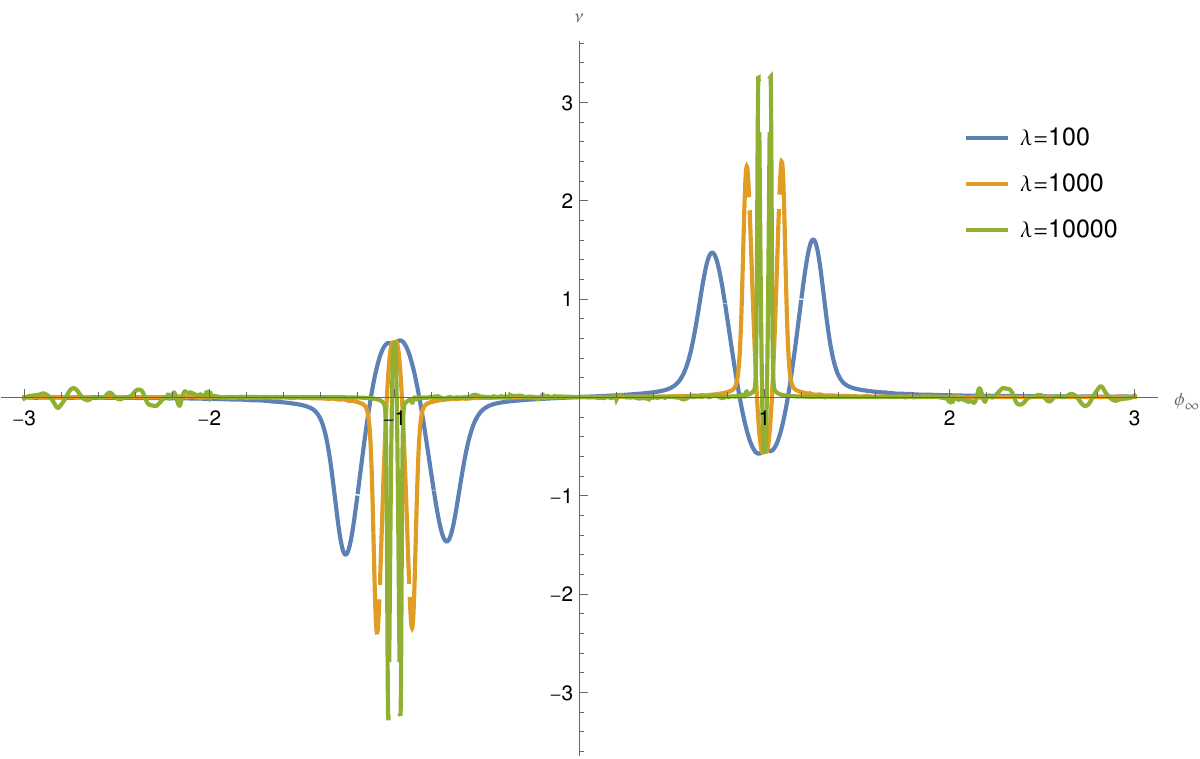}
    \caption{$\nu := \frac{\partial\bras{\log\brap{\zeta_m/\zeta_e}}}{\partial\bras{\log\lambda}}$ as a function of $\phi$ for increasing values of $\lambda$ for the example in \S\ref{sec.sub.non.UV.eg.2} with quadratic $\alpha$. Peaks form around $\phi=\pm1$, while $\nu\rightarrow0$ elsewhere, indicating that the electric and  magnetic charge-to-mass ratios co-scale everywhere except at $\phi = \pm 1$.}
    \label{fig.black.hole.quadratic.nu}
\end{figure}

Co-scaling in limits of type (1) can be expressed as the statement that $\nu := \frac{\partial\bras{\log\brap{\zeta_m/\zeta_e}}}{\partial\bras{\log\lambda}}\rightarrow0$ as $\lambda\rightarrow\infty$.  \autoref{fig.black.hole.quadratic.nu} plots $\nu$ as a function of $\phi$ for various values of $\log\lambda$.
Here, we see that $\nu(\phi)$ peaks at $\phi \approx \pm1$ and tends toward zero elsewhere.
Thus, we observe co-scaling in the large $\lambda$ limit away from $\phi_\infty=\pm1$, but co-scaling is violated at $\phi_\infty=\pm1$.

The violations at $\phi_\infty=\pm1$ can be understood as follows.
In the $\lambda\rightarrow\infty$ limit, we have $\brav{\alpha\brap{\phi}}\rightarrow\infty$ everywhere except at $\phi=\pm1$.
When solving the differential equation for $\zeta_e\brap{\phi}$ near $\phi=1$, we impose boundary conditions $\zeta_e\brap{1}=\tilde\zeta\brap{1} = \xi^{-1/2}$. The solution to $\zeta_m\brap{\phi}$ near $\phi=1$, on the other hand, has its initial conditions set at $\phi\rightarrow+\infty$. Integrating from this initial condition back towards $\phi=1$, we find $\zeta_m\brap{\phi}\lesssim\frac{1}{\brav{\alpha_m\brap{\phi}}}$, which is close to zero except in a small neighborhood around $\phi=1$. In the limit $\lambda \rightarrow \infty$, however, this neighborhood shrinks to zero size, and consequently $\zeta_m\brap{\phi} \ll 1$ even as $\phi \rightarrow 1$. Thus, for $\phi_\infty = 1$, we have
\be
\frac{|z_{\rm el}|}{|z_{\rm mag}|} = \frac{|\zeta_m|}{|\zeta_e|} = \frac{|\zeta_m(1)|}{\xi^{-1/2}} \ll 1\,,
\ee
and thus there is no co-scaling.

The same reasoning may be applied at $\phi=-1$ with the roles of magnetic and electric black branes exchanged, yielding a similar violation of the co-scaling relation. Sufficiently far away from the points $\phi = \pm 1$, we have $\zeta_e(\phi) \approx \zeta_m(\phi) \approx \tilde \zeta(\phi)$, and we observe co-scaling.

\subsection{Example: Infinite peak before asymptotically constant}\label{sec.sub.non.UV.eg.3}

As discussed, the example of the previous subsection is highly unlikely to occur in the landscape due to divergence of $|\alpha(\phi)|$ as $\phi \rightarrow \pm \infty$. As our final example, we consider a theory that presents no obvious violation of well-established swampland principles yet nonetheless violates the co-scaling relation. This suggests that the requirement of co-scaling may place additional constraints on the two-derivative action of a quantum gravity theory.

In particular, let us again consider a theory with a single modulus $\phi$ and a single gauge field, and let us take
\begin{equation}\label{eqn.black.hole.Qe.squared.tanh.text}
    t\brap{\phi} =c\exp\brac{-\int_0^\phi \tanh\bras{\brap{\phi'-2}\brap{\phi'+2}}\bras{1+\lambda^{\brap{1-{\phi'}^2}}} {d}\phi'}\,,
\end{equation}
with $c$ taken to be a fixed constant.

The resulting $t^{-1}\brap{\phi}$ is plotted in \autoref{fig.black.hole.Qe.squared.tanh}. This choice of $t\brap{\phi}$ leads to
\begin{equation}
\alpha\brap{\phi}=\tanh\bras{\brap{\phi-2}\brap{\phi+2}}\bras{1+\exp\brac{\brap{\log\lambda}\bras{1-{\phi}^2}}}\,.
\end{equation}
 Note that $\alpha(\phi)$ approaches a constant in the asymptotic limits, $\phi \rightarrow \pm \infty$.
 
For notational convenience, we define the region $R:=\brap{-1,+1}$.
Then, in the limit $\lambda\rightarrow\infty$, we have
\begin{equation}
 \alpha\brap{\phi} \rightarrow \begin{cases}
                              -\infty & \text{ for }\phi\in R\\
                              2\tanh\bras{\brap{\phi-2}\brap{\phi+2}} & \text{ for }\phi\in\brac{-1,+1}\\
                              \tanh\bras{\brap{\phi-2}\brap{\phi+2}} & \text{ otherwise}\,.
                             \end{cases}
\end{equation}
In other words, for large $\lambda$, $\alpha$ is large but negative inside $R$ and tends to a constant in $\lambda$ outside of it.

We can once again consider two types of limit, namely:
\begin{enumerate}
\item $\lambda\rightarrow\infty$. This limit corresponds to sending the value of $t^{-1}(\phi)$ at its local minimum and maximum to zero and infinity, respectively.
\item $\phi_\infty\rightarrow\pm\infty$. This corresponds to a limit in which the electric black brane charge $Q^2(\phi_\infty)$ vanishes (for $\phi \rightarrow - \infty$) or diverges (for $\phi \rightarrow + \infty$). Conversely, the magnetic black brane charge diverges and vanishes in these respective limits.
\end{enumerate}

\begin{figure}[!t]
    \centering
    \includegraphics[width=0.45\linewidth]{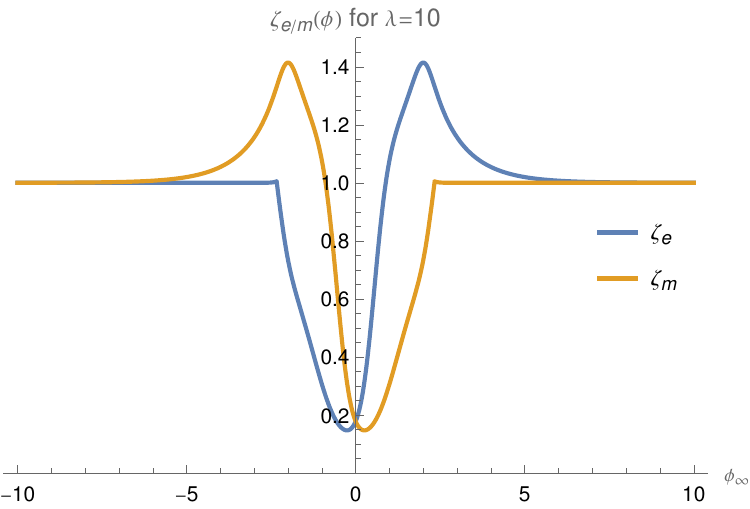}
    \quad
    \includegraphics[width=0.45\linewidth]{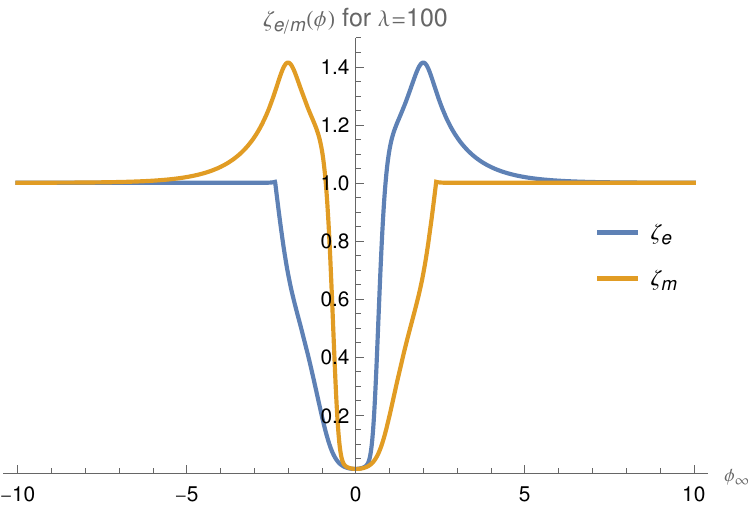}

    \includegraphics[width=0.45\linewidth]{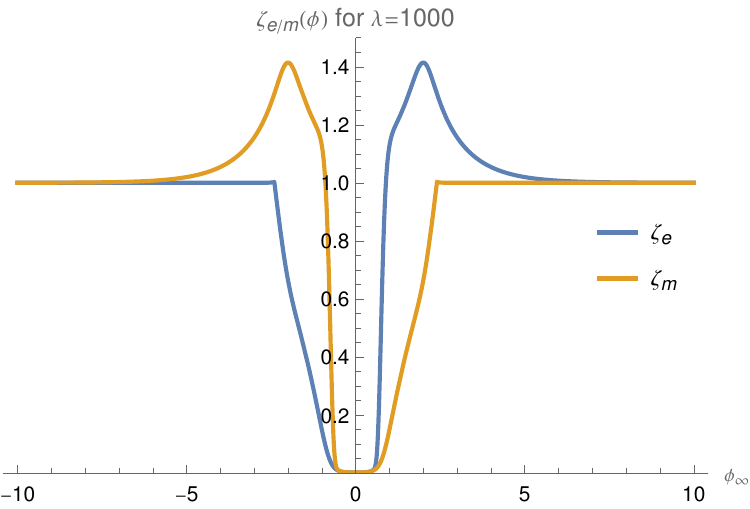}
    \quad
    \includegraphics[width=0.45\linewidth]{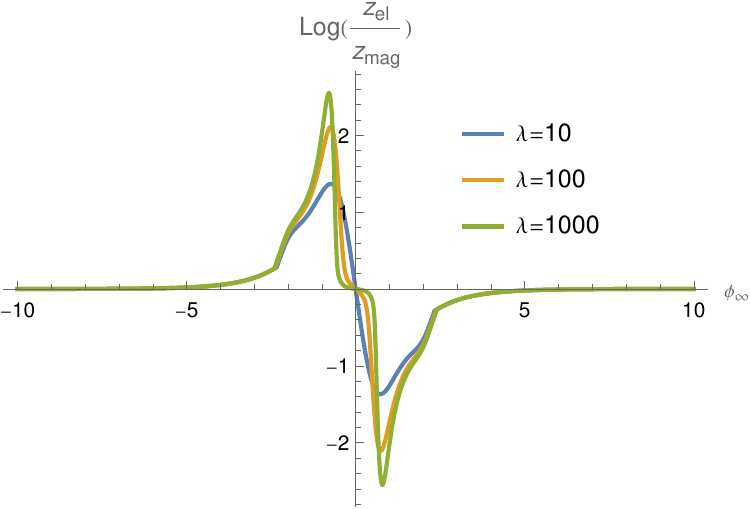}
    \caption{Plots of $\zeta_e$, $\zeta_m$, and the log of their ratio for the example in \S\ref{sec.sub.non.UV.eg.3}.
    The first three sub-figures each depict $\zeta_e$, $\zeta_m$ for a given value of $\lambda$; note the linear scale in the y-axis. The final sub-figure plots the $\log$ (base $e$) of the ratio $\frac{\zeta_m}{\zeta_e}=\frac{|z_{\rm el}|}{|z_{\rm mag}|}$ for the $\lambda$ values from the other sub-figures.}
    \label{fig.black.holes.tanh.zetas}
\end{figure}

\autoref{fig.black.holes.tanh.zetas} shows the numerically obtained solution for some values of $\lambda$. Once again, there is co-scaling in limits of type (2).

\begin{figure}[!t]
    \centering
    \includegraphics[width=\linewidth]{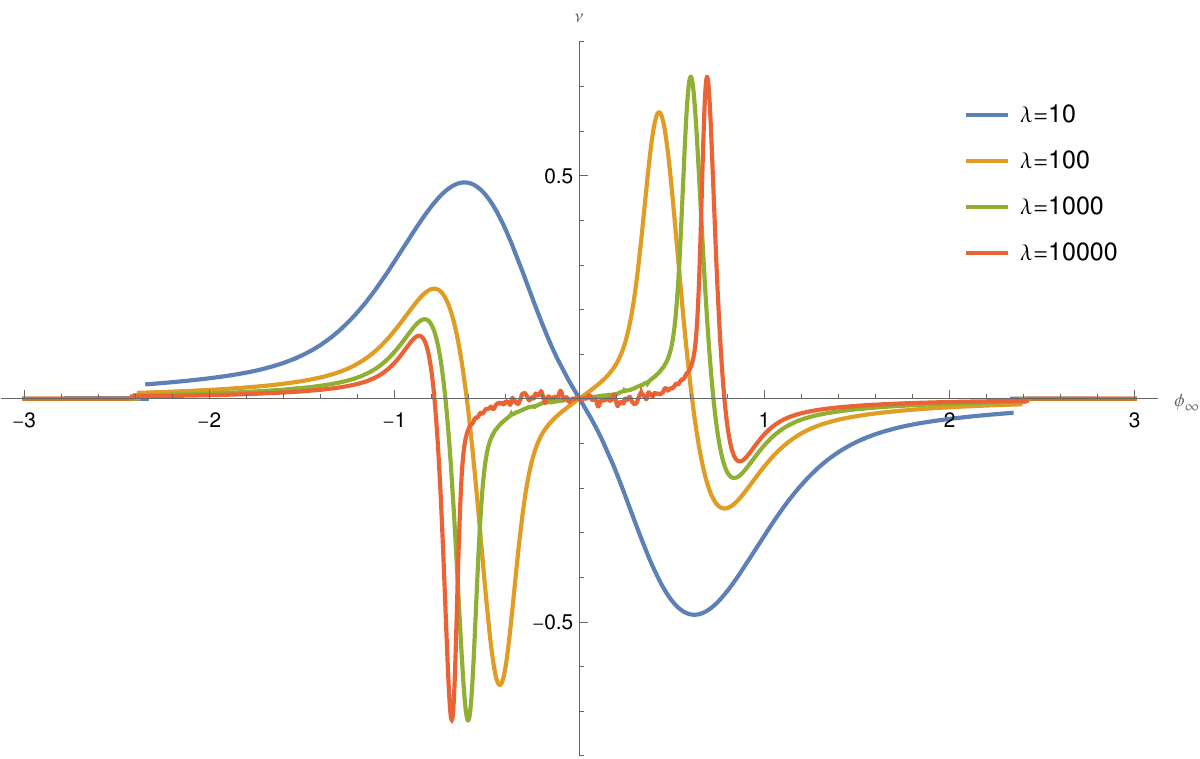}
    \caption{$\nu := \frac{\partial\bras{\log\brap{\zeta_m/\zeta_e}}}{\partial\bras{\log\lambda}}$ as a function of $\phi$ for increasing values of $\lambda$ for the example in \S\ref{sec.sub.non.UV.eg.3}. It may appear that the $\log_{10}\bras{\lambda}=1$ plot terminates at $\sim\pm3.3$, but rather there is a jump to being much closer to zero.}
    \label{fig.black.hole.tanh.nu}
\end{figure}

\autoref{fig.black.hole.tanh.nu} plots $\nu$ as a function of $\phi$ for various values of $\log\lambda$. Once again, there are narrowing peaks, with $\nu\rightarrow0$ elsewhere. Away from these peaks, there is co-scaling in the limit $\lambda \rightarrow \infty$. 

The position of the right narrowing peak appears to be moving to larger $\phi_\infty$ as $\lambda$ increases.
Assuming that the peak location approaches a finite constant $\phi^*$ in the $\lambda\rightarrow\infty$ limit, we see that the co-scaling relation is violated at $\phi_\infty=\pm\phi^*$ but satisfied other values of $\phi_\infty$.

\section{Discussion}\label{s.discussion}

\subsection{Remarks on 4d compactifications}

In this paper, we have focused our attention primarily on 5d supergravity theories, which arise from compactifying M-theory on a Calabi-Yau manifold $X$. By dimensionally reducing this theory, we get a 4d supergravity theory, which may alternatively be viewed as the compactification of Type IIA on the same Calabi-Yau. As a result, much of our 5d analysis above carries over to the 4d case at hand.

There are some important differences between the vector multiplet moduli space of a 4d supergravity theory and its parent 5d supergravity theory.
To begin, in 5d the moduli are restricted to real values, whereas in 4d the moduli are complex. Thus, a codimension-1 boundary in the 5d moduli space may be only a codimension-2 boundary in the 4d moduli space, and it may be possible to skirt such a boundary to access new phases of the theory \cite{Witten:1998qj}.
In addition, there is also the complication of instanton worldsheet corrections to the low-energy effective action; these are negligible when the 2-cycles of the Calabi-Yau manifold are large, but they become important when these cycles become small.

At a conifold locus in moduli space, where a finite number of curves shrink to zero size (and a finite number of charged particles become massless), the instanton contributions to the prepotential can be computed straightforwardly \cite{Strominger:1995cz}. In contrast, an infinite tower of light particles indicates a breakdown of the instanton expansion, as an infinite sum of instanton corrections become unsuppressed. To produce a light tower of BPS particles in 4d with $\lVert \zel \rVert \ll 1$ from D2-branes wrapping curves of homology class $k [C_2] \in H_2(X, \mathbb{Z})$ while maintaining control of the instanton expansion, one must simultaneously decompactify the Calabi-Yau $X$ so that
\begin{align}
    \alpha' \ll \vol(C_2) \ll \vol(X)^{1/3} \,.
\end{align}
Such a limit represents a decompactification to 5d. Consequently, by the dimensional reduction arguments of \S\ref{ss.dimred}, the analysis of co-scaling and alignment in these models reduces to the 5d case studied in detail above. Furthermore, the effects of $B_2$ axions, which lead to the complexification of the K\"ahler cone, are subleading in such a decompactification limit.

It would be worthwhile to explore co-scaling and alignment in 4d Type IIA compactifications further if one could find a way to resum the instanton contributions near a boundary of the extended K\"ahler cone, where an infinite tower of particles become light. We leave this question for future study.

It would also be interesting to consider 4d Type IIB compactifications. In this case, the RR field $C^{(4)}$ gives rise to axions in 4d with instanton actions coming from divisor volumes and axion string tensions from curve volumes, while the 10d 2-form gauge fields $C^{(2)}$ give rise to axions in 4d with the roles of divisors and curves reversed. Similarly, many 10d string theories lead to 4d axions from $B^{(2)}$ reduced on 2-cycles, with instanton actions from curve volumes and axion string tensions from divisor volumes. Thus, the very same geometric data that we studied in 5d can be interpreted in terms of axion physics, provided that we are in the geometric regime where volumes are large in string units (see~\cite{Hebecker:2015zss} for similar remarks about the WGC). These cases are of particular interest in phenomenology, as we will discuss shortly. On the other hand, in Type IIB we obtain ordinary 1-form gauge fields from $C^{(4)}$ reduced on 3-cycles. Understanding the spectrum of electric and magnetic towers for such gauge fields requires a study of complex structure moduli space rather than K\"ahler moduli space, which could be interesting to pursue.

\subsection{Phenomenological outlook}

Here we briefly comment on some possible applications of the idea of electric-magnetic co-scaling to the real world. The notion of co-scaling was introduced in~\cite{Reece:2024wrn} as an element of an argument that, for an extra-dimensional QCD axion (e.g.,~\cite{Witten:1984dg, Choi:2003wr, Conlon:2006tq, Svrcek:2006yi}), an experimental measurement of the decay constant $f$ could serve as a proxy for the quantum gravity cutoff energy $\Lambda_\mathrm{QG}$. In particular, the claim is that for extra-dimensional axions, 
\begin{equation} \label{eq:axioncutoff}
  \Lambda_\mathrm{QG} \lesssim 2\pi \sqrt{S_\mathrm{inst}} f,
\end{equation}
with $S_\mathrm{inst}$ the leading instanton action. For a QCD axion, $S_\mathrm{inst} = 8\pi^2/g^2$ is known (up to modest uncertainty from running the measured QCD coupling up to UV energies), so this gives an upper bound of $\Lambda_\mathrm{QG} \lesssim 100 f$.

To support this claim using co-scaling, consider axion strings, the objects magnetically charged under the axion. If axion strings and instantons exhibit co-scaling, then
\begin{equation} \label{eq:axioncoscaling}
T_\mathrm{axion} \approx 2 \pi S_\mathrm{inst} f^2.
\end{equation}
The notion of alignment introduced in this paper provides a better guide to precisely {\em which} axion strings should exhibit this property, which can be verified in many examples. In extra-dimensional axion theories, the axion strings typically arise from fundamental, ultraviolet objects (e.g., fundamental strings or D-branes) wrapping cycles in the extra dimensions, rather than solitons. At least in the regime where the geometric EFT is under control, their tension is expected to lie at or above the fundamental cutoff scale of the theory. Indeed, one can check in many examples that the fundamental string scale $M_s$, interpreted as the quantum gravity cutoff, obeys the bound~\eqref{eq:axioncutoff}~\cite{Reece:2024wrn, Benabou:2025kgx}.

In this paper, we have found that co-scaling does not always hold. In particular, it can fail in cases where only a finite set of states become light. For example, consider an axion that arises in Type IIB string theory from the RR 2-form field $C^{(2)}$, dimensionally reduced on a shrinking 2-cycle at a conifold singularity. In this case, one can estimate (see~\cite{Svrcek:2006yi}) that the decay constant is at the winding string scale: $f \sim M_\mathrm{Pl}/\cV^{1/3} \sim M_s \cV^{1/6}$ (with $\cV$ the Calabi-Yau volume in string units). Instantons arise from Euclidean D1-branes wrapping the shrinking cycle and thus have small action, whereas axion strings arise from D5-branes wrapping divisors. No divisor shrinks in the conifold limit, so no axion string exhibits the co-scaling relation~\eqref{eq:axioncoscaling}. On the other hand, the relationship~\eqref{eq:axioncutoff} still holds whenever we are in the geometric regime where $S_\mathrm{inst} \gtrsim 1$ (and in particular, for realizations of the QCD axion along these lines), so this exception to co-scaling does not undermine the main phenomenological conclusion. In fact, there are other arguments for~\eqref{eq:axioncutoff} based on naturalness of loop corrections to the axion propagator from towers of KK modes of gauge fields and on perturbative unitarity~\cite{Reece:2024wrn, Seo:2024zzs, Benabou:2025kgx}.  

Along similar lines, examples of co-scaling might lead us to expect that if a magnetic monopole exists in the real world with mass $m_{\rm mag}$, then hypercharge gauge theory should break down by an energy scale of order $m_{\rm el} \sim \alpha m_{\rm mag}$, either because U(1)$_Y$ embeds in a non-abelian group (so that $m_{\rm mag}$ is the mass of an 't~Hooft-Polyakov monopole and $m_{\rm el}$ is the mass of a $W'$ boson) or because a co-scaling tower of hypercharged states begins at $m_{\rm el}$. Such a conclusion could be important, for example, in searches for relatively low-mass magnetic monopoles such as the MoEDAL experiment at the LHC~\cite{MoEDAL:2014ttp, MoEDAL:2019ort}. However, to make this argument one must ask: what if the magnetic monopole is not part of a tower, or we are otherwise in one of the exceptional situations in which we do not find co-scaling and alignment?

These considerations make it apparent that the ideas of electric-magnetic co-scaling and alignment potentially have applications, but these applications require a careful assessment of various exceptional cases and how they might relate to the phenomenology under consideration. We leave a more detailed assessment of phenomenological consequences for a future publication.

\section*{Acknowledgements}

We thank Callum Brodie, Naomi Gendler, Ben Heidenreich, and Jakob Moritz for useful discussions. We also thank Naomi Gendler, Timo Weigand, Max Wiesner, and David Wu for constructive feedback following the first preprint version of this paper. MR is supported in part by the DOE Grant DE-SC0013607. The work of TR was supported in part by STFC through grant ST/T000708/1. The work of CT was supported by a studentship from STFC through grant ST/Y509334/1.

\appendix

\section{Six Dimensions}\label{appendix.6d}

Above, we focused our attention primarily on charged particles and instantons in 4d and 5d supergravity. In this appendix, we show that a similar analysis applies to charged strings in 6d supergravity.
 
 6d supergravity theory features two multiplets with scalar fields: tensor multiplets and hypermultiplets. We are interested here in the tensor branch, parametrized by vevs of massless scalar fields in tensor multiplets. At certain loci on the tensor branch, BPS strings may become tensionless, $T_{\rm el} \rightarrow 0$. In this section, we will argue that there exist magnetically charged strings that satisfy \eqref{rr} in a neighborhood of these loci.

More precisely, as discussed briefly in the Introduction, there are two possibilities for how such BPS strings can become tensionless in 6d: (1) at infinite distance, or (2) at finite distance.

In the former case, the string that becomes asymptotically tensionless has vanishing magnetic charge, i.e., the Dirac pairing between its electric charge and magnetic charge vanishes. The string is extremal, so it saturates the WGC bound with $\zel \sim 1$, and relatedly its gauge coupling $\gel$ vanishes in the limit in question. Meanwhile, the magnetic gauge coupling $\gmag = 2\pi/\gel$ diverges, there is a different BPS string, charged magnetically under this gauge group, which saturates the magnetic WGC with $\zmag \sim 1$. Together, these strings exhibit co-scaling, as $\zel / \zmag \sim 1$.

The latter case of a finite-distance boundary is more subtle. Here, the string in question is dyonic: its electric and magnetic charge have a nonzero Dirac pairing. At the boundary where the string becomes tensionless, an SCFT emerges, and the electric/magnetic gauge charge of the string is order-one. Thus, we have $T_{\rm el} \equiv T_{\rm mag}$ and $\gel^2 \sim 1$, so $\gel^2 T_{\rm mag}/T_{\rm el} \sim 1$, and \eqref{rr} is satisfied.

We begin this section with a lightning review of the salient aspects of 6d supergravity. Next, we specialize to the illustrative case of a single tensor multiplet, demonstrating how both infinite-distance boundaries and finite-distance boundaries of moduli space exhibit co-scaling and alignment.

\subsection{Basics of 6d supergravity}\label{s.basic6d}

In this subsection, we review relevant aspects of the tensor multiplet moduli space of a 6d supergravity theory. Our review follows the unpublished notes of Ben Heidenreich \cite{BenSupergravity}, which in turn are based on \cite{Riccioni:1999xq}.

The kinetic terms for the scalar fields, 2-form gauge fields, and graviton of a 6d supergravity theory are given by
\begin{align}
  \tilde{S} &= \frac{1}{2 \kappa_6^2}  \int d^6 x \sqrt{- g}  [\mathcal{R}-
  \eta_{r s} \nabla X^r (\phi) \cdot \nabla X^s (\phi)] - \frac{1}{8 \pi} 
  \int G_{r s} (\phi)  \tilde{H}^r_3 \wedge \ast \tilde{H}^s_3 \,,.
  \label{6daction}
\end{align}
Here, $r, s = 0, \ldots, n_T$, $g^2_6 = (2\pi)^{3 / 2} \kappa_6$, and the gauge-invariant $\tilde H_3^r$ may in general depend on both the 2-form $B_2^r$ as well as the Chern-Simons couplings for the 1-form gauge fields. For instance, for a theory with $n_V$ abelian gauge fields $A_1^r$, $r=1,...,n_V$, we have
\begin{align}
\tilde{H}_3^r = d B_2^r + \frac{1}{4 \pi} c^r_{\; a b} A_1^a \wedge F_2^b\,,
\end{align}
for integers $c^r_{\; a b}$.

Meanwhile, the scalar manifold is determined by the constraint
\begin{equation}
  \eta_{r s} X^r (\phi) X^s (\phi) = - 1,
  \label{6dcons}
\end{equation}
with $\eta_{r s}$ a constant symmetric matrix of signature $(n_T, 1)$ and
\begin{equation}
  G_{r s} (\phi) = \eta_{r s} + 2 X_r (\phi) X_s (\phi) \,.
  \label{6dgk}
\end{equation}
Here, $r, s$ indices are raised and lowered with $\eta_{r s}$, i.e., $X_r =
\eta_{r s} X^s$. It can be shown that by a suitable basis choice, we may set 
\begin{align}
\eta_{r s} = \rm{diag}(+1, +1, ..., +1, -1)~~~~ \text{or}~~~~
\eta_{r s} = \left(\begin{array}{cc} 0 & 1 \\ 1 & 0 \end{array}\right)\,,
\end{align}
where the latter possibility is unique to the case of $n_T = 1$. As we will see, there may be additional boundaries of the moduli space where strings become tensionless.

The action \eqref{6daction} must be supplemented with the (anti)self-duality
constraint
\begin{equation}
  G^r_{\; s} (\phi) \ast \tilde{H}_3^s = \tilde{H}_3^r,
\end{equation}
for $G^r_{\; s} (\phi) = \eta^{r t} G_{t s} (\phi)$. As a result, the magnetic charge $\tilde q^r$ of a string of electric charge $q_s$ is given by
\begin{align}
\tilde q^r = \eta^{rs} q_s\,.
\end{align}
We will see that this implies that strings whose tension vanishes at finite-distance in moduli space are dyonic, carrying both electric and magnetic charge under the same 2-form gauge field.

A string of charge $q_r$ is constrained to satisfy the BPS bound:
\begin{equation}
  T \geq \frac{\sqrt{2 \pi}}{\kappa_6}  | q_r X^r |,
  \label{6dBPS}
\end{equation}
Strings whose tension saturates this bound are called BPS strings.

\subsection{$n_T = 1$}

For simplicity, we restrict our attention to a theory with $n_T=1$, i.e., a 1-dimensional tensor branch. (Our arguments can be extended to the general case with relative ease.)
We further set $\eta_{rs} = \text{diag}(1,-1)$. With this, the constraint \eqref{6dcons} is solved by
\begin{align}
X^0 = \sinh \phi \,,~~~~ X^1 = \cosh \phi\,,
\end{align}
in which case the scalar kinetic term is given by
\begin{equation}
- \frac{1}{2 \kappa_6^2} \eta_{r s} \nabla X^r (\phi) \cdot \nabla X^s (\phi) =-  \frac{1}{2 \kappa_6^2} (\nabla \phi)^2\,,
\end{equation}
and the scalar field is canonically normalized up to a factor of $\kappa_6$.

By \eqref{6dBPS}, the tension of a BPS string of charge $q_r$ is given by
\begin{align}
T = \frac{\sqrt{2 \pi}}{\kappa_6} | q_0 \sinh \phi + q_1 \cosh  \phi |\,.
\end{align} 
For this to vanish at a point in moduli space, we require $|q_0| \geq |q_1|$. Let us first suppose that $q_ 1 = \mp q_0$, so
\be
q_r := (q_0, q_1) = (q_0, \mp q_0)\,.
\ee
In this case, the tension vanishes exponentially as $T \sim \exp(- |\phi|) / \kappa_6$ in the limit $\phi \rightarrow \pm \infty$. The magnetic charge of this string is given by 
\begin{align}
\tilde q^r = \eta^{rs} q_s = (q_0, \pm q_0)\,.
\end{align}
Thus, the Dirac pairing between electric and magnetic charge of the string is given by
\be
q_r \tilde q^r = q_0^2 - q_0^2 = 0\,.
\ee
In other words, the string does not carry electric and magnetic charge under the same 2-form gauge field. However, a string of charge $q'_r = (q_0, \pm q_0)$ has Dirac pairing
\be
q_r \tilde q^{r,\prime} = \eta^{rs} q_r q_s' = 2 q_0^2 
\ee
with the first string,
which is nonzero. This indicates that the two strings carry electric and magnetic charge, respectively, under the same 2-form gauge field. Furthermore, in the limit where the tension of the first string vanishes as $T \sim \exp(- |\phi| ) / \kappa_6$, the tension of the latter string diverges as $T \sim \exp(+ |\phi|) / \kappa_6$. The gauge coupling of the original string decays as
\begin{equation}
\lVert q \rVert^2 \equiv q_r G^{rs} q_s = q_r \eta^{rt} G_{rv} \eta^{vs} q_s = q_r q^r + 2 (q_r X^r)^2 = \frac{2\kappa_6^2}{2 \pi} T^2 \sim \exp(- 2 |\phi|)\,,
\end{equation}
where we have used \eqref{6dgk}. Meanwhile, the latter string has
\be
\lVert q' \rVert^2 \equiv q'_r G^{rs} q'_s = q'_r \eta^{rt} G_{rv} \eta^{vs} q'_s = q'_r q'^r + 2 (q'_r X^r)^2 = \frac{2\kappa_6^2}{2 \pi} (T')^2 \sim \exp(+ 2 |\phi|)\,,
\ee
Both of these strings have $z \sim 1$, 
so they trivially satisfy the co-scaling relation \eqref{rr}.

For $|q_0| > |q_1|$, on the other hand, the tension of the string $T \propto |q_0 \sinh \phi + q_1 \cosh \phi|$ vanishes at finite distance in moduli space as an SCFT emerges. Here, the Dirac pairing of the electric and magnetic charges is nonzero:
\begin{equation}
q_r \eta^{rs} q_s = q_0^2 - q_1^2 > 0\,,
\end{equation}
so the string is dyonic: it carries electric and magnetic charge under the same 2-form gauge field. Thus, the string plays the role of both the electric and magnetic charged object, so $T_{\rm el} \equiv T_{\rm mag}$. Meanwhile, the charge of the string is given by
\be 
\lVert q \rVert^2 \equiv q_r G^{rs} q_s = q_r \eta^{rs} q_s + 2 (q_r X^r)^2 \underset{T \rightarrow 0}\rightarrow  q_r \eta^{rs} q_s  = q_0^2 - q_1^2\,.
\ee
Assuming that the integer charges $q_0$, $q_1$ are order-one, we have that $\lVert q \rVert^2$ is order-one, and thus
\be
\frac{\lVert q \rVert^2 T_{\rm mag}}{T_{\rm el}} \equiv \lVert q \rVert^2 \sim 1\,,
\ee
satisfying the co-scaling relation.

The angle between the electric and magnetic charge of this string is defined to be
\be
\cos \varphi = \frac{q_r q^r}{\lVert q \rVert \lVert \tilde q \rVert} = \frac{q_r q^r}{\lVert q \rVert^2}\,.
\ee
Here,
\be
\lVert q \rVert^2 = q_r G^{rs} q_s = q_r q^r + 2 (q_r X^r)^2 = q_r q^r + \frac{2 \kappa_6^2}{2 \pi} T^2\,.
\ee
Together with the fact that $\zel = \zmag = \lVert q \rVert/T \sim 1/T$, this yields
\be
1 - \cos \varphi = 1- \frac{q_r q^r}{q_r q^r + (T^2/\pi)} \sim T^2 \sim \frac{1}{\zel^2}\,,
\ee
so the electric charge $q_r$ and the magnetic charge $q^r$ exhibit rapid alignment.

Therefore, we see that BPS strings in 6d theories with $n_T=1$ tensor multiplets that become tensionless at either infinite distance or finite distance exhibit co-scaling and alignment.

It is also interesting to see how the behavior of the species scales differs between the two different types of boundaries using the coefficient of the $R^2$ coefficient in \eqref{speciescoef}, as we did in 5d supergravity in \S\ref{ss.5dspecies}. 
In 6d supergravity, the relevant terms in the action take the form \cite{Grimm:2012yq}:
\begin{equation}
   \mathcal{L} \supset  -\eta_{rs} a^r X^s R_{\mu\nu} R^{\mu\nu}\,,
\end{equation}
where $a^r$ is a constant. By \eqref{speciescoef}, we then have the approximate identification
\be
\frac{1}{\Lambda_{\rm QG}^2} \sim (a_1 X^1 - a_0 X^0) = a_1 \cosh \phi - a_0 \sinh \phi\,. 
\ee
In an asymptotic limit $|\phi| \rightarrow \infty$, we therefore have
\be
\frac{1}{\Lambda_{\rm QG}^2} \sim \frac{1}{2}(a_1 - a_0) \exp(|\phi|) + \frac{1}{2}(a_1 + a_0) \exp(-|\phi|) \,,
\label{LQGrelate}
\ee
which diverges unless $a_1 = a_0$, in which case it tends to zero. This latter possibility would indicate a breakdown of the relationship \eqref{LQGrelate}, since $\Lambda_{\rm QG}$ is bounded above by the Planck scale. Fortunately, explicit examples (see e.g. \cite{Kumar:2009ac}) all have $a_1 \neq a_0$ in this basis, leading to a species scale which vanishes in the asymptotic limit as
\be
\Lambda_{\rm QG} \sim \exp\left(- \frac{1}{2}|\phi|\right)\,.
\ee
This precisely matches the vanishing of the string scale
\be
M_{\rm string} \equiv \sqrt{2 \pi T} \sim \exp\left(- \frac{1}{2}|\phi|\right)\,,
\ee
so (as expected), the species scale is determined by the (fundamental) string scale in an infinite-distance limit.

In contrast, $X^0$ and $X^1$ are order-one near a finite-distance boundary, so any linear combination of them is either order-one or vanishing. The case of a vanishing $R^2$ coefficient $a_1 X^1 -a_0 X_0$ would imply that either the species scale diverges or else the relationship between the $R^2$ coefficient and the species scale \eqref{LQGrelate} breaks down. Rejecting this latter possibility, we must have that $a_1 X^1 -a_0 X_0$ is order-one, and hence the species scale is also order-one. Once again, this conclusion is confirmed by the explicit examples of \cite{Kumar:2009ac}.

The upshot of this is that the species scale decays with the string scale $M_{\rm string
}$ of the fundamental string in an infinite-distance limit, but it remains parametrically above the mass scale of the dyonic string $\sqrt{2 \pi T}$ near an SCFT boundary of the moduli space. In contrast to the fundamental string, this SCFT string decouples from the gravitational dynamics and does not produce a breakdown of non-gravitational EFT below the Planck scale.

\section{Further Details on Extremal Black Holes}\label{appendix.black.hole}

In this appendix, we describe the method used to calculate the charge-to-mass ratios of black branes in \S\ref{sec.non.UV}. To begin, we describe the general attractor mechanism for general number of moduli fields and U(1) gauge fields. We then introduce a recasting of the equations to aid the finding of numerical solutions before specializing to a single modulus and U(1) gauge field and flesh out some of the calculatory details from the examples in \S\ref{sec.non.UV}.

\subsection{Fake superpotential formalism}\label{ss.fake}

\subsubsection{The attractor mechanism}
In this subsection we describe the attractor mechanism, following the method in \cite{Heidenreich:2020upe,Harlow:2022ich}.\footnote{But with a different normalization convention for the metric, so that $G_{ij}^{\brap{\text{there}}}=\frac1{2\kappa^2}{\mathfrak{g}}_{ij}$.}

Consider the low energy Einstein frame effective action
\begin{equation}
 S=\int d^dx\sqrt{-g}\brac{\frac1{2\kappa^2}\bras{\mathcal{R}-\frac12{\mathfrak{g}}_{ij}\nabla\phi^i\cdot\nabla\phi^j} - \frac12t_{IJ}\brap{\phi}F_{p+1}^I \cdot F_{p+1}^J}\,,
\end{equation}
for $\phi^i$ moduli fields, $F_{p+1}^I=dA_p^I$ U(1) $\brap{p-1}$-form vector fields.
Let $\phi_\infty^i$ be the value of $\phi^i$ at $r=\infty$.

For a black brane of electric charge $q^e_I$, (assuming no theta term) and no magnetic charge, we define 
\begin{equation}
 Q_e^2\brap{\phi} := t^{IJ}\brap{\phi}q^e_Iq^e_J\,.
\end{equation}
Then, consider the differential equation
\begin{equation}\label{eqn.de.we}
 \xi W_e^2\brap{\phi} + 2{\mathfrak g}^{ij}\partial_iW_e\brap{\phi}\partial_jW_e\brap{\phi} = \kappa^{-2}Q_e^2\brap{\phi}\,,
\end{equation}
for $\xi:=\frac{p\brap{d-p-2}}{d-2}$.

Suppose $\phi=\phi_h$ is a local (or asymptotic) minimum of $Q_e^2\brap{\phi}$.
Then, solving \eqref{eqn.de.we} with the initial condition that $W_e\brap{\phi}$ has a local minimum at $\phi=\phi_h$ yields a unique solution $W_{e,h}\brap{\phi}$ in some region $R_h$ encompassing $\phi_h$.
Then for any $\phi_\infty\in R_h$, there is a quasiextremal black brane with tension and electric charge given by
\begin{equation}
 T_{\mathrm{el}} = W_{e,h}\brap{\phi_\infty} \quad,\quad \bravv{q_e}^2 = Q_e^2\brap{\phi_\infty}\,,
\end{equation}
where a black brane is said to be quasiextremal if its surface gravity or horizon area is zero \cite{Harlow:2022ich}. An extremal black hole is necessarily quasiextremal, but quasiextremal black holes are not always extremal.

The region of definition $R_h$ need not be the whole of moduli space. Indeed, from \eqref{eqn.de.we}, we must have $\brav{W_{e,\phi_h}\brap{\phi}}\leq\sqrt{\xi}Q_e\brap{\phi}$, so if a solution reaches the value $W_{e,\phi_h}\brap{\phi}/Q_e\brap{\phi}=\sqrt{\xi}$, then $\phi$ must either lie on the boundary of $R_h$ or represent a maximum of $W_{e,\phi_h}\brap{\phi}/Q_e\brap{\phi}$.
There may also be turning points or caustics, indicating that the solution is no longer unique \cite{Harlow:2022ich}.

In order to consider quasiextremal black branes for $\phi_\infty$ outside $R_h$, we can repeat the above process for each local (or asymptotic) minimum, obtaining a collection of such $W_{e,h}\brap{\phi}$'s with corresponding domains $R_h$. We can then patch these functions together to form a continuous (but not necessarily smooth) function
\begin{equation}
 W_e\brap{\phi} := \min_{h:\phi\in R_h}W_{e,\phi_h}\brap{\phi}\,,
\end{equation}
which is defined on $\cup_h R_h$.%
\footnote{
In the examples we consider below, the domain $\cup_h R_h$ is in fact the entirety of the moduli space, though it is not clear to us that this should be true in general.
} The effect of taking the minimum is to make the resulting black brane not only quasiextremal, but extremal.

With this, the tension and electric charge of an extremal black brane for a specified value of $\phi_\infty$ is given by \cite{Harlow:2022ich}
\begin{equation}
 T_{\mathrm{el}} = W_e\brap{\phi_\infty} \quad,\quad \bravv{q_e}^2 = Q_e^2\brap{\phi_\infty}\,.
 \label{TandQ}
\end{equation}

\begin{figure}[h]
 \includegraphics[width=\linewidth]{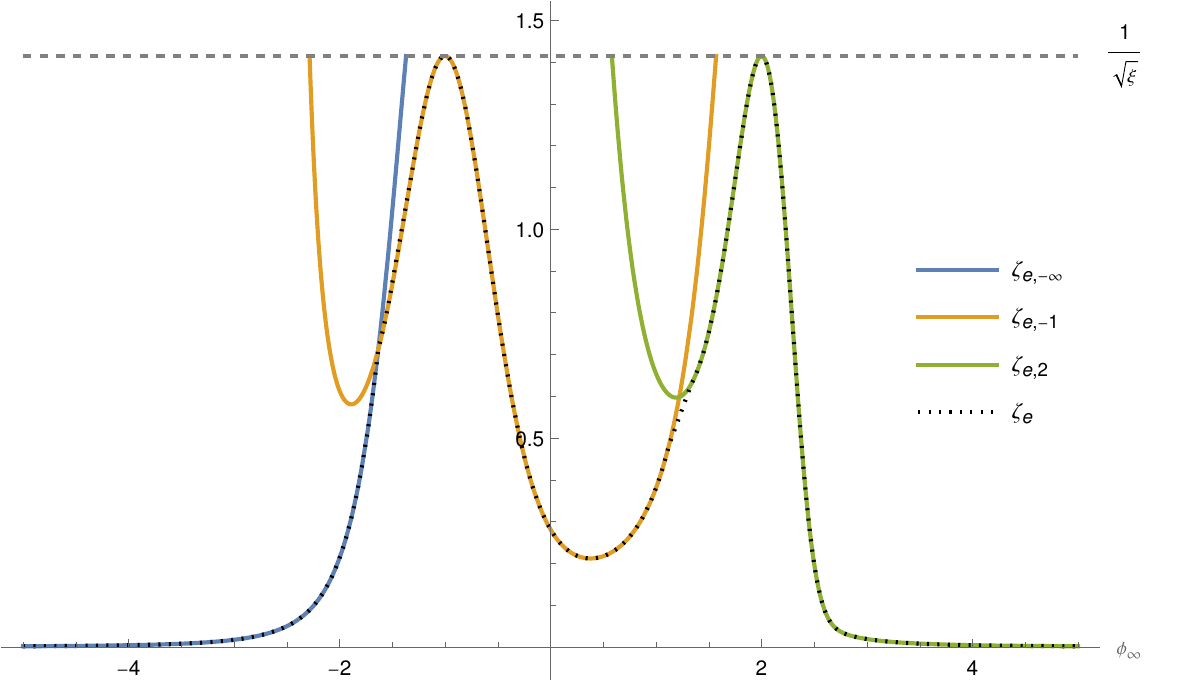}
 \caption{The $\zeta_{e,h}\brap{\phi_\infty}:=W_{e,h}\brap{\phi_\infty}/Q_e\brap{\phi_\infty}$, as well as $\zeta_{e}\brap{\phi_\infty}:=W_{e}\brap{\phi_\infty}/Q_e\brap{\phi_\infty}$, for the system given by \eqref{eqn.Qe.squared.for.black.holes.patching.functions.together}. Note that at any given point $\phi$, the value of the dotted $\zeta_e\brap{\phi}$ is given everywhere by the smallest value amongst the solid $\zeta_{e,h}\brap{\phi}$'s that are defined at $\phi$.}
 \label{fig.black.holes.patching.functions.together}
\end{figure}

\autoref{fig.black.holes.patching.functions.together} demonstrates the above process for finding $W_{e}$ for the case of a single U(1) and modulus, ${\mathfrak{g}}_{11}=1$, $\kappa=1$, $d=4$, $p=1$ and
\begin{equation}\label{eqn.Qe.squared.for.black.holes.patching.functions.together}
 Q_e^2\brap{\phi}=Q^2\exp\brap{\frac25\phi^5 - \frac53\phi^3+4\phi}\,,
\end{equation}
where $Q$ is some constant.
The minima of $Q_e^2\brap{\phi}$ lie at $\phi_h={-\infty,-1,+2}$, producing solutions $W_{e,-\infty}$, $W_{e,-1}$, and $W_{e,+2}$, respectively.
Note that, for ease of presentation, we have divided $W_{e,h}$ and $W_e$ by $Q_e$ to obtain the plotted functions $\zeta_{e,h}$ and $\zeta_e$.

Similarly, for a black hole of magnetic charge $q_m^I$ (and no electric charge), we define
\begin{equation}
 Q_m^2\brap{\phi} = 4\pi^2t^{IJ}\brap{\phi}q_m^Iq_m^J\,.
\end{equation}

Thus, the magnetic story is identical to the electric case after replacing $Q_e^2$, $W_e$, $\bravv{q_e}^2$, etc., with $Q_m^2$, $W_m$, $\bravv{q_m}^2$, etc., respectively.

\subsubsection{Re-casting in terms of $\zeta$}

We are particularly interested in the charge-to-tension ratio of extremal black holes. To this end, we define 
\be
\zeta_e\brap{\phi}:=W_e\brap{\phi}/Q_e{\brap{\phi}}\,.
\ee
Substituting into \eqref{eqn.de.we}, and using units $\kappa=1$, we get 
\begin{equation}\label{eqn.de.zetae}
 \xi \zeta_e^2\brap{\phi}
 + 2{\mathfrak g}^{ij}
    \bras{\partial_i\zeta_e\brap{\phi}+\frac12\alpha_{e,i}\brap{\phi}\zeta\brap{\phi}}
    \bras{\partial_j\zeta_e\brap{\phi}+\frac12\alpha_{e,j}\brap{\phi}\zeta\brap{\phi}}
= 1
\end{equation}
for $\alpha_{e,i}\brap{\phi}:=\frac{\partial\bras{\log\brap{Q_e^2\brap{\phi}}}}{\partial \phi^i}$. Note that, by \eqref{TandQ}, $\zeta_e(\phi_\infty)$ is the reciprocal of the charge-to-tension ratio of the black brane:
\be
z_{\rm el} = \frac{\lVert q_e \rVert}{T_{\rm el}} = \frac{1}{\zeta_e(\phi_\infty)}\,.
\ee

We next reformulate the initial condition for $W_e$ as a condition on $\zeta_e$. At a local minimum of $Q_e^2\brap{\phi}$ (with finite $\phi_h$), the definition of $\alpha_{e,i}$ gives $\alpha_{e,i}\brap{\phi_h}=0$. Since the initial condition for $W_e$ dictates that $\partial_i W_{e}\brap{\phi_h}=0$, we find that $\partial_i \zeta_{e}\brap{\phi_h}=0$. Substituting into \eqref{eqn.de.zetae}, this implies the initial condition $\zeta_e\brap{\phi_h}={\xi}^{-\frac12}$.

It is also possible for the attract point $\phi_h$ to lie at infinite distance in moduli space. Such ``asymptotic attractors'' again have $\partial_i \zeta_e(\phi_h) = 0$, which can be substituted into \eqref{eqn.de.zetae} to obtain the initial condition
\begin{equation}
\zeta_e\brap{\phi}\rightarrow\tilde\zeta_e\brap{\phi}\text{ as }\phi\rightarrow\phi_h\,,
\end{equation}
for
\begin{equation}
 \tilde\zeta_e\brap{\phi}:=\bras{\xi+\frac{{\mathfrak g}^{ij}\brap{\phi}\alpha_{e,i}\brap{\phi}\alpha_{e,j}\brap{\phi}}2}^{-\frac12}\,.
\end{equation}
Note that for a minimum $\phi_h$ at finite distance, we must also have $\zeta_e\brap{\phi_h}=\tilde\zeta_e\brap{\phi_h}$, with the additional restriction that $\alpha_{e,i}(\phi_h) = 0$.

Hence we interpret the initial conditions for $\zeta_e$ as $\zeta_e\brap{\phi}\rightarrow\tilde \zeta_e\brap{\phi_h}$ at all (local or asymptotic) minima $\phi_h$, where ``$\rightarrow$'' can be taken as an equality if $\phi_h$ is finite.

\subsubsection{A single modulus and gauge field}

Let us now specialize to the case of a single modulus $\phi$. Setting ${\mathfrak{g}}_{11}=1$ without loss of generality, \eqref{eqn.de.zetae}  becomes 
\begin{equation}\label{eqn.de.zetae.1.modulus}
 \xi \zeta_e^2\brap{\phi}
 + 2
\bras{\dot\zeta_e\brap{\phi}+\frac12\alpha_{e}\brap{\phi}\zeta_e\brap{\phi}}^2
= 1\,,
\end{equation}
with $^\cdot$ denoting differentiation with respect to $\phi$.
Rearranging, we get
\begin{equation}\label{eqn.de.zetae.1.modulus.for.zetadot}
\dot\zeta_e\brap{\phi} =
-\frac12\alpha_{e}\brap{\phi}\zeta_e\brap{\phi}
+\eta\sqrt{\frac{{1-\xi \zeta_e^2\brap{\phi}}}{2}}\,,
\end{equation}
for some $\eta\brap{\phi}\in\brac{\pm1}$, which encodes whether we take the positive or negative square root.
Returning briefly to the $W_e$ differential equation, we find
\begin{equation}
 \dot W_e\brap{\phi} =
 Q_e\brap{\phi}\cdot\eta\sqrt{\frac{{1-\xi \zeta_e^2\brap{\phi}}}{2}}
\end{equation}
for the same $\eta$.
The condition that $\phi=\phi_h$ represents a minimum of $Q_e$ implies $\eta\rightarrow\pm1$ as $\phi\rightarrow\phi_h^\pm$.
Further, in order for $\dot\zeta_e$ to be continuous when $\eta$ changes between $\pm1$, we must have $\zeta_e(\phi_h
)=\xi^{-\frac12}$.
But since \eqref{eqn.de.zetae} restricts $\brav{\zeta_e}\leq{\xi}^{-\frac12}$, for this to occur $\phi_h$ must be at a local maximum of $\zeta_e$, so $\dot\zeta_e(\phi_h)=0$. Substituting into \eqref{eqn.de.zetae.1.modulus.for.zetadot}, we find $\alpha_e=0$, which is consistent with the fact that $\phi_h$ is a local minimum of $Q_e^2$.

Combining these results, we find that the differential equation to solve is
\begin{equation}\label{eqn.de.zetae.1.modulus.for.zetadot.eta.fixed}
\dot\zeta_e\brap{\phi} =
-\frac12\alpha_{e}\brap{\phi}\zeta\brap{\phi}
+\operatorname{sign}\brap{\phi-\phi_h}\sqrt{\frac{{1-\xi \zeta_e^2\brap{\phi}}}{2}}
\end{equation}
with the initial condition that $\zeta_e\brap{\phi_h}=\xi^{-\frac12}$.
This differential equation is valid in the region surrounding $\phi_h$ with $\zeta_e(\phi) < \xi^{-\frac12}$.

Solving this differential equation (with additional flips of $\eta$ at points with 
$\zeta_{e}=\xi^{-\frac12}$, if necessary)
we get a solution $\zeta_{e,h}\brap{\phi}$ over the domain $R_h$ surrounding $\phi_h$.
Repeating this for every local or asymptotic minimum of $\phi_h$ of $Q_e^2$, we may construct the full $\zeta_e$ by taking
\begin{equation}
    \zeta_e\brap{\phi}=\min_{h \, | \, \phi\in R_h}\zeta_{e,h}\brap{\phi}\,.
\end{equation}

A similar analysis applies to the magnetic case.

Finally, let us restrict our attention to the case of a single gauge field $A_p$.
With this,
using the relation $\gmag = 2 \pi / \gel$ between the electric and magnetic coupling constants, we have
$\alpha\brap{\phi}:=\alpha_e\brap{\phi}=-\alpha_m\brap{\phi}$. Therefore,
\begin{equation}
 \tilde\zeta\brap{\phi}:=\tilde\zeta_e\brap{\phi}=\tilde\zeta_m\brap{\phi}=
 \bras{\xi+\frac{\alpha\brap{\phi}^2}2}^{-\frac12}\,.
\end{equation}

\subsection{Examples}\label{appendix.sub.black.hole.egs}

In what follows, we provided further details on the calculation of $\zeta_e$ and $\zeta_m$ for the examples considered in \S\ref{sec.non.UV}.

\subsubsection{Example: Constant $\alpha$}

First we consider the example in \S\ref{sec.sub.non.UV.eg.1}, namely the simple example of a dilatonic coupling to the Maxwell term, considered previously in e.g. 
\cite{Horowitz:1991cd, Heidenreich:2015nta, Harlow:2022ich}. We set $t_{11}= \frac{1}{\gel^2}= \frac1{e_p^2}e^{-\alpha\phi}$, with $\alpha$ and $e_p$ constant. This gives
\begin{equation}
 Q_e^2=e_p^2 e^{\alpha\phi},\quad
 Q_m^2=4\pi^2e_p^{-2} e^{-\alpha\phi},\quad
 \alpha\brap{\phi}\equiv\alpha,\quad
 \tilde\zeta\brap{\phi}\equiv\bras{\xi+\frac{\alpha^2}2}^{-\frac12}\,.
\end{equation}
The only minimum of the electric/magnetic is at $-\infty$/$+\infty$ respectively, so the differential equations for all $\phi$ are
\begin{eqnarray}
 \dot\zeta_e\brap{\phi} &=&
-\frac12\alpha\zeta_e\brap{\phi}
+\sqrt{\frac{{1-\xi \zeta_e^2\brap{\phi}}}2}\\
 \dot\zeta_m\brap{\phi} &=&
+\frac12\alpha\zeta_m\brap{\phi}
-\sqrt{\frac{{1-\xi \zeta_m^2\brap{\phi}}}2}\,.
\end{eqnarray}
The unique positive solution to these equations over all $\phi \in \mathbb R$ is simply 
\cite{Harlow:2022ich}
\be
\zeta_e=\zeta_m=\tilde\zeta = \bras{\xi+\frac{\alpha\brap{\phi}^2}2}^{-\frac12}\,.
\ee
This indeed satisfies the initial conditions for $\zeta_{e/m}$ from above, which in this case are $\zeta_{e/m}\brap{\phi}\rightarrow\tilde\zeta\brap{\phi}$ as $\phi\rightarrow\mp\infty$.
Notably, $\zeta_e$ and $\zeta_m$ are identical, and both are independent of $\phi$.

\subsubsection{Example: Quadratic $\alpha$}

\begin{figure}
    \centering
    \includegraphics[width=0.5\linewidth]{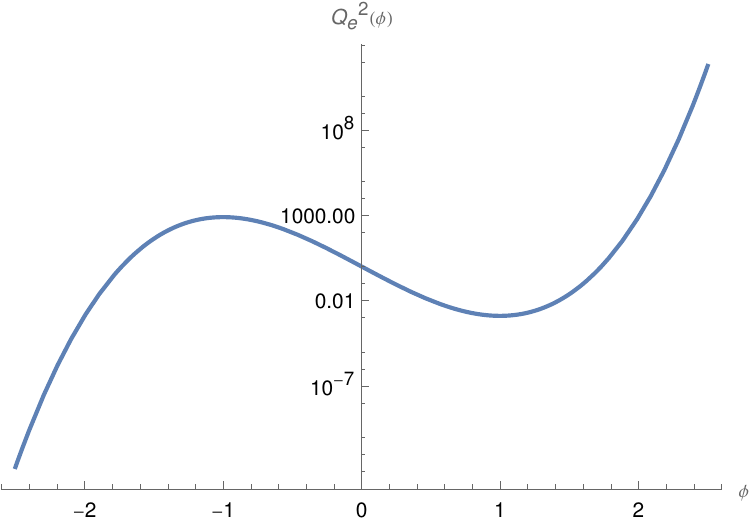}
    \caption{$Q_e^2\brap{\phi}$ given by \eqref{eqn.black.hole.Qe.squared.quadratic}, for $Q_e^2\brap{0}=1$, $\lambda=10$. Note the logarithmic scale on the y-axis.}
    \label{fig.black.hole.Qe.squared.quadratic}
\end{figure}

We now turn to the example in \S\ref{sec.sub.non.UV.eg.2}. We once again consider a theory with a single modulus $\phi$ and a single gauge field, but now we let the electric charge take the form
\begin{equation}\label{eqn.black.hole.Qe.squared.quadratic}
Q_e^2\brap{\phi}=Q_e^2(0) \exp\bras{\lambda\brap{\frac13\phi^3-\phi}}\,,
\end{equation}
with $Q_e^2\brap{0}$ taken to be a fixed constant. This function is plotted in \autoref{fig.black.hole.Qe.squared.quadratic}.
Note that the minima of $Q^2_e(\phi)$ occur for $\phi = -\infty$, and $\phi = +1$, while the maxima occur at $ \phi = -1$ and $\phi = +\infty$.

This choice of $Q^2(\phi)$ leads to a quadratic function for $\alpha(\phi)$:
\be
\alpha\brap{\phi}= \frac{\partial}{\partial \phi} \log Q_e^2 =  \lambda\brap{\phi-1}\brap{\phi+1} \,.
\ee

\subsubsection{Example: Infinite peak before asymptotically constant}

Lastly, we return to the example in \S\ref{sec.sub.non.UV.eg.3} and take
\begin{equation}\label{eqn.black.hole.Qe.squared.tanh}
    Q_e^2\brap{\phi} = Q_e^2\brap{0}\exp\brac{\int_0^\phi \tanh\bras{\brap{\phi'-2}\brap{\phi'+2}}\bras{1+\lambda^{\brap{1-{\phi'}^2}}} {d}\phi'}
\end{equation}
with $Q_e^2\brap{0}$ taken to be a fixed constant.

\begin{figure}
    \centering
    \includegraphics[width=0.5\linewidth]{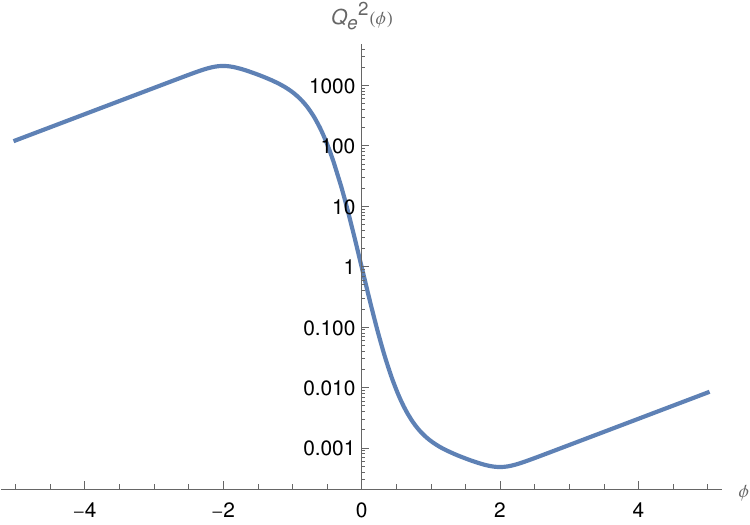}
    \caption{$Q_e^2\brap{\phi}$ given by \eqref{eqn.black.hole.Qe.squared.tanh}, for $Q_e^2\brap{0}=1$, $\lambda=10$. Note the logarithmic scale on the y-axis.}
    \label{fig.black.hole.Qe.squared.tanh}
\end{figure}

This function is plotted in \autoref{fig.black.hole.Qe.squared.tanh}. This choice of $Q_e^2\brap{\phi}$ leads to
\begin{equation}
\alpha\brap{\phi}=\tanh\bras{\brap{\phi-2}\brap{\phi+2}}\bras{1+\exp\brac{\brap{\log\lambda}\bras{1-{\phi}^2}}}\,.
\end{equation}

For notational convenience, we define the region $R:=\brap{-1,+1}$.
Then, in the limit $\lambda\rightarrow\infty$, we have
\begin{equation}
 \alpha\brap{\phi} \rightarrow \begin{cases}
                              -\infty & \text{ for }\phi\in R\\
                              2\tanh\bras{\brap{\phi-2}\brap{\phi+2}} & \text{ for }\phi\in\brac{-1,+1}\\
                              \tanh\bras{\brap{\phi-2}\brap{\phi+2}} & \text{ otherwise}\,.
                             \end{cases}
\end{equation}
In other words, for large $\lambda$, $\alpha$ is large but negative inside $R$ and tends to a constant in $\lambda$ outside of it.

The electric minima and maxima are at $-\infty,+2$ and $-2,+\infty$ respectively, all of which lie outside $R$.

\bibliographystyle{utphys}
\bibliography{ref}
\end{document}